\documentclass[aps,prc,reprint,amsmath,nofootinbib]{revtex4-1}

\usepackage[utf8]{inputenc}

\usepackage{amsmath}
\usepackage{amssymb}

\usepackage{graphicx}
\graphicspath{{../plots/}{./fig/}}

\usepackage{tikz}
\usepackage{subcaption}

\usepackage{color}
\definecolor{theblue}{RGB}{0,50,230}

\usepackage[font=small, justification=raggedright, labelsep=quad]{caption}

\usepackage[inline]{enumitem}

\usepackage{booktabs}

\usepackage[pdfencoding=auto, psdextra]{hyperref}
\hypersetup{
  colorlinks=true,
  linkcolor=theblue,
  citecolor=theblue,
  urlcolor=theblue
}

\newcommand{\trento}{T\raisebox{-0.5ex}{R}ENTo}
\newcommand{\sqrts}{\sqrt{s_\mathrm{NN}}}
\newcommand{\fmc}{\ensuremath{\text{fm}/c}}
\newcommand{\nch}{N_\text{ch}}
\newcommand{\ntrk}{N_\text{trk}^\text{offline}}
\newcommand{\vnk}[2]{v_#1\{#2\}}
\newcommand{\sigmaf}{\sigma_\text{fluct}}
\newcommand{\X}{\chi_\text{struct}}
\newcommand{\taufs}{\tau_\text{fs}}
\newcommand{\dmin}{d_\text{min}}
\newcommand{\Tsw}{T_\text{switch}}
\newcommand{\T}{\tilde{T}}
\newcommand{\x}{\mathbf x}
\newcommand{\y}{\mathbf y}
\newcommand{\z}{\mathbf z}
\newcommand{\paddedhline}{\noalign{\smallskip}\hline\noalign{\smallskip}}
\newcommand{\order}[1]{$\mathcal O(10^{#1})$}

\DeclareMathOperator{\diag}{diag}
\DeclareMathOperator{\cov}{cov}
\DeclareMathOperator{\corr}{corr}
\DeclareMathOperator{\SC}{SC}
\DeclareMathOperator{\NSC}{NSC}

\newenvironment{fullpage}{\onecolumngrid}{\clearpage\twocolumngrid}

\begin{document}

\title{
  Estimating initial state and quark-gluon plasma medium properties \\
  using a hybrid model with nucleon substructure \\
  calibrated to p-Pb and Pb-Pb collisions at \texorpdfstring{$\mathbf{\sqrts=5.02}$}{}~TeV
}

\author{J.\ Scott Moreland}
\author{Jonah E.\ Bernhard}
\author{Steffen A.\ Bass}

\affiliation{Department of Physics, Duke University, Durham, NC 27708-0305}

\date{\today}

\begin{abstract}
We posit a unified hydrodynamic and microscopic description of the quark-gluon plasma (QGP) produced in ultrarelativistic $p$-Pb and Pb-Pb collisions at $\sqrts=5.02$~TeV and evaluate our assertion using Bayesian inference. Specifically, we model the dynamics of both collision systems using initial conditions with parametric nucleon substructure, a pre-equilibrium free-streaming stage, event-by-event viscous hydrodynamics, and a microscopic hadronic afterburner.
Free parameters of the model which describe the initial state and QGP medium are then simultaneously calibrated to fit charged particle yields, mean $p_T$, and flow cumulants.
We argue that the global agreement of the calibrated model with the experimental data strongly supports the existence of hydrodynamic flow in small collision systems at ultrarelativistic energies, and that the flow produced develops at length scales smaller than a single proton.
Posterior estimates for the model's input parameters are obtained, and new insights into the temperature dependence of the QGP transport coefficients and event-by-event structure of the proton are discussed.
\end{abstract}

\maketitle

\section{Introduction}

Ultrarelativistic nuclear collisions between one light-ion and one heavy-ion, e.g.\ $^3$He-Au and $p$-Pb collisions, generate dense, compact sources of nuclear matter which produce long-range multiparticle correlations that are strikingly similar to the correlations observed in heavy-ion collisions where collectivity is commonly explained by the existence of hydrodynamic flow \cite{CMS:2012qk, Abelev:2012ola, Aad:2012gla, Adare:2015ctn}.
This observation suggests that hydrodynamic behavior could be manifest in small droplets of quark-gluon-plasma (QGP) \cite{Bozek:2011if, Bozek:2013uha}, and that flow even might develop at length scales smaller than a single proton \cite{Schenke:2014zha}.

Hydrodynamic models of ultrarelativistic nuclear collisions are complicated by a number of theoretical unknowns, including the detailed geometry of the QGP initial conditions, the strength and duration of pre-equilibrium dynamics, the temperature dependence of QGP transport coefficients, and the boundaries of hydrodynamic applicability \cite{Niemi:2014lha, deSouza:2015ena, Ollitrault:2012cm, Song:2012ua}.
In general, these theoretical uncertainties tend to grow with decreasing system size, where emergent physics at sub-fermi length scales becomes important to describe bulk properties of the produced system.

One method for reducing theoretical uncertainties is to test model calculations by varying the species of colliding nuclei at a single beam energy \cite{Adare:2015bua, Schenke:2014tga, Aidala:2018mcw, Adare:2017wlc, Adamczyk:2015obl, Shen:2016zpp, Aidala:2017ajz, Adare:2006ti}.
Since initial condition and hydrodynamic models generally factorize the structure of the colliding nuclei from the subsequent time dynamics of the collision, a single theory framework can be simultaneously tested and compared to measurements from multiple collision systems using a self-consistent set of model parameters where only the nuclear structure in the model is permitted to vary.

Typically, the macroscopic structure of heavy nuclei, characterized e.g.\ by an atomic mass number and set of Woods-Saxon coefficients \cite{MOLLER1995185, DEVRIES1987495}, is regarded as a known input to hydrodynamic models which contributes negligible uncertainty to simulation predictions, outweighed by large uncertainties in modeling initial energy deposition and off-equilibrium dynamics \cite{Niemi:2014lha, Song:2011hk, Retinskaya:2013gca, Liu:2015nwa, Kurkela:2016vts}.
The geometry of light ions, meanwhile, is naturally more sensitive to the detailed size and shape of individual protons and neutrons inside the nucleus, which may fluctuate event-by-event and differ signficantly from the round blobs typically used to approximate nucleons in heavy-ion collisions \cite{Schenke:2014zha, Welsh:2016siu, Moreland:2017kdx, Schenke:2014gaa, Schlichting:2014ipa}.
These nucleon substructure properties are difficult to measure and calculate from first principles and hence contribute significant uncertainty to model predictions of small systems.

Early substructure studies replaced round protons with composite protons, described by a few salient model parameters, in order to investigate the effect of each parameter on simulated observables \cite{Adler:2013aqf, Mitchell:2016jio, Welsh:2016siu, Broniowski:2016pvx, Bozek:2017jog}.
These sensitivity studies were able to identify cause and effect relationships between model inputs and outputs, but lacked the ability to constrain nucleon substructure parameters in any kind of global or systematic fashion.
It quickly became apparent that numerous substructure implementations might be compatible with available data, and that additional work would be required to identify observables which are particularly sensitive to the average size, shape and fluctuations of the proton.

Several such observables have been identified in proton-proton and proton-lepton scattering data.
Measurements by the TOTEM collaboration at $\sqrt{s}=7$~TeV, for instance, found an unexpected dip in the inelasticity density of $p$-$p$ collisions at zero impact parameter \cite{Antchev:2011zz}.
It was later realized that this depression, or so-called hollowness effect in the $p$-$p$ inelastic collision profile \cite{Arriola2016}, can be explained by the existence of correlated domains inside the proton, and that aspects of these domains, such as their size and correlation strength, may be constrained by comparing model predictions to inelastic $p$-$p$ measurements \cite{Albacete:2016gxu, Albacete:2016pmp}.

Independently, studies of coherent and incoherent $J/\psi$ production based on a color dipole picture of vector meson production were used to simultaneously constrain both the average color charge density of the proton as well as its event-by-event fluctuations in a saturation based framework \cite{Mantysaari:2016ykx, Mantysaari:2016jaz, Aaron:2009aa, Abramowicz:2015mha}.
Initial condition studies using the IP-Glasma model of color-glass condensate effective field theory \cite{Schenke:2012wb} simultaneously demonstrated that these color charge fluctuations leave a lasting imprint on the \mbox{small-x} gluon distribution of the proton and hence the initial geometry of QGP energy deposition \cite{Schlichting:2014ipa}.
In addition, it was recently shown that hydrodynamic simulations using IP-Glasma initial conditions with color charge fluctuations calibrated to fit coherent and incoherent $J/\psi$ diffraction measured by the H1 and Zeus experiments at HERA \cite{Aaron:2009aa, Abramowicz:2015mha} provide a good description of collectivity in small and large collision systems \cite{Schenke:2018fci}.

Model parameters, such as those calibrated by the aforementioned studies, are of course always in some degree of tension.
For instance, fitting one observable may require parameter values that degrade the quantitative description of some other observable.
Similarly, parameters which provide an optimal description of small-system observables may lead to a sub-optimal description of heavy-ion observables or \emph{vice versa}.
It is thus import to look at the experimental data holistically, and to use model calibration methods which
\begin{enumerate*}[label=(\arabic*)]
  \item explore all parameter combinations and
  \item compare model predictions to all relevant experimental measurements in a statistically rigorous fashion.
\end{enumerate*}

With these considerations in mind, we present progress towards a fully global analysis of $p$-Pb and Pb-Pb bulk observables at $\sqrts=5.02$~TeV using a model calibration framework known as Bayesian parameter estimation.
We begin, in Sec.~\ref{sec:model}, by constructing a nuclear collision model for $p$-Pb and Pb-Pb collisions using initial conditions with parametric nucleon substructure, and transport dynamics described by a pre-equilibrium free-streaming stage, viscous hydrodynamics and microscopic Boltzmann transport.
In Sec.~\ref{sec:calibration}, we calibrate free parameters of the model to fit charged particle yields, mean $p_T$, and anisotropic flow cumulants of \emph{both} collision systems at $\sqrts=5.02$~TeV, and finally, in Secs.~\ref{sec:results} and \ref{sec:summary}, we present posterior results for the model input parameters and comment on the implications for hydrodynamic descriptions of small collision systems.

\section{Nuclear collision model}
\label{sec:model}

We employ a multi-stage hybrid transport model that uses relativistic viscous hydrodynamics to describe the QGP medium and microscopic Boltzmann transport to simulate the dynamics of the system after hadronization \cite{Shen:2014vra, Bernhard:2016tnd}.
The hydrodynamic initial conditions are provided by a modified version of the \trento\ model \cite{Moreland:2014oya} with additional parameters to vary the number and size of hot spots inside the nucleon.
Each initial condition profile is free-streamed to the hydrodynamic starting time and matched onto the hydrodynamic energy-momentum tensor using the Landau matching procedure \cite{Broniowski:2008qk, Heinz:2015arc}.
Many of the components of the present model have been documented in previous studies \cite{Moreland:2014oya, Bernhard:2016tnd, Bernhard:2018hnz}; we review each component here for completeness.

\subsection{Initial state}
\label{sec:initial_state}

We model the QGP initial state in $p$-Pb and Pb-Pb collisions at $\sqrts=5.02$~TeV using a simple parametric form for boost-invariant entropy deposition employed by the \trento\ model \cite{Moreland:2014oya}.
Generally speaking, the initial three-dimensional distribution of matter produced in relativistic nuclear collisions is \emph{not} boost-invariant; longitudinal entropy deposition fluctuates both locally point-to-point in the transverse plane as well as globally event-by-event due to asymmetries in the sampled density of participant matter \cite{Ke:2016jrd, Bozek:2010vz}.
Nevertheless, boost-invariance has been shown to be a good approximation for both large and small collision systems when hydrodynamic observables are calculated from particles that are detected close to midrapidity \cite{Shen:2016zpp}.

The \trento\ model operates in the ultrarelativistic limit with a Lorentz factor $\gamma \gg 1$ such that each nucleus appears as a thin sheet of nuclear density in the laboratory frame.
The sheets of colliding nuclear density penetrate and pass through each other in proper time $\Delta \tau \approx D_\text{nucl} / (\gamma\, \beta_z)$ in the laboratory frame, where $D_\text{nucl}$ is the diameter of the nucleus in its rest frame, $\gamma$ is the usual Lorentz factor of the accelerated ions, and $\beta_z$ is their velocity along the beam axis.
The resulting nuclear overlap time $\Delta \tau \lesssim 0.1\ \fmc$ at top RHIC and LHC energies, and thus it is safe to neglect the initial transverse dynamics which occur while the nuclei pass through each other.
We therefore assume that the collision produces all secondary particles at uniform proper time $\tau = 0^+\ \fmc$, and that it deposits entropy (energy) at midrapidity which is a function of the locally varying beam integrated density of each nucleus.

Consider the collision of two protons $A$, $B$ with three-dimensional densities $\rho_{A,B}$ in their local rest frames.
The proton-proton overlap function
\begin{equation}
  \label{eq:tpp}
  T_{pp}(b) \equiv \int dx\, dy \int dz\, \rho_A(\x) \int dz\, \rho_B(\x + \mathbf{b}),
\end{equation}
describes the eikonal overlap of the two proton wave packets at fixed impact parameter $b$, where coordinates $(x, y)$ lie in the transverse plane, and $z$ is parallel to the beam axis.
Here we assume that each proton is comprised of smaller constituents---e.g.\ valence quarks, sea quarks, and small-x gluons---which may collide to produce secondary particles and contribute to the observed inelastic proton-proton cross section.

Within a picture of independent pairwise collisions between the constituents, a Glauber model model may be used to calculate the probability $P_\mathrm{coll}$ that the two protons collide inelastically at impact parameter $b$. In the limit when the number of constituents is large, it yields the particularly simple form
\begin{equation}
  \label{eq:pcoll}
  P_\mathrm{coll} = 1 - \exp[-\sigma_\mathrm{eff}\, T_{pp}(b)], \\
\end{equation}
where $\sigma_\mathrm{eff}$ is an effective cross section for pairwise inelastic collisions between the constituents, and $\sigma_{pp}^\mathrm{inel}$ is the total inelastic proton-proton cross section
\begin{equation}
  \label{eq:sigma_nn}
  \sigma_{pp}^\mathrm{inel} = \int 2 \pi b\, db\, P_\mathrm{coll}(b).
\end{equation}
The proton densities $\rho_{A,B}$ in Eq.~\eqref{eq:tpp} are commonly modeled using a spherically symmetric distribution.
For instance, the original implementation of the \trento\ model uses Gaussian protons, largely because it yields a simple analytic solution to Eq.~\eqref{eq:pcoll}.
Needless to say, such approximations are admittedly crude and may have a significant effect on the dynamics of small collision systems where the proton size is comparable to the size of the produced QGP medium.

A number of previous studies have investigated the effects of deformed or ``lumpy'' protons.
One common implementation is a superposition of three valence quarks, typically described by Gaussian or exponential form factors \cite{Welsh:2016siu, Bozek:2017jog, Schenke:2014zha, Schlichting:2014ipa, Adare:2015bua, Broniowski:2016pvx}.
The corresponding proton density $\rho(\mathbf{x})$ is then assumed to be that of predominantly small-x gluons, seeded by the distribution of color charge in each of the three valence quarks.

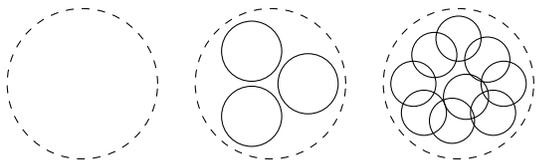
\begin{figure}
  \begin{tikzpicture}
    \draw[dashed, xshift=-2.5cm] (0,0) circle (1cm);
    \draw[dashed] (0,0) circle (1cm);
    \foreach \theta in {0, 120, 240}{
      \draw ({\theta}:.5) circle (.4cm);
    }
    \draw[dashed, xshift=2.5cm] (0,0) circle (1cm);
    \foreach \theta/\radius in {
      0/0.6, 40/0.5, 90/0.6, 130/0.5, 180/0.6, 220/0.6,
      260/0.5, 320/0.6, 300/0.2
    }{
      \draw[xshift=2.5cm] ({\theta}:\radius) circle (.3cm);
    }
  \end{tikzpicture}
  \caption{Schematic of plausible proton shapes. The sketch on the left shows a spherically symmetric proton (dashed line), while the middle and right illustrations depict a fluctuating proton with three and nine constituents respectively (solid lines).}
  \label{fig:substructure}
\end{figure}

\begin{figure}
  \begin{tikzpicture}
    \node {\includegraphics{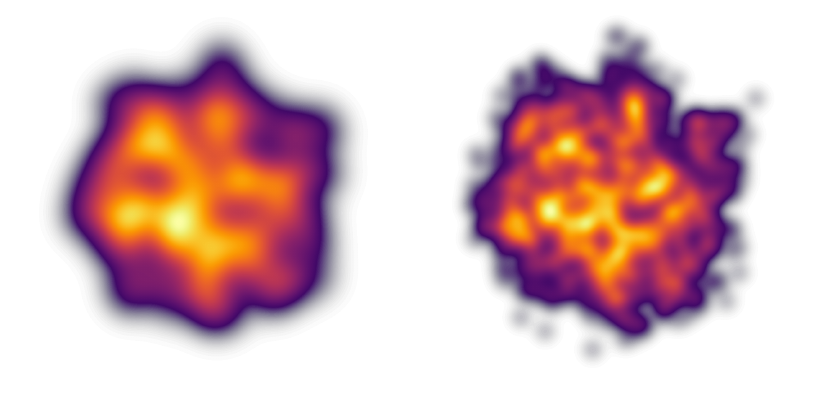}};
    \node[label=above:{10 fm}] (a) at (0, 1.1) {};
    \node (b) at (0, -1.1) {};
    \draw [<->, semithick] (a) to (b);
  \end{tikzpicture}
  \caption{\label{fig:thickness} Effect of nucleon substructure on the nuclear thickness function $T(x, y) \equiv \int dz\, \rho(x, y, z)$ of a $^{208}\mathrm{Pb}$ nucleus. The nucleus on the left has Gaussian nucleons of width $0.8$~fm, while the nucleus on the right has composite nucleons, each containing six constituents of width $0.4$~fm.}
\end{figure}
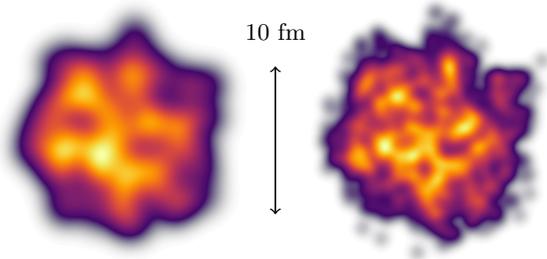

In this work, we pursue a less restrictive and more parametric description of the proton where the number of substructure degrees of freedom are uncertain as depicted in Fig.~\ref{fig:substructure}.
We model each nucleon's density $\rho_{A,B}$ as a sum of $n_c$ independent constituents
\begin{equation}
  \label{eq:rho}
  \rho_{A, B}(\x) = \frac{1}{n_c} \sum\limits_{i=1}^{n_c} \rho_c(\mathbf{x} - \mathbf{x_i}),
\end{equation}
where each constituent density $\rho_c$ is described by a Gaussian distribution of width $v$
\begin{equation}
  \label{eq:constituent_density}
  \rho_c(\mathbf{x}) = \frac{1}{(2\pi v^2)^{3/2}} \exp\left(-\frac{\x^2}{2 v^2}\right),
\end{equation}
and each constituent's position $\mathbf{x_i}$ in Eq.~\eqref{eq:rho} is sampled from a Gaussian radial distribution with standard deviation $r$.
The effect of this additional nucleon substructure on the nuclear thickness functions is visible in Fig.~\ref{fig:thickness}

The two protons $A$, $B$ are assigned a random impact parameter, and Eq.~\eqref{eq:pcoll} is used to sample their inelastic collision probability $P_\mathrm{coll}(b)$.
Note that this proton-proton inelastic collision probability has no direct knowledge of the individual constituent degrees of freedom; it is only \emph{indirectly} sensitive through the geometry of $\rho_{A, B}$ which depends on each of the constituent positions.
This is an important distinction between the present model and a similar nucleon substructure implementation known as the participant or ``wounded'' quark model which allows for a subset of quarks (constituents) to participate inside a single nucleon \cite{ANISOVICH1978477, Broniowski:2016pvx}.
The proton, unlike the nucleus, cannot produce semi-stable spectator fragments in a high-energy collision.
Any spectator quarks produced by a wounded quark model would be colored objects that necessarily contribute to secondary particle production as they fragment and recombine to form color-neutral hadrons.
We correspondingly require that the nucleons in Eq.~\eqref{eq:rho} participate as singular objects, such that all spectator matter discarded by the simulation is appropriately color-neutral and inert.

Assuming our two protons collide at the sampled impact parameter $\mathbf{b}$, we assign each a \emph{fluctuated} thickness
\begin{equation}
  \label{eq:fluctuated_thick}
  \T_{A, B}(\x) \equiv \int dz\, \frac{1}{n_c} \sum\limits_{i=1}^{n_c} \gamma_i\, \rho_c \,(\mathbf{x} - \mathbf{x_i} \pm \mathbf{b}/2),
\end{equation}
equal to the beam-integrated proton density in Eq.~\eqref{eq:rho}, with each constituent shifted by the appropriate impact parameter offset, and multiplied by a gamma random variable $\gamma_i$ with unit mean and variance $1/k$.
These \emph{ad hoc} gamma random weights are necessary to describe the large fluctuations observed in minimum bias proton-proton collisions, although their exact physical origin is not well understood.

Within the eikonal approximation, the initial entropy deposited at midrapidity and proper time $\tau=0^+\ \fmc$ is some function
\begin{equation}
  f: \T_A, \T_B \mapsto \frac{dS}{d^2x_\perp d\eta} \bigg\vert_{\eta=0},
\end{equation}
of the local density of participant matter $\T_A$, $\T_B$ in each nucleus.
A natural first guess for this mapping is the arithmetic mean
\begin{equation}
  \frac{dS}{d^2x_\perp d\eta}\,\bigg\vert_{\eta=0} \propto \frac{\T_A + \T_B}{2},
\end{equation}
which is simply a wounded nucleon model up to meaningless factor of two in the normalization.
The wounded nucleon model was in fact one of the first such mappings used as a proxy for initial particle production and entropy deposition in relativistic heavy-ion collisions \cite{Bialas:1976ed}.
It was quickly realized, however, that the wounded nucleon model predicts the wrong scaling for charged particle production as a function of collision centrality and hence the wrong scaling for initial entropy deposition as a function of participant thicknesses $\T_A$ and $\T_B$ \cite{Kharzeev:2000ph}.

A simple remedy is to replace the arithmetic mean of the wounded nucleon model with a more flexible parametrization
\begin{equation}
  \label{eq:gmean}
  \frac{dS}{d^2x_\perp d\eta}\,\bigg\vert_{\eta=0} \propto \bigg( \frac{\T_A^p + \T_B^p}{2} \bigg)^{1/p},
\end{equation}
based on a family of functions known as the generalized mean(s).
This parametrization introduces a dimensionless parameter $p$ which varies the scaling behavior of initial entropy deposition at midrapidity.
For certain discrete values of $p$, it reduces to well known functional forms such as the arithmetic, geometric, and harmonic means:
\begin{equation}
  \newlength{\extraspace}
  \setlength{\extraspace}{0.5ex}
  \frac{dS}{d\eta}\bigg\vert_{\eta=0} \propto
  \begin{cases}
    \max(T_A, T_B) & p \rightarrow +\infty, \\[\extraspace]
    (T_A + T_B)/2 & p = +1, \hfill \text{ (arithmetic)} \\[\extraspace]
    \sqrt{T_A T_B} & p = 0, \hfill \text{ (geometric)} \\[\extraspace]
    2\, T_A T_B/(T_A + T_B) & p = -1, \hfill \text{ (harmonic)} \\[\extraspace]
    \min(T_A, T_B) & p \rightarrow -\infty.
  \end{cases}
  \label{eq:trento_p}
\end{equation}
Conveniently, it has been shown that the generalized mean ansatz is able to mimic the scaling behavior of certain saturation based initial condition models \cite{Bernhard:2016tnd}, and hence it should serve as a reasonable parametric form for exploring QGP entropy deposition, assuming imperfect knowledge of saturation effects in nature.
Of course the model is \emph{not} a substitute for first principle theory calculations, and it may fail to reproduce nuanced features of \emph{ab initio} models such as the existence of short-range gluon field fluctuations \cite{Schenke:2012wb}.

Equation~\eqref{eq:gmean} is a purely local function of nuclear density in the transverse plane and should (in principle) be equally valid for any pair of colliding nuclei at sufficiently high beam energy.
The model readily generalizes from individual proton-proton collisions to arbitrary nucleus-nucleus collisions by summing the participant thicknesses $\T_{A,B}$ over all nucleons which participate in one or more inelastic collisions.
The only modeling difference between $p$-$p$, $p$-Pb, and Pb-Pb collisions is the number and the position of the nucleons.

When applying the model to heavy-ions, we sample nucleon positions from a Woods-Saxon density distribution subject to a minimum distance criteria $|\mathbf{x_i} - \mathbf{x_j}| > d_\mathrm{min}$ between all pairs of nucleons $i$, $j$.
The minimum distance algorithm, first described in Ref.~\cite{Bernhard:2018hnz}, uses a simple trick to resample the nucleon positions without modifying the target Woods-Saxon radial distribution.
We first presample the radii of all nucleons in a given nucleus and sort them in ascending order.
We then sample the solid angles of each nucleon one-by-one, starting with the nucleon closest to the center of the nucleus and working our way outwards.
If a sampled nucleon position is too close to any of its previously placed neighbors, its solid angle is resampled until the minimum distance criteria is satisfied.
Similar methods could be used to model correlations between individual constituents inside each nucleon, although the numerical implementation would be somewhat tedious.

\subsection{Pre-equilibrium dynamics}

\begin{figure}[b]
  \includegraphics{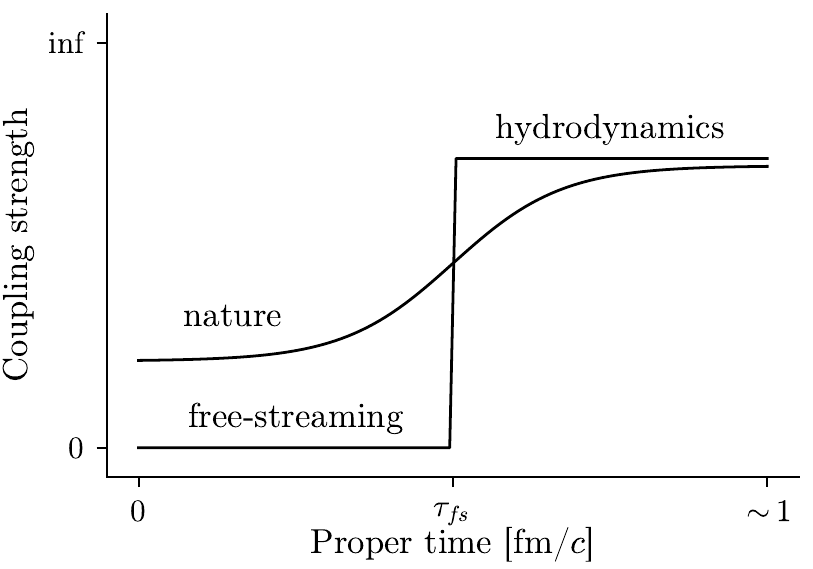}
  \caption{Cartoon of the free-streaming approximation for hydrodynamic initialization. The initial state is free-streamed for proper time $\taufs$ (zero coupling) before it is matched to hydrodynamics (strong coupling). This piecewise evolution approximates the more realistic scenario expected in nature where the medium's coupling strength smoothly changes as a function of time.}
  \label{fig:coupling}
\end{figure}

There are of course two limiting cases for the strength of interactions inside the QGP medium immediately at the collision: infinitely weak coupling where the secondary partons free-stream without interacting, and infinitely strong coupling where the fluid's inter-particle mean free path effectively vanishes.
Realistically, one expects the initial parton interactions to lie somewhere between these two extremes.
We therefore choose to model the QGP's initial off-equilibrium dynamics using a simple step-function approximation, depicted in Fig.~\ref{fig:coupling}, which free-streams the initial state for proper time $\taufs$ (zero coupling) before instantaneously switching to viscous hydrodynamics (strong coupling) \cite{Liu:2015nwa, Broniowski:2008qk}.
The free parameter $\taufs$ allows us to parametrically vary the \emph{time averaged} coupling strength in the approximate window $0 < \taufs \lesssim1\ \fmc$.

The parametric entropy deposition ansatz in Eq.~\eqref{eq:gmean} does not provide any information about the initial masses or momenta of particles produced in the collision.
In general, these details will affect the dynamics predicted by the collisionless Boltzmann equation
\begin{equation}
  \label{freestream}
  p^\mu \partial_\mu f(x, p) = 0,
\end{equation}
through its dependence on the underlying distribution function $f(x, p)$, and hence are necessary inputs for any free-streaming implementation.
Equation~\eqref{freestream}, however, simplifies for massless partons with momentum distributions that are locally isotropic.
Under this assumption, it was shown in Refs.~\cite{Broniowski:2008qk, Liu:2015nwa} that the energy-momentum tensor $T^{\mu\nu}$ of partons at midrapidity and time $\tau$ only depends on its spatial distribution at some earlier time $\tau_0$, but not its $p_\perp$-distribution, which could vary as a function of position.

The entropy density for a gas of massless noninteracting particles is very nearly proportional to its particle density, and so we are able to recast the parametrization in Eq.~\eqref{eq:gmean} in the form
\begin{equation}
  \frac{dN}{d^2x_\perp d\eta}\,\bigg\vert_{\eta=0} = \, \text{Norm} \times \bigg(\frac{\T_A^p + \T_B^p}{2}\bigg)^{1/p},
\end{equation}
where the left-side is the initial density of free-streaming partons at midrapidity.
The resulting free-streamed energy-momentum tensor $T^{\mu\nu}(x, y, \eta, \tau)$ at transverse coordinate $(x, y)$, midrapidity $\eta=0$, and proper time $\tau > \tau_0$ is then given by
\begin{multline}
  \label{energy-momentum}
  T^{\mu\nu}(x, y, 0, \tau) = \frac{1}{\tau} \int_{0}^{2\pi} d\phi\, n(x - \Delta \tau \cos \phi, y - \Delta \tau \sin \phi) \\ \times
  \begin{bmatrix}
    1 & \cos\phi & \sin\phi  & 0\\
    \cos\phi & \cos^2\phi & \cos\phi\sin\phi & 0 \\
    \sin\phi & \sin\phi \cos\phi & \sin^2\phi & 0\\
    0 & 0 & 0 & 0
  \end{bmatrix},
\end{multline}
where $n$ is the local density of massless partons, and $\Delta\tau$ is the free-streaming time $\Delta\tau = \tau - \tau_0$.
The solution \eqref{energy-momentum} can then be decomposed in hydrodynamic form
\begin{equation}
  \label{hydro-eqn}
  T^{\mu\nu} = e u^\mu u^\nu - (P + \Pi) \Delta^{\mu\nu} + \pi^{\mu\nu},
\end{equation}
where $e$ and $P$ are the energy density and pressure in the local fluid rest frame, $u^\mu$ is the local fluid velocity, ${\Delta^{\mu\nu} \equiv g^{\mu\nu} - u^\mu u^\nu}$ is the projector onto the space orthogonal to $u^\mu$, and $\Pi$ and $\pi^{\mu\nu}$ are the bulk pressure and shear stress tensor respectively.
We then solve for the energy density $e$ and fluid velocity $u^\mu$ using the Landau matching condition which defines the fluid rest frame velocity as the time-like eigenvector of $T^{\mu\nu}$ with energy density $e$ as its eigenvalue,
\begin{equation}
  T^{\mu\nu} u_\nu = e u^\mu.
\end{equation}
The initial bulk and shear corrections are finally solved for by subtracting the ideal pressure from the total pressure to find $\Pi$, then solving for $\pi^{\mu\nu}$ using Eq.~\eqref{hydro-eqn}
\begin{align}
  \Pi &= -\frac{1}{3} \mathrm{Tr}(\Delta_{\mu\nu} T^{\mu\nu}) - P,\\
  \pi^{\mu\nu} &= T^{\mu\nu} - e u^\mu u^\nu + (P + \Pi) \Delta^{\mu\nu}.
\end{align}

This procedure provides initial values for $T^{\mu\nu}$, $u^\mu$, $\Pi$, and $\pi^{\mu\nu}$ which conserve energy and are consistent with the underlying hydrodynamic equation of state.
We therefore expect it to provide a more realistic description of the initial stages of the collision as compared to a previous study using the \trento\ initial condition model which set $\Pi$, $\pi^{\mu\nu}$ and $u^\mu$ initially to zero \cite{Bernhard:2016tnd}.

\subsection{Hydrodynamics}

After free-streaming for proper time $\taufs$, we transition to viscous hydrodynamics which solves the conservation equations
\begin{equation}
  \label{eq:continuity}
  \partial_\mu T^{\mu\nu} = 0
\end{equation}
for the hydrodynamic energy-momentum tensor $T^{\mu\nu}$ expressed in Eq.~\eqref{hydro-eqn} using a set of second-order Israel-Stewart equations formulated in the 14-moment approximation
\cite{Israel:1979wp, Israel:1976aa, Denicol:2012cn, Denicol:2010xn}.
This produces a pair of relaxation-type equations
\begin{subequations}
  \label{eq:relaxation}
  \begin{align}
    \tau_\Pi \Pi + \dot{\Pi} &=
      - \zeta \theta - \delta_{\Pi\Pi} \Pi\theta
      + \lambda_{\Pi\pi} \pi^{\mu\nu} \sigma_{\mu\nu}, \\[1ex]
    \tau_\pi \dot{\pi}^{\langle \mu\nu \rangle} + \pi^{\mu\nu} &=
      2\eta\sigma^{\mu\nu} - \delta_{\pi\pi} \pi^{\mu\nu} \theta
      + \phi_7 \pi_\alpha^{\langle \mu} \pi^{\nu \rangle \alpha} \nonumber \\
      &\qquad {} - \tau_{\pi\pi} \pi_\alpha^{\langle \mu}\sigma^{\nu \rangle \alpha}
      + \lambda_{\pi\Pi} \Pi \sigma^{\mu\nu},
  \end{align}
\end{subequations}
for the bulk pressure $\Pi$ and shear-stress $\pi^{\mu\nu}$.
We model the shear viscosity $\eta$ and bulk viscosity $\zeta$ as unknown temperature dependent quantities and fix the remaining transport coefficients $\{\tau_\Pi, \delta_{\Pi\Pi}, \lambda_{\Pi\pi}, \tau_\pi, \delta_{\pi\pi}, \phi_7, \tau_{\pi\pi}, \lambda_{\pi\Pi}\}$ using analytic results derived in the limit of small but finite masses \cite{Denicol:2014vaa}.

\begin{figure}[t]
  \includegraphics{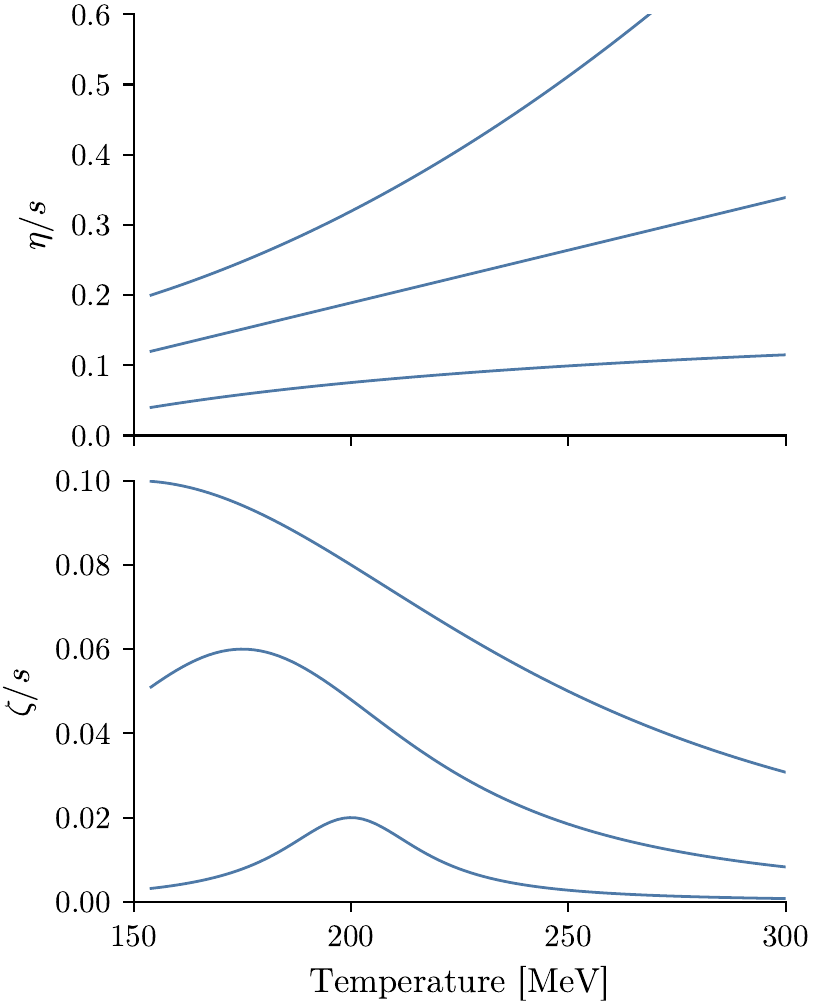}
  \caption{Degrees of freedom in the temperature dependent shear and bulk viscosity parametrizations. Lines are chosen for illustrative purposes only and do not represent all possible variability. For instance, $\eta/s$ could have a large slope and negative curvature, or $\zeta/s$ could have a large max and narrow width, neither of which are depicted above.}
  \label{fig:viscosity_dof}
\end{figure}

The hydrodynamic equations of motion are necessarily closed using an equation of state (EoS) to relate the energy density $e$ and pressure $P$ of the produced medium.
We use a parametrization for $P(e)$ that matches a hadron resonance gas EoS at low temperature to a lattice QCD EoS at high temperature by smoothly connecting their trace anomalies in the interval
$165 \le T \le 200$~MeV \cite{Bernhard:2018hnz}.
For the lattice EoS, we use a calculation by the HotQCD collaboration for (2+1)-flavor QCD which was extrapolated to the continuum limit \cite{Bazavov:2014pvz}. Recent developments in lattice QCD now enable calculations in (2+1+1)-flavors \cite{Borsanyi:2016ksw}, i.e.\ with thermalized charm quarks, and the additional charm flavor has been shown to visibly affect predictions of $p_T$-differential flow observables \cite{Noronha-Hostler:2018zxc}.
Investigating this sensitivity would thus be a natural target for future improvements to the present work.

We parametrize the temperature dependence of the QGP viscosities in order to marginalize over their uncertainty when calibrating to data.
For the specific shear viscosity $\eta/s$, we use a modified linear ansatz
\begin{equation}
  \label{eq:shear_viscosity}
  (\eta/s)(T) = (\eta/s)_\mathrm{min} + (\eta/s)_\mathrm{slope}\cdot(T - T_c)\cdot(T/T_c)^{(\eta/s)_\mathrm{crv}},
\end{equation}
where $\eta/s$ min, slope, and curvature are tunable parameters, and $T_c=0.154$~GeV is the pseudocritical transition temperature of the HotQCD EoS.
While for the specific bulk viscosity $\zeta/s$, we use an unnormalized Cauchy distribution
\begin{equation}
  \label{eq:bulk_viscosity}
  (\zeta/s)(T) = \frac{(\zeta/s)_\mathrm{max}}{1 + \left(\dfrac{T - (\zeta/s)_{T_0}}{(\zeta/s)_\mathrm{width}}\right)^2},
\end{equation}
described by tunable maximum, width, and location ($T_0$) parameters.
Figure~\ref{fig:viscosity_dof} shows several of the possible curves parametrized by Eqs.~\eqref{eq:shear_viscosity} and \eqref{eq:bulk_viscosity}, although many more are possible.

The aforementioned hydrodynamic equations are solved numerically using the boost-invariant VISH2+1 viscous hydrodynamics code \cite{Song:2007ux, Shen:2014vra}.
We simulate each hydrodynamic event on a spacetime grid with transverse extent $x_\mathrm{max}$, spatial grid step $dx$, and time step $d\tau$ which are optimized event-by-event to balance trade-offs between numerical accuracy and computation time (see Appendix~\ref{app:adaptive_grid}).
Although these details are somewhat mundane, they are critically important to the present study, since the computation time scales with the number of spacetime cells $n_x^2\, n_\tau$, and $n_x$ and $n_\tau$ typically have to be quite large to resolve the small length scales associated with nucleon substructure.

\subsection{Particlization and Boltzmann transport}

We evolve the system hydrodynamically down to a pre-specified switching isotherm $T_{sw}$ at which point the medium is converted into particles using the Cooper-Frye formula \cite{PhysRevD.10.186}
\begin{equation}
  \label{cooper-frye}
  E \frac{dN_i}{d^3p} = \frac{g_i}{(2\pi)^3} \int_\Sigma f_i(x, p)\, p^\mu\, d^3\sigma_\mu,
\end{equation}
where $i$ is an index over species, $f_i$ is the distribution function of that species, and $d^3\sigma_\mu$ is a volume element of the isothermal hypersurface $\Sigma$ defined by $\Tsw$.
Thermal particles are then sampled in the rest frame of each fluid cell according to a Bose-Einstein or Fermi-Dirac distribution at zero baryon chemical potential
\begin{equation}
  \label{distribution}
  f(m, p) = \frac{1}{\exp(\sqrt{m^2 + p^2}/T) \mp 1},
\end{equation}
where $m$ is the mass of the sampled particle, $p$ is its momentum, and $T$ is the temperature of the fluid cell.

Traditionally, particlization models have sampled resonances using each particle's pole mass in Eq.~\eqref{distribution}.
This approximation, however, is somewhat crude and has been known to underpredict pion production, particularly at low $p_T$ \cite{Sollfrank:1991xm, Huovinen:2016xxq, Vovchenko:2018fmh}.
We thus follow Ref.~\cite{Bernhard:2018hnz}, and sample particles with a \emph{distribution} of masses
\begin{equation}
  f(p) = \int dm\, \mathcal{P}(m)\, f(m, p),
\end{equation}
where $\mathcal{P}(m)$ is modeled by a Breit-Wigner distribution
\begin{equation}
  \mathcal{P}(m) \propto \frac{\Gamma(m)}{(m - m_0)^2 + \Gamma(m)^2/4}.
\end{equation}
Here $m_0$ is the resonance's Breit-Wigner mass, and $\Gamma(m)$ is its mass-dependent width, for which we use a simple form:
\begin{equation}
  \Gamma(m) = \Gamma_0 \sqrt{\frac{m - m_\mathrm{min}}{m_0 - m_\mathrm{min}}},
\end{equation}
where $\Gamma_0$ is the usual Breit-Wigner width, and $m_\mathrm{min}$ is a production threshold equal to the total mass of the lightest decay products.
We tabulate the values of $\{\Gamma_0, m_0, m_\mathrm{min}\}$ for all particles and sample the masses of each particle during particlization \cite{PDG:2017}.
The resonances are then passed to a hadronic transport model, described shortly, which simulates subsequent scatterings and decays.

When the viscous terms $\pi^{\mu\nu}$ and $\Pi$ are nonzero in Eq.~\eqref{hydro-eqn}, the distribution function $f$ must be modified to preserve the continuity of $T^{\mu\nu}$ as the system transitions from hydrodynamics to Boltzmann transport.
We perform the appropriate modification using a general method which transforms the momentum vector \emph{inside} the distribution function \cite{Pratt:2010jt}
\begin{align}
  \label{viscous_correction}
  p_i \rightarrow p'_i &= p_i + \sum\limits_j \lambda_{ij}\, p_j,\\
  \lambda_{ij} &= (\lambda_\mathrm{shear})_{ij} + \lambda_\mathrm{bulk}\, \delta_{ij},
\end{align}
where $\lambda_{ij}$ is a linear transformation matrix consisting of a traceless shear part and a bulk part which is proportional to the identity matrix.

We use for the shear viscous correction the form \cite{Pratt:2010jt}
\begin{equation}
  (\lambda_\mathrm{shear})_{ij} = \frac{\tau}{2 \eta} \pi_{ij},
\end{equation}
with a value for $\eta/\tau$ obtained from the noninteracting hadron resonance gas model
\begin{equation}
  \frac{\eta}{\tau} = \frac{1}{15 T} \sum\limits_{sp} g \int \frac{d^3p}{(2\pi)^3}\frac{p^4}{E^2} f_0 (1 \pm f_0),
\end{equation}
where the sum runs over all species in the hadron gas, and $g$ and $f_0$ are the degeneracy factor and equilibrium distribution function of each species respectively.

For the bulk viscous correction, we use a novel procedure developed in Ref.~\cite{Bernhard:2018hnz}.
The total kinetic pressure of the system is
\begin{equation}
  \label{kinetic-pressure}
  P + \Pi = \sum\limits_\mathrm{sp} g \int \frac{d^3p}{(2\pi)^3} \frac{p^2}{3E} f(p).
\end{equation}
For a given bulk pressure, we rescale the momentum $p$ inside the distribution function $f(p) \rightarrow f(p + \lambda_\mathrm{bulk}\, p)$ and adjust the parameter $\lambda_\mathrm{bulk}$ to match the total pressure on the left side of Eq.~\eqref{kinetic-pressure}.
This substitution of course also modifies the energy density
\begin{equation}
  e = \sum\limits_\mathrm{sp} g \int \frac{d^3p}{(2 \pi)^3} E f(p),
\end{equation}
and so a fugacity term $z_\mathrm{bulk}$ is introduced which modifies the yield of all particles by the same overall factor to compensate.
The full transformation is then given by $f(p) \rightarrow z_\mathrm{bulk}\, f(p + \lambda_\mathrm{bulk}\, p)$, where the parameters $\lambda_\mathrm{bulk}$ and $z_\mathrm{bulk}$ are determined numerically for each value of the bulk pressure.

Once the fluid is converted into particles, we simulate its subsequent microscopic dynamics using the Ultra-relativistic Quantum modifies Dynamics (UrQMD) transport model \cite{Bass:1998ca, Bleicher:1999xi}.
It solves the microscopic Boltzmann equation
\begin{equation}
  \frac{df_i(x, p)}{dt} = \mathcal{C}_i(x, p)
\end{equation}
where $f_i$ is the distribution function for species $i$, and $\mathcal{C}_i$ is its microscopic collision kernel.
The model propagates all produced hadrons along classical trajectories, and simulates their scatterings, resonance formations and decays until the last interactions cease.

One primary advantage of using a microscopic transport model such as UrQMD as an afterburner, is that it realistically simulates the system break-up when the mean free path becomes large relative to the system size.
This dilute limit is expected to play a significant role in small collision systems where the produced medium is smaller and shorter lived.

\section{Parameter estimation}
\label{sec:calibration}

The nuclear collision model constructed in Sec.~\ref{sec:model} includes a number of free parameters $\x$ which describe the initial state, pre-equilibrium dynamics, and hydrodynamic medium.
Given values for the parameters $\x$, the model may be used to predict a vector of simulated observables $\y_m$.
For example, $\y_m$ might be a vector consisting of charged particle yields in different centrality bins.
The physics model thus represents a vector-valued function $f(\x) = \y_m$ which maps the parameter values $\x$ to the simulated observables $\y_m$.

The goal of this work is to estimate the true model parameters $\x_\star$ provided some evidence that our model predictions $\y_m$ describe experimental measurements $\y_e$.
The problem involves three distinct components:
\begin{enumerate}[itemsep=0pt, leftmargin=2\parindent]
  \item $H_f$: the hypothesis that the nuclear collision model $f$ formulated in this work provides a realistic description of reality,
  \item $H_\x$: the hypothesis that the model parameters $\x$ are the true model parameters $\x_\star$ of $f$, and
  \item $E$: the evidence provided by the experimental data $\y_e$ and its corresponding uncertainties.
\end{enumerate}
As a practical matter, we always assume that hypothesis $H_f$ is correct, meaning there are no glaring flaws in our chosen theoretical framework.
This is a significant assumption; the application of hydrodynamic simulations to small collision systems is speculative, and our conclusions are conditional on the framework making sense.

Subject to this assumption, we can apply Bayes' theorem to evaluate the hypothesis $H_\x$ for the true model parameters,
\begin{equation}
  \label{eq:bayes}
  P(H_\x | E) \propto P(E | H_\x)\, P(H_\x).
\end{equation}
The left-side of this expression is the \emph{posterior}: the probability for hypothesis $H_\x$ given the experimental evidence $E$.
On the right-side there are two separate terms.
The first term $P(E | H_\x)$ is the \emph{likelihood} function: the probability of observing the experimental evidence $E$ given our model and the hypothesis $H_\x$ for the true model parameters $\x_\star$, and the second term $P(H_\x)$ is the \emph{prior}: an estimate of the probability of hypothesis $H_\x$ in the absence of evidence $E$.

We assume that the likelihood function in Eq.~\eqref{eq:bayes} is described by a multivariate normal distribution
\begin{equation}
  \label{eq:likelihood}
  P(E | H_\x) = \frac{1}{\sqrt{(2\pi)^m \det \Sigma}} \exp \left ( -\frac{1}{2}\Delta\y^\intercal \Sigma^{-1} \Delta\y \right ),
\end{equation}
where $\Delta\y = \y_m(\x) - \y_e$ is the discrepancy of the model and experiment, and $\Sigma = \Sigma_m(\x) + \Sigma_e$ is the \emph{total} covariance matrix, equal to the sum of a modeling component $\Sigma_m(\x)$ and an experimental component $\Sigma_e$ which account for all known sources of uncertainty in the simulated and measured observables.

\begin{table}[t]
  \caption{Input parameter ranges for the physics model.}
  \begin{ruledtabular}
  \begin{tabular}{lll}
    Parameter         & Description                        & Range             \\
    \paddedhline
    Norm              & Normalization factor                 & 9--28           \\
    $p$               & Entropy deposition parameter         & $-1$ to $+1$    \\
    $\sigmaf$         & Nucleon fluctuation std.\ dev.\      & 0--2            \\
    $w$               & Nucleon width parameter              & 0.4--1.2 fm     \\
    $n_c$             & Number of nucleon constituents       & 1--9            \\
    $\X$              & Nucleon structure parameter          & 0--1            \\
    $\dmin$           & Minimum inter-nucleon distance       & 0--1.7 fm       \\
    $\taufs$          & Free-streaming time                  & 0.1--1.5 \fmc   \\
    $\eta/s$ min      & Minimum value of $\eta/s$ (at $T_c$) & 0--0.2          \\
    $\eta/s$ slope    & Slope of $\eta/s$ above $T_c$        & 0--8 GeV$^{-1}$ \\
    $\eta/s$ crv      & Curvature of $\eta/s$ above $T_c$    & $-1$ to $+1$    \\
    $\zeta/s$ max     & Maximum value of $\zeta/s$           & 0--0.1          \\
    $\zeta/s$ width   & Width of $\zeta/s$ peak              & 0--0.1 GeV      \\
    $\zeta/s$ $T_0$   & Temperature of $\zeta/s$ maximum     & 150--200 MeV    \\
    $\Tsw$            & Switching/particlization temp.       & 135--165 MeV    \\
  \end{tabular}
  \end{ruledtabular}
  \label{tab:design}
\end{table}

\subsection{Parameter design}

\begin{table*}
  \caption{
    \label{tab:observables}
    Experimental data used to calibrate the model.
  }
  \begin{ruledtabular}
  \begin{tabular}{cc}
    Pb-Pb $\sqrts=5.02$~TeV & $p$-Pb $\sqrts=5.02$~TeV \\
    \paddedhline
    Charged-particle multiplicity $d\nch/d\eta$, $|\eta| < 0.5$ \cite{Adam:2015ptt} & Charged-particle multiplicity $d\nch/d\eta$, $|\eta| < 1.4$ \cite{Adam:2014qja} \\
    \noalign{\smallskip}
  Two-particle flow cumulants  $\vnk{n}{2}$ for $n=2,3,4$, $|\eta| < 0.8$,  & Two-particle flow cumulants $\vnk{n}{2}$ for $n=2,3$, $|\eta| < 2.4$, \\
    charged-particles, $|\Delta\eta| > 1$,\, $0.2 < p_T < 5.0$~GeV \cite{Adam:2016izf} & charged-particles, $|\Delta\eta| > 2$,\, $0.3 < p_T < 3.0$~GeV \cite{Chatrchyan:2013nka}\\
    \noalign{\smallskip}
    & Charged-particle mean $p_T$, $0.15 < p_T < 10$~GeV, $|\eta| < 0.3$ \cite{Abelev:2013bla}\\
  \end{tabular}
  \end{ruledtabular}
\end{table*}

For the prior $P(H_\x)$, we specify ranges, i.e.\ minimum and maximum values, for each parameter which are listed in Table~\ref{tab:design}.
We assume the prior distribution is constant and nonzero within each specified range and zero otherwise.
The selected parameter ranges are intentionally wide to avoid clipping the calibrated posterior; for example, a previous analysis of the \trento\ model \cite{Bernhard:2016tnd} found $p \sim 0$, but we use a prior range $p \in [-1, 1]$ to account for differences in the present model, e.g.\ nucleon substructure, which could modify its posterior.
Several of the model parameters require special care and are reparametrized accordingly:

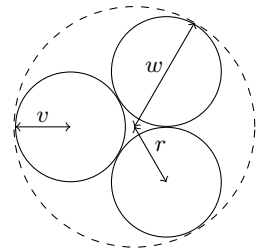
\begin{figure}
  \begin{tikzpicture}
    \draw[dashed] (0,0) circle (1.6cm);
    \foreach \theta in {60, 180, 300}{
      \draw ({\theta}:.85) circle (.73cm);
    }
    \coordinate (a) at (0, 0);
    \coordinate (b) at (.8, 1.385);
    \coordinate (c) at (.425, -.736);
    \coordinate (d) at (-1.58, 0);
    \coordinate (e) at (-.85, 0);
    \draw[<->] (a) -- (b) node[midway, xshift=-1ex, yshift=.8ex]{$w$};
    \draw[<->] (a) -- (c) node[midway, xshift=1ex, yshift=.8ex]{$r$};
    \draw[<->] (d) -- (e) node[midway, yshift=1ex]{$v$};
  \end{tikzpicture}
  \caption{
    \label{fig:nucleon_schematic}
    Schematic illustrating the constituent sampling radius $r$, constituent width $v$, and nucleon width parameter $w = \sqrt{r^2 + v^2}$ for a nucleon with $n_c=3$ constituents.
  }
\end{figure}

\begin{itemize}[leftmargin=1\parindent]
  \item
    The constituent fluctuations, modeled by the gamma random weights $\gamma_i$ in Eq.~\eqref{eq:fluctuated_thick}, generate overall nucleon fluctuations which are suppressed by the number of constituents $n_c$ inside the nucleon.
    The observed nucleon fluctuation variance falls like $1 / n_c$ which means the natural range for the constituent fluctuations is larger when the number of constituents is larger and \emph{vice versa}.
    We therefore reparametrize the constituent fluctuations using the standard deviation of the resulting nucleon fluctuations
    \begin{equation}
      \sigma_\mathrm{fluct} = 1 / \sqrt{k\, n_c},
    \end{equation}
    where $k$ is the shape parameter of the gamma distribution used to fluctuate each individual constituent.
  \item
    In Sec.~\ref{sec:initial_state} we parametrized nucleon substructure using three degrees of freedom:
    \begin{enumerate*}[label=(\roman*)]
      \item a parameter $v$ for the Gaussian width of each constituent,
      \item a parameter $r$ for the Gaussian width of the radial distribution used to sample the constituent centers, and
      \item a parameter $n_c$ to vary the number of constituents inside the nucleon.
    \end{enumerate*}
    In the limit $n_c \to \infty$, the composite nucleon's root-mean-square (RMS) radius is simply the convolution of its sampling radius $r$ and constituent width $v$ which add together in quadrature
    \begin{equation}
      \lim_{n_c \to\infty} r_\mathrm{RMS}\{\rho\} = \sqrt{r^2 + v^2}.
    \end{equation}
    We therefore choose to reparametrize the sampling radius $r$ in terms of a new variable
    \begin{equation}
      \label{eq:nucleon_width}
      w = \sqrt{r^2 + v^2},
    \end{equation}
    which approximates the RMS radius of the nucleon when the number of sampled constituents is large (see Fig.~\ref{fig:nucleon_schematic} for an example proton).
    We thus call $w$ a nucleon ``width'', although for smaller numbers of constituents, the \emph{actual} RMS radius of our sampled nucleons can be significantly smaller than our width parameter $w$ due to fluctuations in the nucleon's center of mass, and hence one should account for the difference when discussing the nucleon's posterior RMS radius.
  \item
    Equation~\eqref{eq:nucleon_width} requires the nucleon width to be larger than the constituent width, i.e. $w > v$, lest the sampling radius $r$ turn imaginary, and thus we cannot vary the nucleon width $w$ and the constituent width $v$ independently.
    We therefore reparametrize the constituent width $v$ using a new variable $\X$ to interpolate between minimum and maximum allowed values:
    \begin{equation}
      \label{eq:struct_param}
      v = v_\mathrm{min} + \X (v_\mathrm{max} - v_\mathrm{min}).
    \end{equation}
    Here we choose a minimum constituent width $v_\mathrm{min} = 0.2$~fm determined by computational limits and a maximum width $v_\mathrm{max} = w$ equal to the nucleon width.
    Thus for $\X=0$, the nucleons consist of small, distinct hot spots, whereas for $\X=1$ the nucleon is a single Gaussian blob of width $w$.
    The parameter $\X$ thus varies the granularity of the nucleon while keeping the number of constituents $n_c$ and approximate size of the nucleon $w$ fixed.
\end{itemize}

\subsection{Observables}
\label{sec:observables}

The likelihood function \eqref{eq:likelihood} provides evidence for \mbox{(or against)} the model parameters $\x$ by comparing the model predictions $\y_m(\x)$ to experimental data $\y_e$.
We focus on simple experimental observables in the present study which are sensitive to the bulk properties of the produced medium.
We calculate for each set of model parameters the following observables at midrapidity:
\begin{itemize}[leftmargin=1\parindent, itemsep=0pt]
  \item Charged-particle multiplicity $d\nch/d\eta$.
  \item Identified particle yields $dN/dy$ of pions, kaons, and protons.
  \item Transverse energy production $dE_T/d\eta$.
  \item Charged particle mean transverse momentum $\langle p_T \rangle$ ($0.15 < p_T < 10$~GeV).
  \item Identified particle mean transverse momentum $\langle p_T \rangle$ of pions, kaons, and protons.
  \item Mean transverse momentum fluctuations $\delta p_T / \langle p_T \rangle$ (charged particles, $0.15 < p_T < 2.0$~GeV).
  \item Two-particle flow cumulants $\vnk{n}{2}$ for $n=2,3,4$\\ (charged particles, $0.2 < p_T < 5.0$~GeV for ALICE, and $0.3 < p_T < 3.0$~GeV for CMS).
  \item Four-particle flow cumulant $\vnk{2}{4}$ \\(charged particles, $0.2 < p_T < 5.0$~GeV).
  \item Symmetric cumulants $\SC(4, 2)$ and $\SC(3,2)$.
\end{itemize}
Each observable is calculated from the list of final state particles produced by UrQMD using the same methods applied by experiment.
We generally match the kinematic cuts of all measurements with two exceptions: we use a larger rapidity interval $|\eta| < 0.8$ than experiment for some boost-invariant observables to improve our finite particle statistics, and we do not apply a rapidity gap, e.g.\ $|\Delta \eta| > 1$, between pairs of particles when calculating the two-particle cumulant $\vnk{n}{2}$ since we already oversample particles from each hydrodynamic event, and this oversampling suppresses non-flow correlations.

At the time of this writing, many of the aforementioned experimental observables are not yet published for $p$-Pb and Pb-Pb collisions at $\sqrts=5.02$~TeV.
We therefore restrict our calibration to the subset of measured and published observables listed in Table~\ref{tab:observables}.
Notably absent from this list are the four-particle cumulants $\vnk{n}{4}$ at $\sqrts=5.02$~TeV despite being measured and published.
Unfortunately, the four-particle cumulants require minimum-bias event statistics an order of magnitude larger than those used in this work.
We therefore refrain from \emph{calibrating} on the four-particle cumulants, although we do show calculations of the four-particle cumulant $\vnk{2}{4}$ later in the text, using a single set of calibration parameters.

\begin{figure*}[t]
  \includegraphics{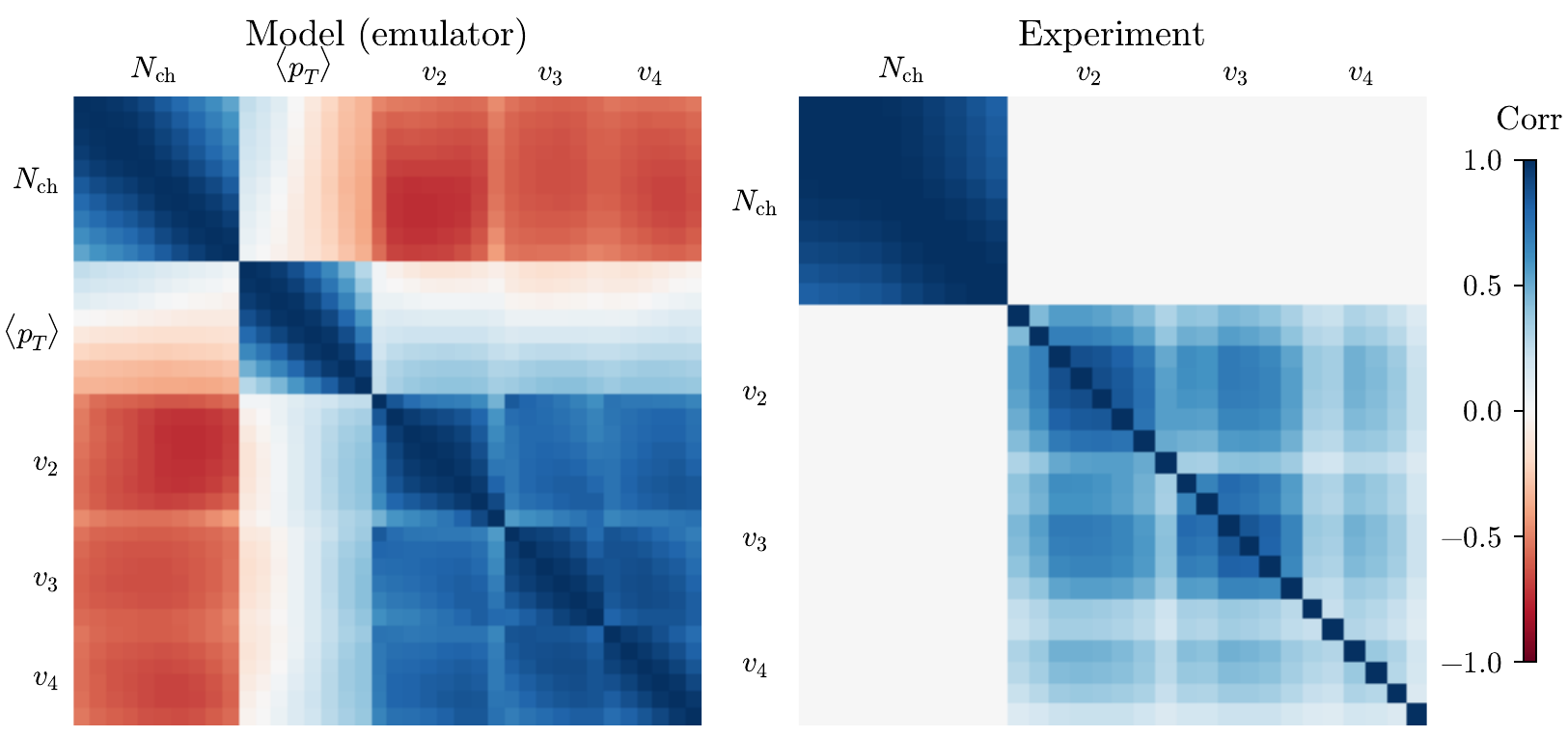}
  \caption{
    \label{fig:correlation}
    Visualization of the Pb-Pb correlation matrix $\corr(y_i, y_j) = \cov(y_i,y_j)/(\sigma_i \sigma_j)$ for the model (emulator) at a random point in parameter space (left-side) and for the experimental data (right-side).
    Each cell represents an observable in a single centrality bin. Experimental statistical and systematic errors are from ALICE \cite{Adam:2015ptt, Adam:2016izf}.
    The experimental correlation structure is modeled using Eq.~\eqref{eq:corr}.
  }
\end{figure*}

Most of the calibration observables listed in Table~\ref{tab:observables} are calculated as a function of collision centrality, where centrality is defined using some measure of the underlying event activity, e.g.\ the charged particle yield in a given rapidity window.
When calculating these observables, we generate \order{4} minimum-bias events at each design point and divide the events into centrality bins using the charged particle yield at midrapidity, similar to the procedure used by experiment.

However, for some observables such as $p$-Pb mean $p_T$ \cite{Abelev:2013bla} and flow cumulants $\vnk{n}{k}$ \cite{Chatrchyan:2013nka}, the experiments use a special high-multiplicity trigger to select rare, ultra-central events according to the number of charged particles produced or detector tracks offline.
These high-multiplicity bins are too selective for our modest minimum-bias event sample, and so a different procedure is required.
We exploit, for this purpose, the approximate correspondence
\begin{equation}
  \label{eq:bin_equiv}
  \ntrk \sim \nch \propto N_\text{parton}
\end{equation}
between each event's initial secondary-parton density ${N_\text{parton} = \int d^2x\,n(\x)}$ in Eq.~\eqref{energy-momentum} and its final-state event activity, characterized by $\nch$ or $\ntrk$ \cite{Abelev:2013bla, Chatrchyan:2013nka}.

Consider, for example, an experimental multiplicity bin $(\nch^\text{low}, \nch^\text{high})$ with some kinematic cuts on $p_T$ and $\eta$.
We rescale this bin by the average multiplicity $\langle \nch \rangle$ of the corresponding minimum-bias event sample, i.e.\
\begin{equation}
  (\nch^\text{low}, \nch^\text{high}) \rightarrow (\nch^\text{low} / \langle \nch \rangle,\, \nch^\text{high} / \langle \nch \rangle)
\end{equation}
in order to reexpress each bin edge as a unitless variable.
These rescaled bins are then used to select initial condition events using the equivalence \eqref{eq:bin_equiv}:
\begin{equation}
  \label{eq:mult_trigger}
  \left (\frac{N^\text{low}_\text{parton}}{\langle N_\text{parton} \rangle}, \frac{N^\text{high}_\text{parton}}{\langle N_\text{parton} \rangle} \right) \leftrightarrow \left (\frac{\nch^\text{low}}{\langle \nch \rangle}, \frac{\nch^\text{high}}{\langle \nch \rangle} \right ).
\end{equation}

Finally, we mimic the method used by experiment and apply \eqref{eq:mult_trigger} to select rare high-multiplicity events from a continuous stream of minimum-bias \trento\ events satisfying the correct relative multiplicity bin edges.
This of course means that, in addition to running a large sample of minimum-bias events for centrality binned observables, we must also generate (much like experiment) a separate sample of multiplicity triggered events.
In practice, we use a few hundred to a few thousand events per multiplicity bin, depending on the type of observable.

We also take stock of the statistical and systematic errors reported by each experiment and incorporate their uncertainty into the likelihood covariance matrix
\begin{equation}
  \Sigma = \Sigma_m + \Sigma_e
\end{equation}
in Eq.~\eqref{eq:likelihood}, which includes uncertainty contributions from both the model $\Sigma_m$ and experimental data $\Sigma_e$.
The experimental contribution to the covariance $\Sigma_e$ is further broken down into statistical and systematic components,
\begin{equation}
  \Sigma_e = \Sigma_e^\text{stat} + \Sigma_e^\text{sys}.
\end{equation}
The statistical errors in $\Sigma_e^\text{stat}$ are uncorrelated, and thus its covariance matrix is diagonal:
\begin{equation}
  \Sigma_e^\text{stat} = \diag[(\sigma^\text{stat}_1)^2, (\sigma^\text{stat}_2)^2, \dots (\sigma^\text{stat}_m)^2 ],
\end{equation}
where $\sigma^\text{stat}_i$ is the statistical uncertainty of observable $y_i$ in the experimental observable vector $\y_e = (y_1, \dots, y_m)$.
The systematic errors, meanwhile, are typically correlated, but the correlation structure is not reported by the experiments so we assert a reasonable form.
We can expand the systematic covariance matrix as
\begin{equation}
  (\Sigma_e^\text{sys})_{ij} =  \rho_{ij} \sigma_i \sigma_j,
\end{equation}
where $\sigma_i$ and $\sigma_j$ are the systematic errors of observables $y_i$ and $y_j$ respectively, and $\rho_{ij}$ is the Pearson correlation coefficient between observable $y_i$ and $y_j$:
\begin{equation}
  \rho_{ij} = \frac{\cov(y_i, y_j)}{\sigma_i \sigma_j},
\end{equation}
which satisfies $\rho_{ij}=1$ for $i=j$ and $|\rho_{ij}| \le 1$ for $i \ne j$.
We assume that each observable is correlated across different centrality/multiplicity bins, and uncorrelated with observables of a different type, e.g.\ correlations between yields and flows.
This is a crude simplifying assumption but it is better than neglecting the correlation structure of the experimental data entirely.

For the correlation structure between different observable bins, we assert a simple Gaussian form
\begin{equation}
  \label{eq:corr}
  \rho_{ij}^\text{sys} = \exp \left[ -\frac{1}{2} \left(\frac{b_i - b_j}{l} \right)^2 \right],
\end{equation}
where $b_i$ and $b_j$ are the midpoints of two observable bins of a single type (centrality or relative multiplicity), and $l$ is a correlation length which describes how quickly the observable bins decorrelate as the distance between the bins increases.
We use centrality correlation lengths $l=100$ for all of the centrality binned Pb-Pb observables and $l=30$ for the centrality binned $p$-Pb charged particle yield $d\nch/d\eta$.
The \mbox{$p$-Pb} mean $p_T$ and flow observables, meanwhile, use relative multiplicity bins $\nch / \langle \nch \rangle$ and $\ntrk / \langle \ntrk \rangle$ which necessitate a smaller correlation length $l=5$.
We show an example correlation matrix
\begin{equation}
  \mathrm{corr}(y_i, y_j) = \cov(y_i, y_j)/(\sigma_i \sigma_j)
\end{equation}
for the Pb-Pb experimental data constructed using Eq.~\eqref{eq:corr} on the right-side of Fig.~\ref{fig:correlation}.
Here $y_i$ denotes an element of the experimental data $\y_e$ and $\sigma_i$ its corresponding uncertainty.
The correlation matrix is block diagonal, with each block representing the correlations within a single class of observable.

\subsection{Model emulator}

In principle, one could calculate the likelihood function in Eq.~\eqref{eq:likelihood} directly, e.g.\ by running the model a large number of times at a given parameter point $\x$ to calculate the model observables $\y_m(\x)$ from the ensemble of simulated events, but in practice such a procedure would be intractable.
The model is computationally intensive to evaluate, and thousands of events are required to calculate the simplest observables at a single parameter point.
Moreover, we need to evaluate Eq.~\eqref{eq:likelihood} \emph{numerous} times in order to sample the multidimensional posterior distribution so that the samples may be histogrammed and visualized.

We therefore follow an established framework for computationally intensive models and train an emulator to act as a fast surrogate for the full physics simulation \cite{OHagan:2006ba, Higdon:2008cmc, Higdon:2014tva}.
The emulator enables essentially instantaneous predictions for $\y_m = f(\x)$ and allows us to sample the posterior distribution millions of times.
In order to train the emulator, we first generate a scaffolding of the parameter space using a maximin Latin hypercube design \cite{Morris:1995lh} to distribute 500 points uniformly throughout our 15-dimensional parameter space according to the parameter ranges in Table~\ref{tab:design}.
We then run minimum-bias and multiplicity triggered $p$-Pb and Pb-Pb events at each design point and calculate the model observables from the ensemble of events.

Specifically, let $X$ denote the ${d \times n}$ design matrix of $d=500$ training points, where each training point is a vector $\x = (x_1, x_2, \dots, x_n)$ of the $n=15$ model parameters in Table~\ref{tab:design}.
Similarly, let $Y$ denote the corresponding ${d \times m}$ observables matrix, where each row of $Y$ is a vector $\y_m = (y_1, y_2, \dots, y_m)$ of $m$ different simulated observables (here we overload our notation so the subscript $m$ means both \emph{model} and \emph{observable number}).
Our goal is to train an emulator for the physics model $f$ using the discrete observations $f: X \mapsto Y$.

We use for this purpose a specific type of emulator known as a Gaussian process emulator (GPE) \cite{Rasmussen:2006gp}.
The advantage of using GPE's is that they provide an estimate of their own uncertainty which allows us to account for this uncertainty when constructing the covariance matrix $\Sigma$ in Eq.~\eqref{eq:likelihood}.
One quirk of GPE's is that they are restricted to scalar-valued functions, i.e.\ functions of one output, whereas we require an emulator for vector-valued functions with multiple outputs.
This restriction is commonly circumvented using principal component analysis (PCA): a general procedure which transforms a set of correlated variables $\y = (y_1, y_2, \dots, y_m)$ into a new basis representation $\z = (z_1, z_2, \dots, z_m)$ where the linear correlations between $z_i$, $z_j$ vanish for all $i\ne j \in m$ \cite{Tipping:1999}.
Independent GPE's can then be used to emulate each $z \in \z$ since the variables $(z_1, \dots, z_m)$ are linearly uncorrelated.
The emulated vector $\z$ is then easily reexpressed in the basis of $\y$ through its inverse transformation.

We first preprocess our model observables by centering and scaling each column of $Y$ (single observable) to zero mean and unit variance to generate a standardized observable matrix $\tilde{Y}$.
PCA is then used to reexpress each row-vector $\tilde{\y}$ of $\tilde{Y}$ (all standardized observables at a single design point) in the new principal component basis:
\begin{equation}
  \tilde{\y}_i = \sum\limits_{j=1}^m z_{ij} \mathbf{v_j},
\end{equation}
where $\tilde{\y}_i$ are the standardized observables of the $i$-th row-vector (design point) of matrix $\tilde{Y}$, and $z_{ij}$ and $\mathbf{v}_j$ are the coefficients and vectors of its $j$-th principal component.

The principal components are reported in order of explained variance, with the first principal component vector $\mathbf{v}_1$ accounting for the most variance in $\tilde{Y}$, and the last principal component vector $\mathbf{v}_m$ accounting for the least.
We then train a set of independent GPE's $\{z_i=\mathrm{gp}_i(\x)\}$ to predict the first $k$ principal components $(z_1, \dots, z_k)$ as a function of the model parameters $\x$ which vary across the design $X$.
For the present study, we use $k=7$ principal components when emulating the $p$-Pb system and $k=8$ principal components when emulating Pb-Pb, chosen to describe 99.5\% of the total observed variance of each system.

\begin{fullpage}
  {\centering\includegraphics{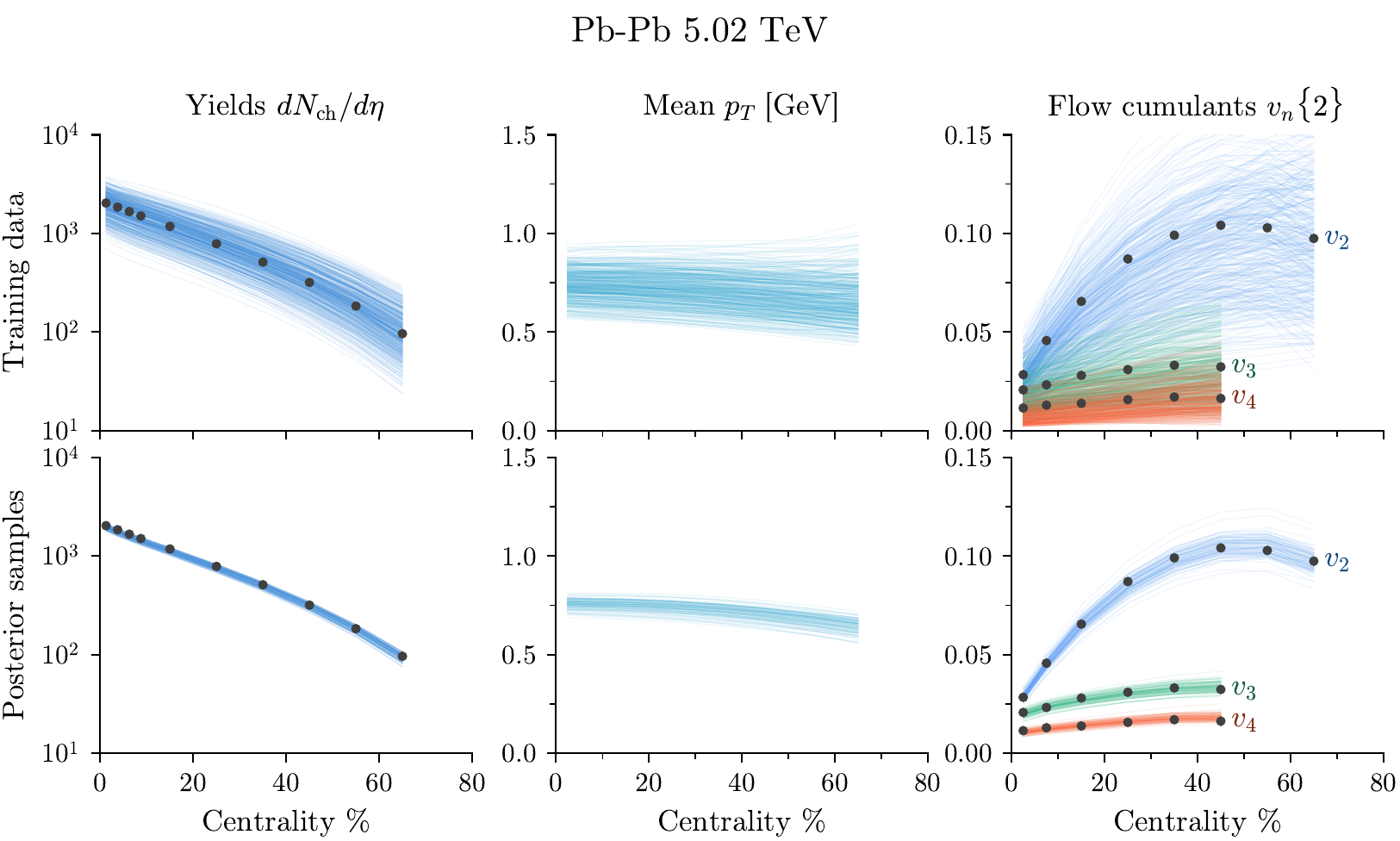}\par}
  \captionof{figure}{
    \label{fig:obs_pbpb}
    Simulated observables compared to experimental data for Pb-Pb collisions at $\sqrts=5.02$~TeV.
    Top row: explicit model calculations (no emulator) for each of the $d=500$ design points; bottom row: emulator predictions for $n=100$ random samples drawn from the posterior.
    Points with error bars are experimental data from ALICE with statistical and systematic errors added in quadrature \cite{Adam:2015ptt, Adam:2016izf}.
  }
  \vspace{.5cm}
  {\centering\includegraphics{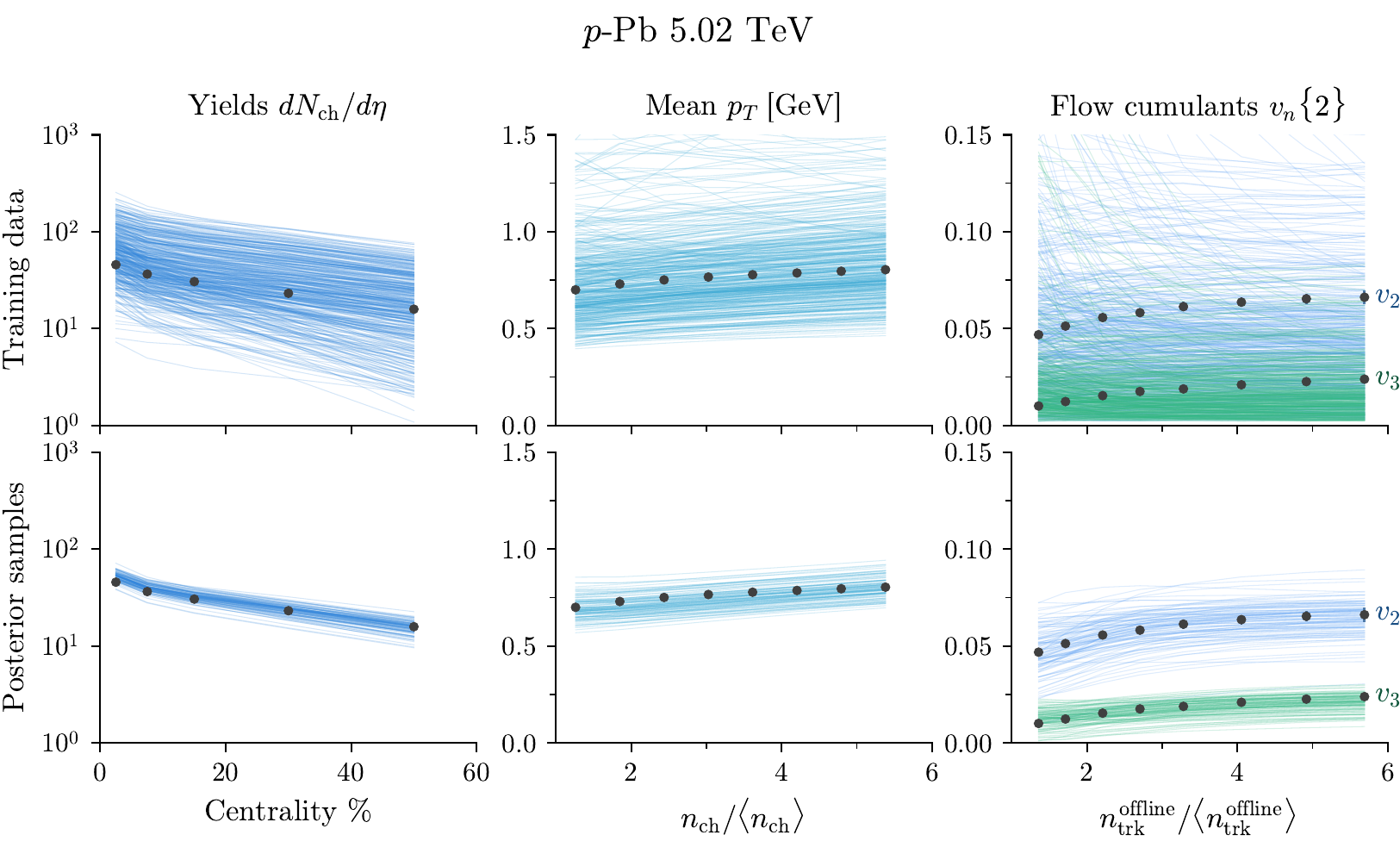}\par}
  \captionof{figure}{
    \label{fig:obs_ppb}
    Same as Fig.~\ref{fig:obs_pbpb} but for $p$-Pb collisions at $\sqrts=5.02$~TeV.
    Note that multiplicity bins are used for mean $p_T$ and flow cumulant observables to match the bins used by experiment.
    Experimental data are from ALICE \cite{Adam:2014qja, Abelev:2013bla} and CMS \cite{Chatrchyan:2013nka}.
  }
\end{fullpage}

The GPE's are essentially fancy interpolators applied to the model's training points and PCA transformed observables.
Each GPE reports a mean value $z(\x)$ as well as an estimated error $\delta z(\x)$ which accounts for statistical noise in the training data and interpolation error between the design points.
Once the GPE's are trained, we can predict the observables $\y_m$ at parameter point $\x$ by transforming the vector of principal components
\begin{equation}
  \mathbf{z}(\x) = (z_1(\x), z_2(\x), \dots, z_k(\x))
\end{equation}
\emph{back} to physical space.
Similarly, we can construct the covariance matrix of the observables in PCA space,
\begin{equation}
  \cov(z_i, z_j) = \diag[(\delta z_{1})^2, (\delta z_2)^2, \dots, (\delta z_k)^2 ],
\end{equation}
and transform it back to physical space as well to obtain the covariance matrix of the model observables $\y_m$ at a given parameter point $\x$.

The resulting emulator therefore predicts both a mean prediction $\y_m(\x)$ and an uncertainty covariance matrix $\Sigma_m(\x)$ which accounts for multiple sources of model and emulator uncertainty, including the truncation error expected from using a finite number of principal components $k < m$.
The model covariance matrix $\Sigma_m$ includes so-called \emph{known}-unknowns such as statistical error and emulator interpolation error, but not \emph{unknown}-unknowns such as the overall validity of small-system hydrodynamics, i.e.\ things which lack a unified consensus or are difficult to quantify.
We show in Fig.~\ref{fig:correlation} the resulting Pb-Pb correlation matrix $\mathrm{corr}(y_i,y_j)$ for the model (emulator) at a random parameter point $\x$ in the design space (left-side), along side the same correlation matrix for the experimental data (right-side) discussed previously.
For additional information on the model emulator, we direct the reader to Appendix~\ref{app:validation} which includes several validation tests of the emulator prediction accuracy.

\subsection{Bayesian calibration}

In order to calibrate the model on two different collision systems, we expand the likelihood function \eqref{eq:likelihood} into a joint likelihood
\begin{equation}
  \label{eq:joint_likelihood}
  P(E | H_\x) = P(E_\text{Pb-Pb} | H_\x) \cdot P(E_\text{$p$-Pb} | H_\x),
\end{equation}
where $E$ subsumes all evidence from the $p$-Pb and Pb-Pb collision systems and $H_\x$ is our hypothesis that $\x$ equals the true parameters $\x_\star$.
We then perform Markov-chain Monte Carlo (MCMC) importance sampling on the posterior distribution in Eq.~\eqref{eq:bayes} to draw random samples for $P(H_\x | E)$, the estimate of the true model parameters given the model and the experimental data \cite{Goodman:2010en, FM:2013mc}.
For this we use an affine-invariant ensemble sampler which uses a large ensemble of interdependent walkers \cite{Goodman:2010en, FM:2013mc} and allow the MCMC chain to ``burn-in'' before generating \order{7} posterior samples.

\section{Results}
\label{sec:results}

We show the simulated and emulated model observables (thin colored lines) for Pb-Pb collisions in Fig.~\ref{fig:obs_pbpb} and for \mbox{$p$-Pb} collisions in Fig.~\ref{fig:obs_ppb} at $\sqrts=5.02$~TeV compared to experimental data from CMS \cite{Chatrchyan:2013nka} and ALICE \cite{Adam:2015ptt, Adam:2016izf, Adam:2014qja, Abelev:2013bla}.
The top row of each figure shows explicit model calculations at each of the $d=500$ design points (training data), while the bottom row shows emulator predictions for $n=100$ random parameter samples drawn from the Bayesian posterior (sampled from the MCMC chain).
Each column shows a different class of observable.
The charged-particle yield $d\nch/d\eta$ is shown on the left, mean $p_T$ is in the middle, and two-particle flow cumulants $\vnk{n}{2}$ for $n=2,3,4$ are on the right.
The Pb-Pb mean $p_T$ and $p$-Pb $\vnk{4}{2}$ datasets are missing and hence are omitted from the present calibration.

Notice the large spread of the observables calculated at the training points (top row of each figure).
The design is constructed to vary each parameter across a wide range of values, specified in Table~\ref{tab:design}, and hence the corresponding model calculations are equally uncertain.
We also point out that there is considerably more variance in the $p$-Pb training data than the Pb-Pb training data.
The $p$-Pb yields, mean $p_T$, and flow cumulants all vary wildly within the chosen parameter ranges.
For instance, we can turn the $p$-Pb flows completely \emph{off} with suitably chosen parameters which is not possible in the Pb-Pb system.
Evidently the $p$-Pb model predictions are far more sensitive to modeling uncertainties.

Conversely, the calibrated (posterior sampled) emulator predictions (bottom row of Figs.~\ref{fig:obs_pbpb} and \ref{fig:obs_ppb}) are far better constrained and nicely track the experimental data points.
We emphasize here that the posterior parameter values are obtained from a \emph{simultaneous} calibration to $p$-Pb and Pb-Pb data, and thus they are self-consistent between the two systems.
The spread in the posterior samples reflects different sources of model and experimental uncertainty as well as tension in the optimal fit parameters which describe each observable.
We demonstrate later in the text that a single set of model parameters well describes all of the calibration data, and thus we believe that much of the spread in the posterior samples is uncertainty contributed by our emulator.
We also note that although the $p$-Pb posterior samples have a somewhat larger spread than the Pb-Pb samples, the percentage uncertainty of the $p$-Pb emulator is similar to that of the Pb-Pb emulator, and thus the difference is likely due to the larger variance of the $p$-Pb training data.
The uncertainty in the posterior distribution could thus be improved by running the calibration with more design points or with a narrower range of parameter values to increase the density of the training points and reduce interpolation uncertainty.

\begin{fullpage}
  \begin{figure}
    \makebox[\textwidth]{\includegraphics[width=.95\paperwidth]{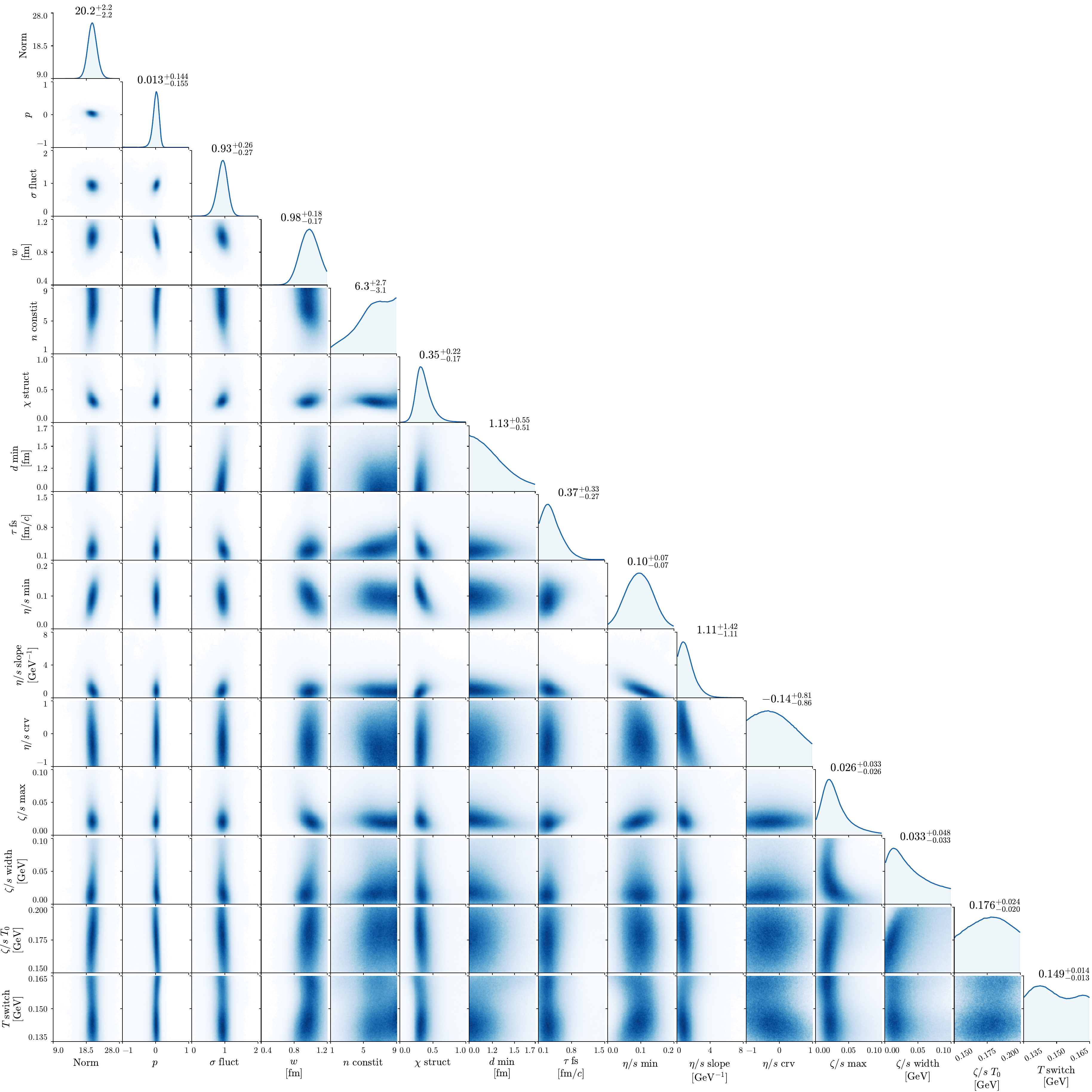}}
  \end{figure}
  \captionof{figure}{
    \label{fig:posterior}
    Bayesian posterior distribution of the model input parameters.
    The diagonal panels show the marginalized distributions of individual model parameters, while off-diagonal panels show the joint distributions for pairs of model parameters, visualizing their correlations.
    The marginalized distribution medians and 90\% credible intervals are annotated along the diagonal.
  }
\end{fullpage}

We now direct our attention to Fig.~\ref{fig:posterior} which shows the main result of this work, the posterior distribution of the model input parameters.
Recall that the posterior $P(H_\x | E)$ is the probability that our hypothesis $H_\x$ for the true model parameters $\x_\star$ is correct, given the evidence $E$ provided by experiment.
The present posterior has 15 dimensions, one for each parameter listed in Table~\ref{tab:design}, and thus its joint distribution cannot be summarized by one figure alone.
We therefore sample the distribution and histogram the samples to project the distribution onto one or two dimensions at a time.
Each diagonal panel is the distribution of a \emph{single} model parameter (marginalized over all others), and each lower-diagonal panel is the joint distribution of a \emph{pair} of model parameters (marginalized over all others).
We also report numeric estimates for each parameter's median value and 90\% credible interval and annotate their values along the distribution diagonal (see Table~\ref{tab:post_param}).
For example, the fictitious estimate $x=2.45_{-0.15}^{+0.20}$ reports a median value $\tilde{x}=2.45$ and 90\% credible interval $2.30 < x < 2.65$.

\begin{table}
  \caption{
    \label{tab:post_param}
    Posterior parameter estimates corresponding to Fig.~\ref{fig:posterior}.
    The reported values are for the distribution median and 90\% highest posterior density credible interval.
  }
  \begin{ruledtabular}
    \newlength{\cellwidth}
    \settowidth{\cellwidth}{$-0.00$}
    \newcommand{\est}[3]{\parbox{\cellwidth}{\hfill$#1$}$_{-#2}^{+#3}$}
    \begin{tabular}{llll}
      \toprule
      \multicolumn{2}{c}{Initial condition / Pre-eq}     & \multicolumn{2}{c}{QGP medium}              \\
      \cmidrule(r){1-2}                                    \cmidrule(l){3-4}
      \addlinespace[.4ex]
      Norm       & \est{20.2}{2.2}{2.2}          & $\eta/s$ min     & \est{0.10}{0.07}{0.07}            \\[1.1ex]
      $p$        & \est{0.013}{0.155}{0.144}     & $\eta/s$ slope   & \est{1.11}{1.11}{1.42} GeV$^{-1}$ \\[1.1ex]
      $\sigmaf$  & \est{0.93}{0.27}{0.26}        & $\eta/s$ crv     & \est{-0.14}{0.86}{0.81}           \\[1.1ex]
      $w$        & \est{0.98}{0.17}{0.18} fm     & $\zeta/s$ max    & \est{0.026}{0.026}{0.033}         \\[1.1ex]
      $n_c$      & \est{6.3}{3.1}{2.7}           & $\zeta/s$ width  & \est{0.033}{0.033}{0.048} GeV     \\[1.1ex]
      $\X$       & \est{0.35}{0.17}{0.22}        & $\zeta/s$ $T_0$  & \est{0.176}{0.020}{0.024} GeV     \\[1.1ex]
      $\dmin$    & \est{1.13}{0.51}{0.55} fm     & $\Tsw$           & \est{0.149}{0.013}{0.014} GeV     \\[1.1ex]
      $\taufs$   & \est{0.37}{0.27}{0.33} \fmc   & \\
      \addlinespace[.4ex]
      \bottomrule
    \end{tabular}
  \end{ruledtabular}
\end{table}

\subsection{Initial condition properties}

The \trento\ normalization factor ${\text{Norm} = 20.2^{+2.2}_{-2.2}}$ and generalized mean parameter $p=0.013^{+0.144}_{-0.155}$ are well constrained by the present analysis.
Moreover, their posterior values nicely describe the $p$-Pb and Pb-Pb calibration observables in Figs.~\ref{fig:obs_pbpb} and \ref{fig:obs_ppb}.
While it would not be surprising, for example, to fit one or two of these observables using such a narrow range of values, the quality of the combined fit (more on this later) and the number of observables described is highly non-trivial.
For example, consider the ratio of the $p$-Pb yield over the Pb-Pb yield, which imposes a strong constraint on physically reasonable initial condition models.
As the entropy deposition parameter $p$ trends toward positive (negative) infinity, particle production scales like the local maximum (minimum) of the nuclear overlap density.
The parameter $p$ thus strongly affects the $p$-Pb and Pb-Pb yield ratio.
It just so happens that the small range of $p$-values needed to describe this yield ratio are the same values needed to describe all of the calibration observables in the present study, \emph{and} numerous Pb-Pb observables at $\sqrts=2.76$~TeV \cite{Bernhard:2018hnz}.
This work thus reaffirms an empirical scaling law reported by several previous studies \cite{Moreland:2014oya, Bernhard:2016tnd, Ke:2016jrd, Bernhard:2018hnz} for the initial transverse entropy density (or massless parton density):
\begin{equation}
  \label{eq:geometric_mean}
  \frac{dS}{dy}\bigg\vert_{y=0} \propto \sqrt{\T_A \T_B},
\end{equation}
where $\T_A$ and $\T_B$ are the participant thickness functions \eqref{eq:fluctuated_thick} of each nucleus.
We emphasize that this specific analytic form should not be interpreted too literally.
For instance, a generalized mean described by $p=0.05$ is well within our 90\% credible interval, but it is not equal to the geometric mean in Eq.~\eqref{eq:geometric_mean}.
This approximate form nevertheless has been shown to mimic the scaling behavior of several saturation-based theory calculations in high-energy QCD \cite{Bernhard:2016tnd}, and thus appears consistent with general theoretical expectations.

\begin{figure}
  \includegraphics{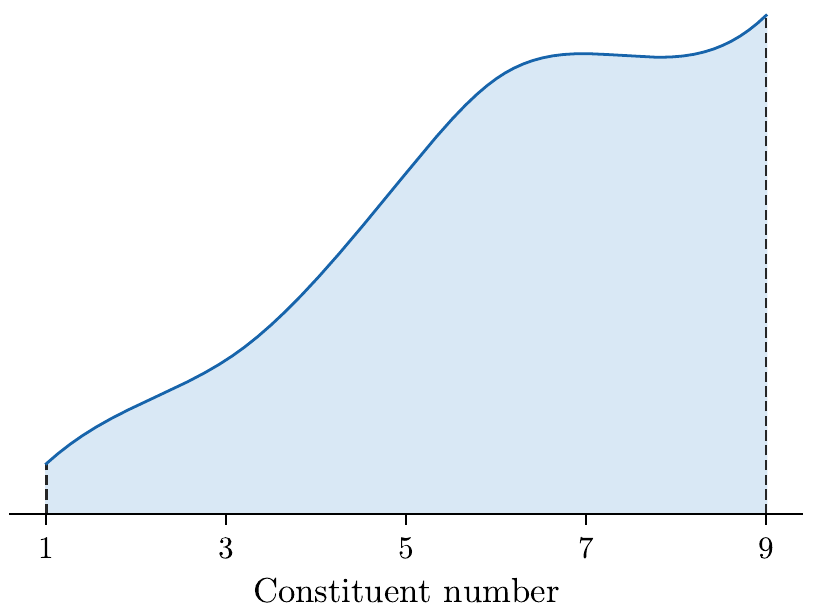}
  \caption{
    \label{fig:posterior_parton_number}
    Posterior distribution for the number of nucleon constituents $n_c$ determined by the analysis.
    The parameter $n_c$ is an integer (discrete) variable at every design point, but the emulator interpolation produces a posterior distribution which is continuous.
  }
\end{figure}

Continuing down the diagonal in Fig.~\ref{fig:posterior}, we see that the nucleon width parameter $w=0.98_{-0.17}^{+0.18}$~fm.
We caution that this parameter $w$ is \emph{not} the RMS radius of our nucleons due to idiosyncrasies of our constituent sampling procedure.
We can, however, easily calculate the RMS radius for a specific nucleon width $w$, constituent width $v$, and constituent number $n_c$.
For example, the single highest posterior probability region of the parameter space prefers a nucleon width $w=0.92$~fm, constituent width $v=0.43$~fm, and constituent number $n_c=6$.
The corresponding RMS nucleon radius for these parameters is $R_n = 0.86$~fm, conspicuously close the proton's RMS electric charge radius $R_p = 0.879(8)$~fm \cite{Bernauer:2010wm}.
This is perhaps the single largest difference between our work and the conclusions of recent saturation-based calculations which constrained the event-by-event fluctuations of the proton using a color-dipole picture of vector meson production \cite{Mantysaari:2016ykx, Mantysaari:2016jaz}.
Those studies find that the measured coherent and incoherent $J/\Psi$ spectra at HERA prefer a compact gluon distribution inside each nucleon, with an RMS radius $R_g \approx 0.4$~fm, which is roughly \emph{half} the analogous RMS nucleon width preferred by our analysis.

\begin{figure}
  \includegraphics{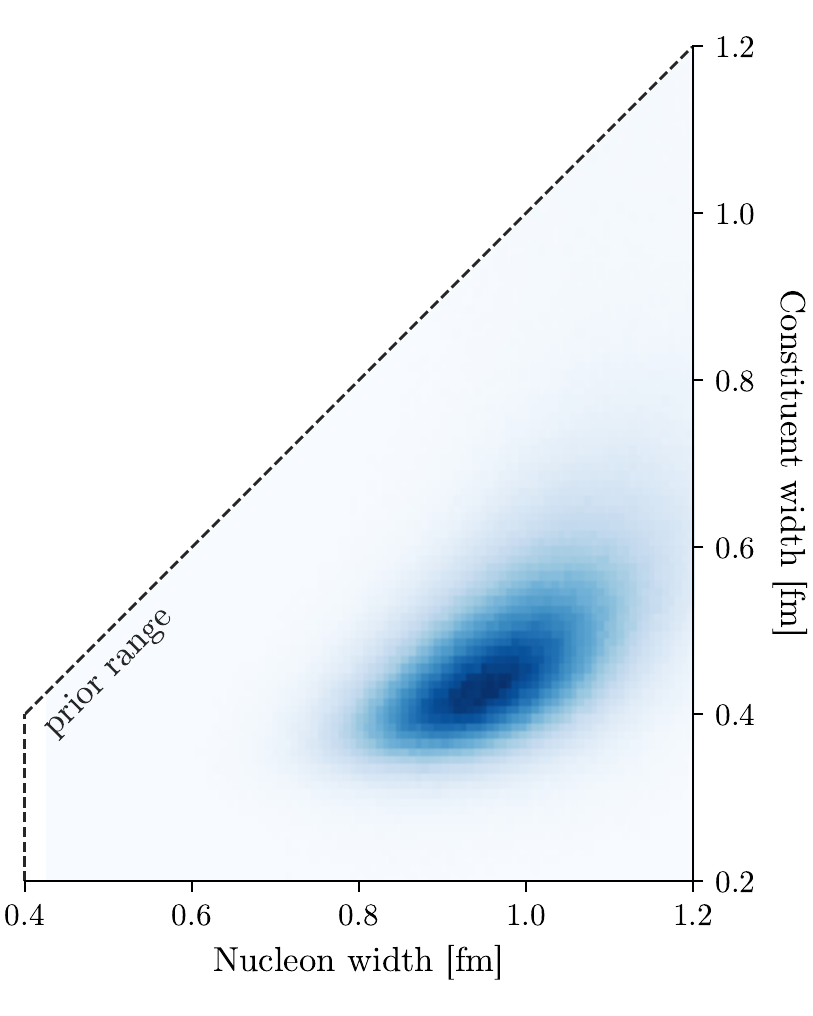}
  \caption{
    \label{fig:posterior_proton_shape}
    Posterior distribution for the nucleon width $w$ and constituent width $v$.
    The enclosed trapezoidal region covers the prior range of allowed values for $w$ and $v$.
    The posterior distribution, shown in blue, indicates the preferred values for $w$ and $v$ determined by the analysis.
  }
\end{figure}

The posterior on the constituent number $n_c$, shown enlarged in Fig.~\ref{fig:posterior_parton_number}, is not sharply peaked.
We therefore refrain from quoting a distribution median and 90\% credible interval, although we do note that the distribution clearly favors $n_c > 1$ constituents.
This is not surprising.
The \trento\ model mimics saturation-based initial condition models \cite{Bernhard:2016tnd}, and saturation models tend to produce ``proton-sized'' fireballs in $p$-Pb collisions \cite{Bzdak:2013zma}.
When the proton is spherically symmetric, the resulting proton-sized QGP is also largely symmetric and thus produces very little flow.
Saturation-based models are therefore unable to describe the significant flow measured in high-multiplicity $p$-Pb collisions without nucleon substructure, or alternatively, some other source of additional correlations \cite{Schenke:2017bog}.

The posterior on the nucleon substructure parameter $\X=0.35_{-0.17}^{+0.22}$, on the other hand, is particularly sharply peaked.
Recall that this parameter, defined in Eq.~\eqref{eq:struct_param}, interpolates between the minimum and maximum widths of each constituent and hence
two different limits for the granularity of the nucleon.
When $\X=0$, the nucleon is populated by $n_c$ small compact hot-spots, and when $\X=1$ the hot-spots are large and fully overlapping, restoring spherical symmetry.

The nucleon structure parameter $\X$ is somewhat awkward to conceptualize, but we can easily transform its value back into a constituent width $v$.
Figure~\ref{fig:posterior_proton_shape} shows the resulting joint posterior distribution for the nucleon width parameter $w$ and constituent width $v$.
The posterior distribution, shown in blue, is remarkably well constrained.
We report a posterior estimate for the constituent width $v=0.47_{-0.15}^{+0.20}$~fm which is roughly the same magnitude as the \emph{nucleon} hot-spots used in recent saturation-based substructure studies that employed IP-Glasma initial conditions \cite{Schenke:2018fci}.
Evidently, it may be necessary to place an informative prior on our nucleon substructure parameters in order to resolve the apparent tension between our parameter values and those needed to describe DIS measurements at HERA.
Alternatively, it is also possible that the fluctuations probed by coherent and incoherent $J/\Psi$ production are different than those probed by minimum-bias particle production.

\subsection{Transport properties}

\begin{figure*}
  \includegraphics{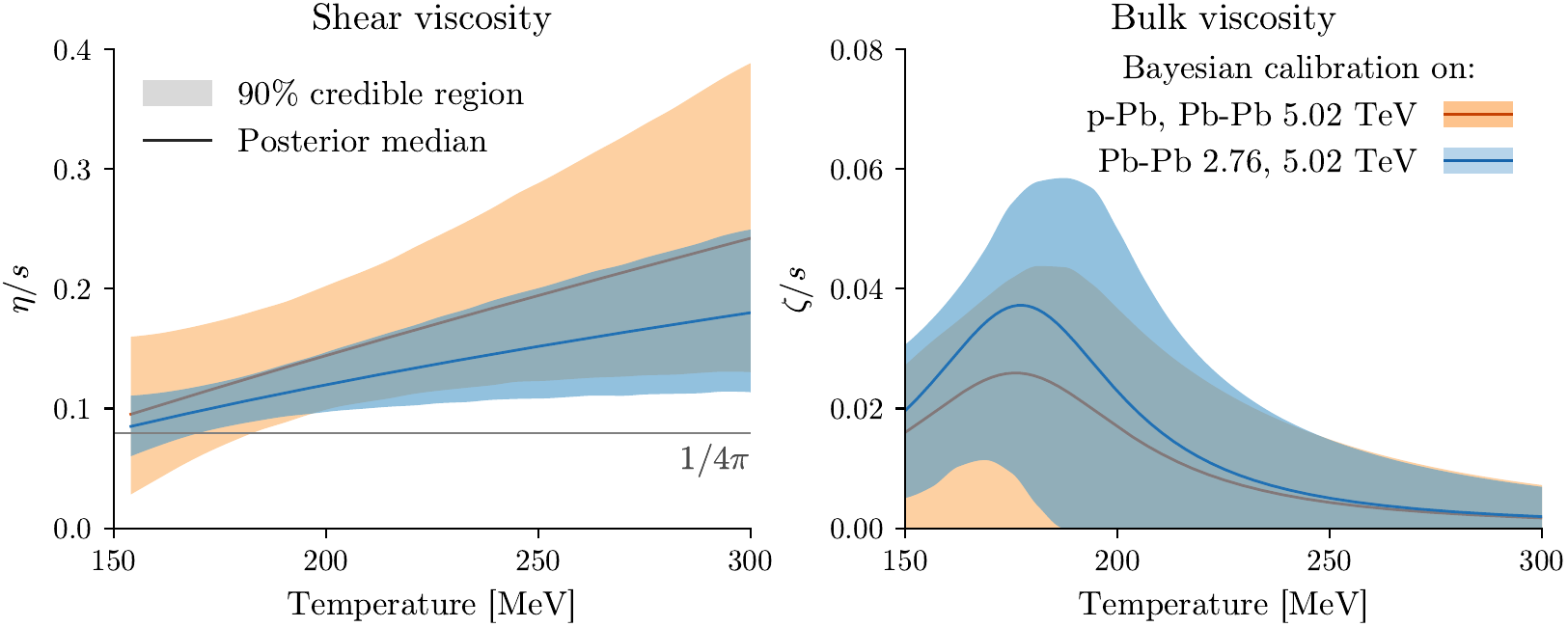}
  \caption{
    \label{fig:region_shear_bulk}
    Left figure: estimated temperature dependence of the QGP specific shear viscosity $(\eta/s)(T)$ determined by the present Bayesian analysis of $p$-Pb and Pb-Pb collisions at $\sqrts=5.02$~TeV (orange line/band) compared to a previous Bayesian analysis of Pb-Pb collisions at $\sqrts=2.76$ and 5.02~TeV (blue line/band) \cite{Bernhard:2018hnz}.
    The lines are the medians of each posterior distribution, and the bands are their 90\% credible regions.
    Right figure: same as before, but for the temperature dependence of the QGP specific bulk viscosity $(\zeta/s)(T)$.
  }
\end{figure*}

In this section, we compare several of our posterior estimates to those obtained from a similar Bayesian analysis in Ref.~\cite{Bernhard:2018hnz} which used an (almost) identical version of the present physics model.
The only modeling difference is the inclusion of nucleon substructure in the present study which was absent in Ref.~\cite{Bernhard:2018hnz}.
Several calibration details, however, are different between the two analyses.
This work used a modest number of $p$-Pb and Pb-Pb observables at $\sqrts=5.02$~TeV (limited by availability), whereas Ref.~\cite{Bernhard:2018hnz} calibrated on a much larger number of Pb-Pb observables at $\sqrts=2.76$ and 5.02~TeV.

The posterior free-streaming time $\taufs=0.37_{-0.27}^{+0.33}\ \fmc$ obtained in this work is significantly smaller than our previous estimate $\taufs=1.16_{-0.25}^{+0.29}\ \fmc$ in Ref.~\cite{Bernhard:2018hnz}.
We point out that the present study is missing several important observables which could affect the estimated free-streaming time, e.g.\ the Pb-Pb mean $p_T$ and mean $p_T$ fluctuations at $\sqrts=5.02$~TeV.
Nevertheless, it appears that the inclusion of nucleon substructure significantly reduces the maximum allowed free-streaming time, although more work is needed to establish if this is indeed the case.

We also compare in Fig.~\ref{fig:region_shear_bulk} our estimates for the temperature dependence of the QGP specific shear viscosity $(\eta/s)(T)$ and bulk viscosity $(\zeta/s)(T)$ with those of Ref.~\cite{Bernhard:2018hnz}.
The lines are the distribution medians, and the bands are their 90\% credible regions.
The results of this work are shown in orange, and the results of Ref.~\cite{Bernhard:2018hnz} are shown in blue.
In general, our estimates are broader and less certain but otherwise self-consistent.
Evidently, the combined analysis of Pb-Pb data at $\sqrts=2.76$ and 5.02~TeV in Ref.~\cite{Bernhard:2018hnz} provides a better constraint on the QGP viscosities which is not surprising given the additional observables and multiple beam energies studied.
The $p$-Pb data, meanwhile, does not appear to provide any unique viscous constraints.

\subsection{Verification of high-probability parameters}

We verified the emulator and tested the accuracy of our physics model framework using a single set of high-probability parameters selected from the Bayesian posterior.
These parameters, listed in Table~\ref{tab:mode_params}, are the approximate ``best fit'' values of the calibrated model, commonly referred to as the \emph{maximum a posteriori} (MAP) estimate:
\begin{equation}
  \x_\mathrm{MAP} \equiv \operatorname*{arg\, max}_{\x} P(H_\x | E).
\end{equation}
We then ran \order{6} minimum-bias and multiplicity triggered events using the MAP estimate $\x_\mathrm{MAP}$ and computed all of the model observables listed in Sec.~\ref{sec:observables}.
The resulting model calculations are shown in Fig.~\ref{fig:obs_map} alongside experimental data from CMS \cite{Chatrchyan:2013nka} and ALICE \cite{Adam:2015ptt, Adam:2016izf, Adam:2014qja, Abelev:2013bla}.
The left and right columns show the results for the $p$-Pb and Pb-Pb collision systems respectively, and each row shows a different group of related observables.

The global agreement of the MAP model calculations with the experimental data is very good.
The largest tension is observed in the two-particle cumulants $\vnk{2}{2}$ and $\vnk{3}{2}$ of the $p$-Pb system, although even that tension is only about 10--15\%.
Quite remarkably, the model perfectly describes the shape of the $p$-Pb and Pb-Pb two-particle correlations which is strong evidence that these correlations are hydrodynamic in origin.
Moreover, we obtain an excellent description of the \mbox{$p$-Pb} mean $p_T$, although this fit is somewhat less meaningful since we are unable to calibrate on the Pb-Pb mean $p_T$ simultaneously (data is not yet available).
Additionally, the model provides a simultaneous description of the \mbox{$p$-Pb} and Pb-Pb charged-particle yields using a single entropy deposition parameter $p=0$.
This is the \emph{exact same} generalized mean $p$-value supported by multiple previous studies \cite{Moreland:2014oya, Bernhard:2016tnd, Ke:2016jrd, Bernhard:2018hnz}.
Evidently, this scaling continues to hold for initial conditions with sizable nucleon substructure.

We also present calculations for several observables which were omitted from the calibration due to the statistical limitations of our training data.
Here our MAP event sample is several orders-of-magnitude larger so the statistics are no issue.
The bottom-right panel of Fig.~\ref{fig:obs_map} shows our model calculation for the four-particle elliptic flow cumultant $\vnk{2}{4}$ along with the measured data points from \mbox{ALICE} \cite{Adam:2016izf}.
We see that the MAP estimate nicely describes the measured $\vnk{2}{4}$ data which is encouraging since this particular observable was never used to calibrate the model.

The relative mean $p_T$ fluctuation $\delta p_T / \langle p_T \rangle$ is another important bulk observable to test the predictions of the calibrated model.
It measures the \emph{dynamical} component of event-by-event mean $p_T$ fluctuations, quantified by the two-particle correlator
\begin{equation}
  \label{eq:mean_pT_corr}
  (\delta p_T)^2 = \langle \langle (p_{T,i} - \langle p_T \rangle) (p_{T,j} - \langle p_T \rangle) \rangle \rangle.
\end{equation}

\begin{fullpage}
  {\centering\includegraphics{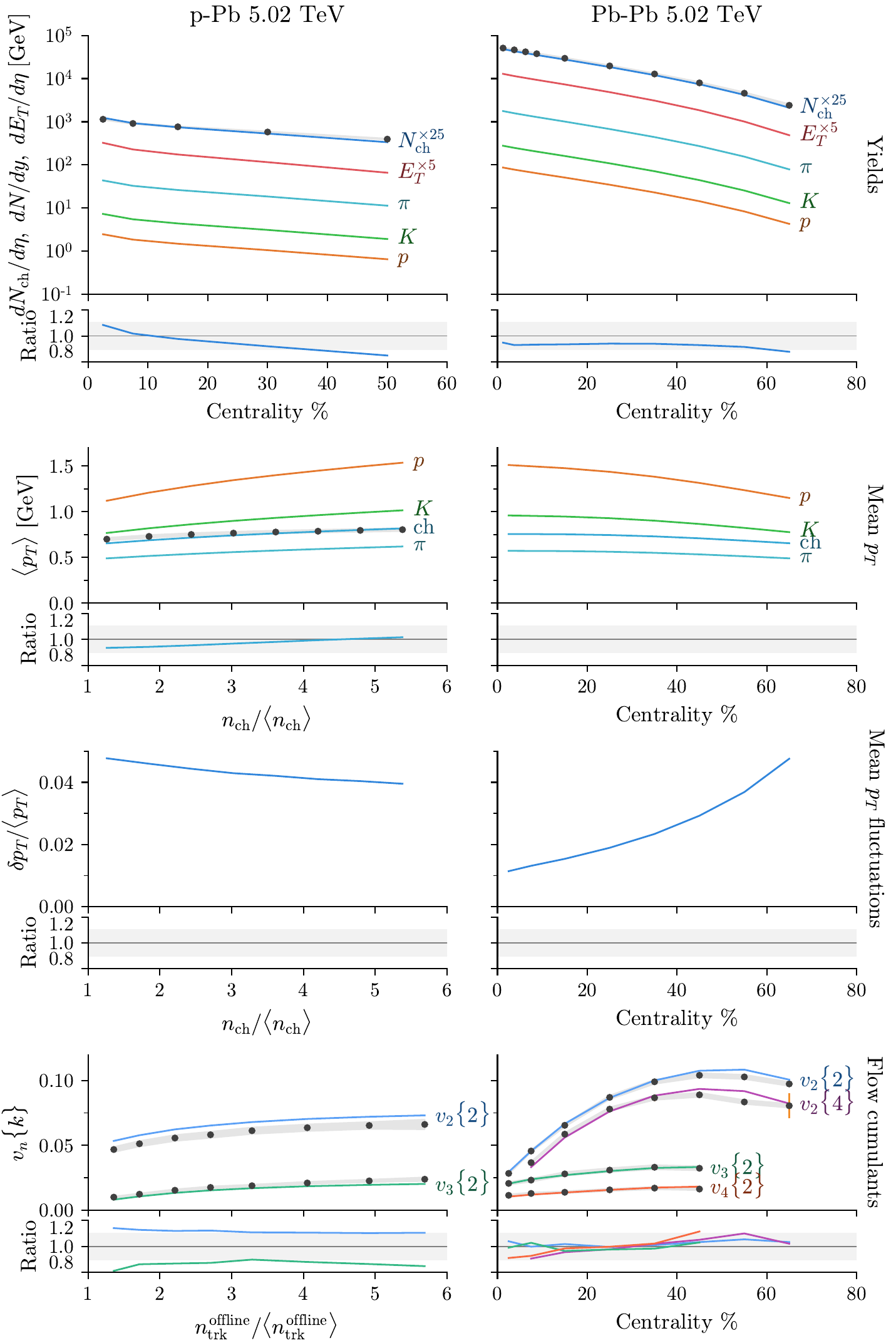}\par}
  \captionof{figure}{
    \label{fig:obs_map}
    Model calculations using the \emph{maximum a posterior} (MAP) parameters compared to experiment.
    Colored lines are model calculations for $p$-Pb collisions (left) and Pb-Pb collisions (right) at $\sqrts=5.02$~TeV.
    Points with error bars are the experimental data with statistical uncertainties, and gray bands their corresponding systematic uncertainties, from CMS \cite{Chatrchyan:2013nka} and ALICE \cite{Adam:2015ptt, Adam:2016izf, Adam:2014qja, Abelev:2013bla}.
    The sub-axes show the ratio of model over data where available with gray bands indicating $\pm 10\%$.
  }
\end{fullpage}

\noindent The inner-average in Eq.~\eqref{eq:mean_pT_corr} runs over all pairs of particles $i,j$ in the same event, the outer average runs over all events in a given bin (centrality or multiplicity), and the symbol $\langle p_T \rangle$ denotes the usual mean transverse momentum of particles in the bin.
The observable is typically presented in terms of the dimensionless ratio $\delta p_T / \langle p_T \rangle$, which quantifies the strength of the dynamical fluctuations in units of the average transverse momentum $\langle p_T \rangle$.

We show the MAP estimate predictions for the $p$-Pb and Pb-Pb relative mean $p_T$ fluctuations $\delta p_T / \langle p_T \rangle$ at $\sqrts=5.02$~TeV in the third row of Fig.~\ref{fig:obs_map}.
For the Pb-Pb system, we use centrality bins and for the $p$-Pb system we use the same relative multiplicity bins used for the $p$-Pb charged-particle mean $p_T$.
The relative mean $p_T$ fluctuations have been shown to be particularly sensitive to the existence of nucleon substructure \cite{Bozek:2017elk}, and thus it would be interesting to ultimately include this observable in the calibration when the data becomes available.

\begin{table}
  \caption{
    \label{tab:mode_params}
    High-probability parameters selected from the posterior distribution and used to generate Fig.~\ref{fig:obs_map}.
    The posterior distribution on the particlization temperature $\Tsw$ is flat (agnostic), so we fix it's value using Ref.~\cite{Bernhard:2018hnz}.
  }
  \begin{ruledtabular}
    \begin{tabular}{ll@{\hspace{2em}}ll}
      \multicolumn{2}{c}{Initial condition / Pre-eq} & \multicolumn{2}{c}{QGP medium} \\
      \paddedhline
      Norm     & 20.            & $\eta/s$ min      & 0.11           \\
      $p$      & 0.0            & $\eta/s$ slope    & 1.6 GeV$^{-1}$ \\
      $k$      & 0.19           & $\eta/s$ curv     & -0.29          \\
      $n_c$    & 6              & $\zeta/s$ max     & 0.032          \\
      $w$      & 0.92 fm        & $\zeta/s$ width   & 0.024 GeV      \\
      $v$      & 0.43 fm        & $\zeta/s$ $T_0$   & 175 MeV        \\
      $\dmin$  & 0.81 fm        & $\Tsw$            & 151 MeV        \\
      $\taufs$ & 0.37 \fmc
    \end{tabular}
  \end{ruledtabular}
\end{table}

\begin{figure}
  \includegraphics{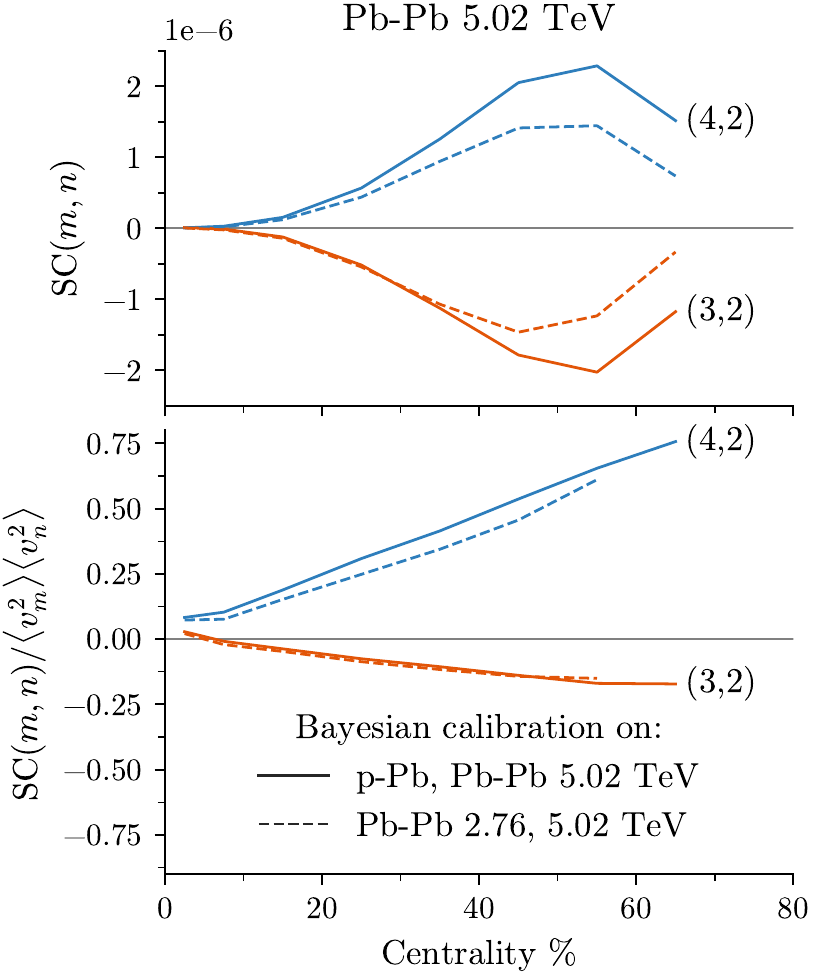}
  \caption{
    \label{fig:flow_corr}
    Model calculations of the symmetric cumulants (top) and normalized symmetric cumulants (bottom) for Pb-Pb collisions at $\sqrts=5.02$~TeV using the \emph{maximum a posteriori} (MAP) parameters.
    The solid lines are the MAP estimate of the present analysis (with nucleon substructure), and the dashed lines are the MAP estimate of Ref.~\cite{Bernhard:2018hnz} (without nucleon substructure) which was calibrated on Pb-Pb observables at $\sqrts=2.76$ and 5.02~TeV.
    In general, most model parameters are somewhat different between the two studies.
  }
\end{figure}

Lastly, we compute the symmetric cumulants $\SC(m,n)$ for the Pb-Pb collision system at $\sqrts=5.02$~TeV which quantify correlations between event-by-event fluctuations of the flow harmonics of different order \cite{Bilandzic:2013kga, ALICE:2016kpq},
\begin{align}
  \SC(m, n) &= \langle\langle \cos[m(\phi_1 - \phi_3) + n(\phi_2-\phi_4)]\rangle\rangle \nonumber \\
  \nonumber &- \langle\langle\cos[m(\phi_1-\phi_2)]\rangle\rangle\langle\langle\cos[n(\phi_1-\phi_2)]\rangle\rangle \label{eq:scmn}\\
  &\approx \langle v_m^2 v_n^2 \rangle - \langle v_m^2\rangle\langle v_n^2\rangle.
\end{align}
We show these model predictions in Fig.~\ref{fig:flow_corr} along with the \emph{normalized} symmetric cumulants
\begin{equation}
  \NSC(m,n) = \SC(m,n)/\langle v_m^2\rangle\langle v_n^2\rangle,
\end{equation}
which are expected to be less sensitive to the medium response and more sensitive to the properties of the initial state.
The solid lines are the MAP estimate of the present study and the dashed lines are the MAP estimate of Ref.~\cite{Bernhard:2018hnz} which did not include nucleon substructure and was calibrated on Pb-Pb collisions at $\sqrts=2.76$ and 5.02~TeV.
We observe that the gap between $\SC(3,2)$ and $\SC(4,2)$ is generally wider in the present analysis than in Ref.~\cite{Bernhard:2018hnz}, as is the gap between the normalized symmetric cumulants $\NSC(3,2)$ and $\NSC(4,2)$.

We emphasize that multiple aspects of the two analyses are different such as the collision systems and beam energies considered, the observables which were included in each calibration, and the existence of nucleon substructure in the model.
Thus we can only speculate what might have caused the large difference in the MAP estimate for the symmetric flow cumulants.
Two reasonable culprits would be the inclusion of nucleon substructure and the large difference in the preferred pre-equilibrium free-streaming time determined by the two studies.

\section{Summary and conclusions}
\label{sec:summary}

Relativistic heavy-ion collisions produce long-range multiparticle correlations which are commonly explained by the existence of hydrodynamic flow \cite{deSouza:2015ena}.
This narrative is evidenced by the global, self-consistent and highly non-trivial quantitative agreement of hydrodynamic models with a large number of heavy-ion bulk observables \cite{Niemi:2015qia, Bernhard:2016tnd, Gale:2012rq}.
Naturally, such descriptions rely on the validity of hydrodynamic approximations, and these approximations begin to break down in the so-called dilute limit where discrete particle degrees-of-freedom dominate and continuous field descriptions of the medium cease to make sense.
Tell-tale signatures of hydrodynamic collectivity were thus always expected to vanish in smaller nuclear collision systems, e.g.\ $p$-$p$ and $p$-Pb collisions, where the number of produced particles is orders of magnitude smaller than a typical Pb-Pb collision.

These expectations were upended, however, when long-range multiparticle correlations were detected in high-multiplicity $p$-Pb collisions and found to be similar in magnitude to those observed in Pb-Pb collisions \cite{CMS:2012qk, Abelev:2012ola, Aad:2012gla}.
Nuclear collision systems which were previously thought to be too small for hydrodynamic flow, were subsequently found to generate the same collectivity used to justify hydrodynamic flow in heavy-ion collisions.
It is thus natural to wonder if a single unified hydrodynamic framework might be able to describe $p$-Pb and Pb-Pb bulk observables simultaneously.

In this work, we performed a semi-exhaustive search for a unified description of $p$-Pb and Pb-Pb collisions at $\sqrts=5.02$~TeV using Bayesian methods to rigorously calibrate and constrain free parameters of a flexible nuclear collision model based on viscous hydrodynamics.
The goal of our study was two fold.
First, we aimed to establish whether or not our hydrodynamic framework was able to describe both collision systems simultaneously.
And second, in the event that the former was true, we wished to obtain estimates for the \emph{true} parameters of our model given the assumptions of our framework and the evidence provided by the experimental data.

We built, for this purpose, a flexible multi-stage nuclear collision model characterized by a number of free parameters which vary theoretically uncertain aspects of the framework such as the QGP initial conditions and hydrodynamic transport properties.
For the QGP initial conditions, we employed a modified version of the \trento\ model \cite{Moreland:2014oya} which adds new parameters to vary the fluctuating size and shape of each nucleon.
Specifically, we modeled each nucleon as a cluster of $n_c$ constituents (hot spots), where each constituent is described by a Gaussian density profile of width $v$.
Each constituent position inside the nucleon was sampled randomly (without correlations) according to a Gaussian radial distribution of variable width, controlled indirectly by an effective nucleon width parameter $w$.

The transport dynamics of the collision were simulated using a pre-equilibrium free-streaming stage of variable duration followed by boost-invariant viscous hydrodynamics for hot and dense regions of the fireball and a microscopic hadronic afterburner for the relatively dilute corona.
We parametrized various sources of uncertainty in each stage of the collision including the duration of the pre-equilibrium free-streaming stage, the temperature dependence of the QGP shear and bulk viscosities, and the particlization temperature used to switch from a hydrodynamic description to microscopic Boltzmann transport.

With the full evolution model in hand, we applied Bayesian methods which were developed to estimate the parameters of computationally intensive models \cite{OHagan:2006ba, Higdon:2008cmc, Higdon:2014tva}.
We first constructed a scaffolding of $n=500$ parameters points distributed throughout our 15-dimensional parameter space and evaluated the nuclear collision model using \order{4} events at each parameter point.
The ensemble of events was then used to calculate a large number of experimental observables at each design point and train Gaussian process emulators to interpolate the model predictions as a function of the input parameters.
Finally, we used Markov chain Monte Carlo (MCMC) importance sampling to explore the parameter space and draw samples from the Bayesian posterior distribution for the \emph{true} values of the model parameters, given our emulated model predictions, the experimental data, and their associated uncertainties.
The model calibration process is summarized by Figs.~\ref{fig:obs_pbpb} and \ref{fig:obs_ppb}, and the resulting posterior distribution for the model input parameters is shown in Fig.~\ref{fig:posterior}.
We also evaluated the model predictions using a single set of high-probability parameters in Fig.~\ref{fig:obs_map}.
With these results, we are able to address the two primary goals of the study.

First, we demonstrated in Fig.~\ref{fig:obs_map} the existence of a single set of model parameters which can simultaneously describe $p$-Pb and Pb-Pb charged-particle yields, mean $p_T$, and flow cumulants at $\sqrts=5.02$~TeV.
The excellent quantitative agreement of the model calculation with the experimental data is strong evidence for a unified hydrodynamic description of $p$-A and A-A collisions at ultrarelativistic energies.
Moreover, the modifications to the physics model which were required to obtain this agreement are generally modest; one must simply replace Gaussian nucleons with composite nucleons of several or more constituents.

Second, we obtained a posterior distribution for the model input parameters in Fig.~\ref{fig:posterior} and reported quantitative estimates for these parameters in Table~\ref{tab:post_param}.
Here we summarize our key findings about the model:
\begin{enumerate}[itemsep=0pt, leftmargin=2\parindent]
  \item
    Using \trento\ initial conditions, we find that initial entropy deposition (or massless noninteracting particle production) scales like the geometric mean of local participant nuclear densities, in agreement with a number of previous estimates \cite{Moreland:2014oya, Bernhard:2016tnd, Bernhard:2018hnz}.
  \item
    Nucleon substructure is necessary to simultaneously describe $p$-Pb and Pb-Pb bulk observables. However, there is no strong preference for a specific number of constituents inside the nucleon.
  \item
    Bulk observables in $p$-Pb and Pb-Pb collisions prefer larger nucleons, similar in size to the proton's RMS electric charge radius.
    This agrees with a similar Bayesian analysis which was calibrated to Pb-Pb observables at $\sqrts=2.76$ and 5.02~TeV using a hybrid model without nucleon substructure \cite{Bernhard:2018hnz}.
    We note that our result is in significant tension with an estimate for the effective nucleon width based on the gluon distribution extracted from HERA data \cite{Rezaeian:2012ji}.
  \item
    We obtain an estimate $v=0.47_{-0.15}^{+0.20}$~fm for the Gaussian width of the constituent hot spots inside each nucleon.
    This is considerably larger than the length scales typically associated with nucleon substructure, and it may help explain the seemingly unreasonable success of hydrodynamics in small collision systems.
  \item
    Our posterior estimate for the pre-equilibrium free-streaming time is $\taufs=0.37_{-0.27}^{+0.33}\ \fmc$.
    This is significantly shorter than the estimate obtained from a similar Bayesian study in Ref.~\cite{Bernhard:2018hnz}, which reported $\taufs=1.16_{-0.25}^{+0.29}\ \fmc$.
    It is not clear whether the difference is a result of nucleon substructure or the different observables used to calibrate each analysis.
  \item
    We compare in Fig.~\ref{fig:region_shear_bulk} our estimate for the temperature dependence of the QGP specific shear and bulk viscosities to those of Ref.~\cite{Bernhard:2018hnz}, which performed a Bayesian calibration to Pb-Pb bulk observables at $\sqrts=2.76$ and 5.02~TeV using a physics model without nucleon substructure.
    The two studies are in good agreement, although Ref.~\cite{Bernhard:2018hnz} obtains a more precise estimate for $(\eta/s)(T)$, likely due to the additional beam energies and observables included, and an enhanced sensitivity of larger collision systems to the QGP viscosity.
  \item
    We make predictions in Figs.~\ref{fig:obs_map} and \ref{fig:flow_corr} for several quantities which were not included in the model calibration, including the identified yields, transverse energy, symmetric cumulants, and mean $p_T$ fluctuations at $\sqrts=5.02$~TeV.
    Interestingly, our MAP estimate for the Pb-Pb symmetric cumulants at $\sqrts=5.02$~TeV are significantly different than those estimated in Ref.~\cite{Bernhard:2018hnz}.
    This could be a direct (or indirect) result of including nucleon substructure in the model calibration.
\end{enumerate}

The present study would benefit from a number of improvements.
Perhaps the most obvious target for improvement is the absence of several important experimental datasets.
Specifically, we are missing the transverse energy, identified particle yields, and the mean $p_T$ fluctuations of both collision systems, as well the charged-particle mean $p_T$ for the Pb-Pb system.
These observables would certainly influence the quality of the combined fit and correspondingly our estimates for the model parameters.

Similarly, the results would greatly benefit from additional beam energies and collision systems.
Notably absent is data from RHIC which includes measurements for $p$-Au, $d$-Au, $^3$He-Au, Cu-Cu, Cu-Au, Au-Au, and \mbox{U-U} collision systems at $\sqrts=200$~GeV.
This data would undoubtedly provide an additional constraint on the model parameters and would enable more stringent tests of the calibrated model predictions.
The RHIC data may also help elucidate the beam-energy dependence of the model parameters which would be worth investigating.
We leave these improvements for future studies.

\medskip

All software used in this work is open source:
\begin{itemize}[leftmargin=2\parindent, itemsep=0pt, topsep=5pt]
  \item \trento\ with nucleon substructure (C++) \cite{trento:code}
  \item Pre-equilibrium free-streaming (Python) \cite{freestream:code}
  \item VISH2+1 hydrodynamics (Fortran) \cite{osuhydro:code}
  \item FRZOUT particle sample (Python) \cite{frzout:code}
  \item UrQMD microscopic transport model (Fortran) \cite{urqmd:code}
  \item DukeQCD event generator wrapper (Python) \cite{eventgen:code}
  \item Bayesian parameter estimation (Python) \cite{bayesian:code}
\end{itemize}

\medskip

\begin{acknowledgments}
  JSM thanks Berndt M\"uller, Weiyao Ke, Bj\"orn Schenke, and Heikki M\"antysaari for helpful discussions and clarifying comments.
  This research was completed using 6 million CPU hours provided by the National Energy Research Scientific Computing Center (NERSC), a U.S. Department of Energy Office of Science User Facility operated under Contract No. DE-AC02-05CH11231.
  JSM and SAB are supported by the U.S.\ Department of Energy Grant No.\ DE-FG02-05ER41367 and JEB by NSF Grant No.\ NSF-ACI-1550225.
\end{acknowledgments}

\bibliography{substructure}

\begin{thebibliography}{100}%
\makeatletter
\providecommand \@ifxundefined [1]{%
 \@ifx{#1\undefined}
}%
\providecommand \@ifnum [1]{%
 \ifnum #1\expandafter \@firstoftwo
 \else \expandafter \@secondoftwo
 \fi
}%
\providecommand \@ifx [1]{%
 \ifx #1\expandafter \@firstoftwo
 \else \expandafter \@secondoftwo
 \fi
}%
\providecommand \natexlab [1]{#1}%
\providecommand \enquote  [1]{``#1''}%
\providecommand \bibnamefont  [1]{#1}%
\providecommand \bibfnamefont [1]{#1}%
\providecommand \citenamefont [1]{#1}%
\providecommand \href@noop [0]{\@secondoftwo}%
\providecommand \href [0]{\begingroup \@sanitize@url \@href}%
\providecommand \@href[1]{\@@startlink{#1}\@@href}%
\providecommand \@@href[1]{\endgroup#1\@@endlink}%
\providecommand \@sanitize@url [0]{\catcode `\\12\catcode `\$12\catcode
  `\&12\catcode `\#12\catcode `\^12\catcode `\_12\catcode `\%12\relax}%
\providecommand \@@startlink[1]{}%
\providecommand \@@endlink[0]{}%
\providecommand \url  [0]{\begingroup\@sanitize@url \@url }%
\providecommand \@url [1]{\endgroup\@href {#1}{\urlprefix }}%
\providecommand \urlprefix  [0]{URL }%
\providecommand \Eprint [0]{\href }%
\providecommand \doibase [0]{http://dx.doi.org/}%
\providecommand \selectlanguage [0]{\@gobble}%
\providecommand \bibinfo  [0]{\@secondoftwo}%
\providecommand \bibfield  [0]{\@secondoftwo}%
\providecommand \translation [1]{[#1]}%
\providecommand \BibitemOpen [0]{}%
\providecommand \bibitemStop [0]{}%
\providecommand \bibitemNoStop [0]{.\EOS\space}%
\providecommand \EOS [0]{\spacefactor3000\relax}%
\providecommand \BibitemShut  [1]{\csname bibitem#1\endcsname}%
\let\auto@bib@innerbib\@empty
\bibitem [{\citenamefont {Chatrchyan}\ \emph
  {et~al.}(2013{\natexlab{a}})\citenamefont {Chatrchyan} \emph
  {et~al.}}]{CMS:2012qk}%
  \BibitemOpen
  \bibfield  {author} {\bibinfo {author} {\bibfnamefont {S.}~\bibnamefont
  {Chatrchyan}} \emph {et~al.} (\bibinfo {collaboration} {CMS}),\ }\href
  {\doibase 10.1016/j.physletb.2012.11.025} {\bibfield  {journal} {\bibinfo
  {journal} {Phys. Lett.}\ }\textbf {\bibinfo {volume} {B718}},\ \bibinfo
  {pages} {795} (\bibinfo {year} {2013}{\natexlab{a}})},\ \Eprint
  {http://arxiv.org/abs/1210.5482} {arXiv:1210.5482 [nucl-ex]} \BibitemShut
  {NoStop}%
\bibitem [{\citenamefont {Abelev}\ \emph
  {et~al.}(2013{\natexlab{a}})\citenamefont {Abelev} \emph
  {et~al.}}]{Abelev:2012ola}%
  \BibitemOpen
  \bibfield  {author} {\bibinfo {author} {\bibfnamefont {B.}~\bibnamefont
  {Abelev}} \emph {et~al.} (\bibinfo {collaboration} {ALICE}),\ }\href
  {\doibase 10.1016/j.physletb.2013.01.012} {\bibfield  {journal} {\bibinfo
  {journal} {Phys. Lett.}\ }\textbf {\bibinfo {volume} {B719}},\ \bibinfo
  {pages} {29} (\bibinfo {year} {2013}{\natexlab{a}})},\ \Eprint
  {http://arxiv.org/abs/1212.2001} {arXiv:1212.2001 [nucl-ex]} \BibitemShut
  {NoStop}%
\bibitem [{\citenamefont {Aad}\ \emph {et~al.}(2013)\citenamefont {Aad} \emph
  {et~al.}}]{Aad:2012gla}%
  \BibitemOpen
  \bibfield  {author} {\bibinfo {author} {\bibfnamefont {G.}~\bibnamefont
  {Aad}} \emph {et~al.} (\bibinfo {collaboration} {ATLAS}),\ }\href {\doibase
  10.1103/PhysRevLett.110.182302} {\bibfield  {journal} {\bibinfo  {journal}
  {Phys. Rev. Lett.}\ }\textbf {\bibinfo {volume} {110}},\ \bibinfo {pages}
  {182302} (\bibinfo {year} {2013})},\ \Eprint {http://arxiv.org/abs/1212.5198}
  {arXiv:1212.5198 [hep-ex]} \BibitemShut {NoStop}%
\bibitem [{\citenamefont {Adare}\ \emph {et~al.}(2015)\citenamefont {Adare}
  \emph {et~al.}}]{Adare:2015ctn}%
  \BibitemOpen
  \bibfield  {author} {\bibinfo {author} {\bibfnamefont {A.}~\bibnamefont
  {Adare}} \emph {et~al.} (\bibinfo {collaboration} {PHENIX}),\ }\href
  {\doibase 10.1103/PhysRevLett.115.142301} {\bibfield  {journal} {\bibinfo
  {journal} {Phys. Rev. Lett.}\ }\textbf {\bibinfo {volume} {115}},\ \bibinfo
  {pages} {142301} (\bibinfo {year} {2015})},\ \Eprint
  {http://arxiv.org/abs/1507.06273} {arXiv:1507.06273 [nucl-ex]} \BibitemShut
  {NoStop}%
\bibitem [{\citenamefont {Bozek}(2012)}]{Bozek:2011if}%
  \BibitemOpen
  \bibfield  {author} {\bibinfo {author} {\bibfnamefont {P.}~\bibnamefont
  {Bozek}},\ }\href {\doibase 10.1103/PhysRevC.85.014911} {\bibfield  {journal}
  {\bibinfo  {journal} {Phys. Rev.}\ }\textbf {\bibinfo {volume} {C85}},\
  \bibinfo {pages} {014911} (\bibinfo {year} {2012})},\ \Eprint
  {http://arxiv.org/abs/1112.0915} {arXiv:1112.0915 [hep-ph]} \BibitemShut
  {NoStop}%
\bibitem [{\citenamefont {Bozek}\ and\ \citenamefont
  {Broniowski}(2013)}]{Bozek:2013uha}%
  \BibitemOpen
  \bibfield  {author} {\bibinfo {author} {\bibfnamefont {P.}~\bibnamefont
  {Bozek}}\ and\ \bibinfo {author} {\bibfnamefont {W.}~\bibnamefont
  {Broniowski}},\ }\href {\doibase 10.1103/PhysRevC.88.014903} {\bibfield
  {journal} {\bibinfo  {journal} {Phys. Rev.}\ }\textbf {\bibinfo {volume}
  {C88}},\ \bibinfo {pages} {014903} (\bibinfo {year} {2013})},\ \Eprint
  {http://arxiv.org/abs/1304.3044} {arXiv:1304.3044 [nucl-th]} \BibitemShut
  {NoStop}%
\bibitem [{\citenamefont {Schenke}\ and\ \citenamefont
  {Venugopalan}(2014{\natexlab{a}})}]{Schenke:2014zha}%
  \BibitemOpen
  \bibfield  {author} {\bibinfo {author} {\bibfnamefont {B.}~\bibnamefont
  {Schenke}}\ and\ \bibinfo {author} {\bibfnamefont {R.}~\bibnamefont
  {Venugopalan}},\ }\href {\doibase 10.1103/PhysRevLett.113.102301} {\bibfield
  {journal} {\bibinfo  {journal} {Phys. Rev. Lett.}\ }\textbf {\bibinfo
  {volume} {113}},\ \bibinfo {pages} {102301} (\bibinfo {year}
  {2014}{\natexlab{a}})},\ \Eprint {http://arxiv.org/abs/1405.3605}
  {arXiv:1405.3605 [nucl-th]} \BibitemShut {NoStop}%
\bibitem [{\citenamefont {Niemi}(2014)}]{Niemi:2014lha}%
  \BibitemOpen
  \bibfield  {author} {\bibinfo {author} {\bibfnamefont {H.}~\bibnamefont
  {Niemi}},\ }\bibfield  {booktitle} {\emph {\bibinfo {booktitle}
  {{Proceedings, 24th International Conference on Ultra-Relativistic
  Nucleus-Nucleus Collisions (Quark Matter 2014): Darmstadt, Germany, May
  19-24, 2014}}},\ }\href {\doibase 10.1016/j.nuclphysa.2014.09.100} {\bibfield
   {journal} {\bibinfo  {journal} {Nucl. Phys.}\ }\textbf {\bibinfo {volume}
  {A931}},\ \bibinfo {pages} {227} (\bibinfo {year} {2014})}\BibitemShut
  {NoStop}%
\bibitem [{\citenamefont {Derradi~de Souza}\ \emph {et~al.}(2016)\citenamefont
  {Derradi~de Souza}, \citenamefont {Koide},\ and\ \citenamefont
  {Kodama}}]{deSouza:2015ena}%
  \BibitemOpen
  \bibfield  {author} {\bibinfo {author} {\bibfnamefont {R.}~\bibnamefont
  {Derradi~de Souza}}, \bibinfo {author} {\bibfnamefont {T.}~\bibnamefont
  {Koide}}, \ and\ \bibinfo {author} {\bibfnamefont {T.}~\bibnamefont
  {Kodama}},\ }\href {\doibase 10.1016/j.ppnp.2015.09.002} {\bibfield
  {journal} {\bibinfo  {journal} {Prog. Part. Nucl. Phys.}\ }\textbf {\bibinfo
  {volume} {86}},\ \bibinfo {pages} {35} (\bibinfo {year} {2016})},\ \Eprint
  {http://arxiv.org/abs/1506.03863} {arXiv:1506.03863 [nucl-th]} \BibitemShut
  {NoStop}%
\bibitem [{\citenamefont {Ollitrault}\ and\ \citenamefont
  {Gardim}(2013)}]{Ollitrault:2012cm}%
  \BibitemOpen
  \bibfield  {author} {\bibinfo {author} {\bibfnamefont {J.-Y.}\ \bibnamefont
  {Ollitrault}}\ and\ \bibinfo {author} {\bibfnamefont {F.~G.}\ \bibnamefont
  {Gardim}},\ }\bibfield  {booktitle} {\emph {\bibinfo {booktitle}
  {{Proceedings, 23rd International Conference on Ultrarelativistic
  Nucleus-Nucleus Collisions : Quark Matter 2012 (QM 2012): Washington, DC,
  USA, August 13-18, 2012}}},\ }\href {\doibase
  10.1016/j.nuclphysa.2013.01.047} {\bibfield  {journal} {\bibinfo  {journal}
  {Nucl. Phys.}\ }\textbf {\bibinfo {volume} {A904-905}},\ \bibinfo {pages}
  {75c} (\bibinfo {year} {2013})},\ \Eprint {http://arxiv.org/abs/1210.8345}
  {arXiv:1210.8345 [nucl-th]} \BibitemShut {NoStop}%
\bibitem [{\citenamefont {Song}(2013)}]{Song:2012ua}%
  \BibitemOpen
  \bibfield  {author} {\bibinfo {author} {\bibfnamefont {H.}~\bibnamefont
  {Song}},\ }\bibfield  {booktitle} {\emph {\bibinfo {booktitle} {{Proceedings,
  23rd International Conference on Ultrarelativistic Nucleus-Nucleus Collisions
  : Quark Matter 2012 (QM 2012): Washington, DC, USA, August 13-18, 2012}}},\
  }\href {\doibase 10.1016/j.nuclphysa.2013.01.052} {\bibfield  {journal}
  {\bibinfo  {journal} {Nucl. Phys.}\ }\textbf {\bibinfo {volume} {A904-905}},\
  \bibinfo {pages} {114c} (\bibinfo {year} {2013})},\ \Eprint
  {http://arxiv.org/abs/1210.5778} {arXiv:1210.5778 [nucl-th]} \BibitemShut
  {NoStop}%
\bibitem [{\citenamefont {Adare}\ \emph {et~al.}(2016)\citenamefont {Adare}
  \emph {et~al.}}]{Adare:2015bua}%
  \BibitemOpen
  \bibfield  {author} {\bibinfo {author} {\bibfnamefont {A.}~\bibnamefont
  {Adare}} \emph {et~al.} (\bibinfo {collaboration} {PHENIX}),\ }\href
  {\doibase 10.1103/PhysRevC.93.024901} {\bibfield  {journal} {\bibinfo
  {journal} {Phys. Rev.}\ }\textbf {\bibinfo {volume} {C93}},\ \bibinfo {pages}
  {024901} (\bibinfo {year} {2016})},\ \Eprint
  {http://arxiv.org/abs/1509.06727} {arXiv:1509.06727 [nucl-ex]} \BibitemShut
  {NoStop}%
\bibitem [{\citenamefont {Schenke}\ \emph {et~al.}(2014)\citenamefont
  {Schenke}, \citenamefont {Tribedy},\ and\ \citenamefont
  {Venugopalan}}]{Schenke:2014tga}%
  \BibitemOpen
  \bibfield  {author} {\bibinfo {author} {\bibfnamefont {B.}~\bibnamefont
  {Schenke}}, \bibinfo {author} {\bibfnamefont {P.}~\bibnamefont {Tribedy}}, \
  and\ \bibinfo {author} {\bibfnamefont {R.}~\bibnamefont {Venugopalan}},\
  }\href {\doibase 10.1103/PhysRevC.89.064908} {\bibfield  {journal} {\bibinfo
  {journal} {Phys. Rev.}\ }\textbf {\bibinfo {volume} {C89}},\ \bibinfo {pages}
  {064908} (\bibinfo {year} {2014})},\ \Eprint {http://arxiv.org/abs/1403.2232}
  {arXiv:1403.2232 [nucl-th]} \BibitemShut {NoStop}%
\bibitem [{\citenamefont {Aidala}\ \emph
  {et~al.}(2018{\natexlab{a}})\citenamefont {Aidala} \emph
  {et~al.}}]{Aidala:2018mcw}%
  \BibitemOpen
  \bibfield  {author} {\bibinfo {author} {\bibfnamefont {C.}~\bibnamefont
  {Aidala}} \emph {et~al.} (\bibinfo {collaboration} {PHENIX}),\ }\href@noop {}
  {\  (\bibinfo {year} {2018}{\natexlab{a}})},\ \Eprint
  {http://arxiv.org/abs/1805.02973} {arXiv:1805.02973 [nucl-ex]} \BibitemShut
  {NoStop}%
\bibitem [{\citenamefont {Adare}\ \emph {et~al.}(2018)\citenamefont {Adare}
  \emph {et~al.}}]{Adare:2017wlc}%
  \BibitemOpen
  \bibfield  {author} {\bibinfo {author} {\bibfnamefont {A.}~\bibnamefont
  {Adare}} \emph {et~al.} (\bibinfo {collaboration} {PHENIX}),\ }\href
  {\doibase 10.1103/PhysRevC.97.064904} {\bibfield  {journal} {\bibinfo
  {journal} {Phys. Rev.}\ }\textbf {\bibinfo {volume} {C97}},\ \bibinfo {pages}
  {064904} (\bibinfo {year} {2018})},\ \Eprint
  {http://arxiv.org/abs/1710.09736} {arXiv:1710.09736 [nucl-ex]} \BibitemShut
  {NoStop}%
\bibitem [{\citenamefont {Adamczyk}\ \emph {et~al.}(2015)\citenamefont
  {Adamczyk} \emph {et~al.}}]{Adamczyk:2015obl}%
  \BibitemOpen
  \bibfield  {author} {\bibinfo {author} {\bibfnamefont {L.}~\bibnamefont
  {Adamczyk}} \emph {et~al.} (\bibinfo {collaboration} {STAR}),\ }\href
  {\doibase 10.1103/PhysRevLett.115.222301} {\bibfield  {journal} {\bibinfo
  {journal} {Phys. Rev. Lett.}\ }\textbf {\bibinfo {volume} {115}},\ \bibinfo
  {pages} {222301} (\bibinfo {year} {2015})},\ \Eprint
  {http://arxiv.org/abs/1505.07812} {arXiv:1505.07812 [nucl-ex]} \BibitemShut
  {NoStop}%
\bibitem [{\citenamefont {Shen}\ \emph {et~al.}(2017)\citenamefont {Shen},
  \citenamefont {Paquet}, \citenamefont {Denicol}, \citenamefont {Jeon},\ and\
  \citenamefont {Gale}}]{Shen:2016zpp}%
  \BibitemOpen
  \bibfield  {author} {\bibinfo {author} {\bibfnamefont {C.}~\bibnamefont
  {Shen}}, \bibinfo {author} {\bibfnamefont {J.-F.}\ \bibnamefont {Paquet}},
  \bibinfo {author} {\bibfnamefont {G.~S.}\ \bibnamefont {Denicol}}, \bibinfo
  {author} {\bibfnamefont {S.}~\bibnamefont {Jeon}}, \ and\ \bibinfo {author}
  {\bibfnamefont {C.}~\bibnamefont {Gale}},\ }\href {\doibase
  10.1103/PhysRevC.95.014906} {\bibfield  {journal} {\bibinfo  {journal} {Phys.
  Rev.}\ }\textbf {\bibinfo {volume} {C95}},\ \bibinfo {pages} {014906}
  (\bibinfo {year} {2017})},\ \Eprint {http://arxiv.org/abs/1609.02590}
  {arXiv:1609.02590 [nucl-th]} \BibitemShut {NoStop}%
\bibitem [{\citenamefont {Aidala}\ \emph
  {et~al.}(2018{\natexlab{b}})\citenamefont {Aidala} \emph
  {et~al.}}]{Aidala:2017ajz}%
  \BibitemOpen
  \bibfield  {author} {\bibinfo {author} {\bibfnamefont {C.}~\bibnamefont
  {Aidala}} \emph {et~al.} (\bibinfo {collaboration} {PHENIX}),\ }\href
  {\doibase 10.1103/PhysRevLett.120.062302} {\bibfield  {journal} {\bibinfo
  {journal} {Phys. Rev. Lett.}\ }\textbf {\bibinfo {volume} {120}},\ \bibinfo
  {pages} {062302} (\bibinfo {year} {2018}{\natexlab{b}})},\ \Eprint
  {http://arxiv.org/abs/1707.06108} {arXiv:1707.06108 [nucl-ex]} \BibitemShut
  {NoStop}%
\bibitem [{\citenamefont {Adare}\ \emph {et~al.}(2007)\citenamefont {Adare}
  \emph {et~al.}}]{Adare:2006ti}%
  \BibitemOpen
  \bibfield  {author} {\bibinfo {author} {\bibfnamefont {A.}~\bibnamefont
  {Adare}} \emph {et~al.} (\bibinfo {collaboration} {PHENIX}),\ }\href
  {\doibase 10.1103/PhysRevLett.98.162301} {\bibfield  {journal} {\bibinfo
  {journal} {Phys. Rev. Lett.}\ }\textbf {\bibinfo {volume} {98}},\ \bibinfo
  {pages} {162301} (\bibinfo {year} {2007})},\ \Eprint
  {http://arxiv.org/abs/nucl-ex/0608033} {arXiv:nucl-ex/0608033 [nucl-ex]}
  \BibitemShut {NoStop}%
\bibitem [{\citenamefont {Moller}\ \emph {et~al.}(1995)\citenamefont {Moller},
  \citenamefont {Nix}, \citenamefont {Myers},\ and\ \citenamefont
  {Swiatecki}}]{MOLLER1995185}%
  \BibitemOpen
  \bibfield  {author} {\bibinfo {author} {\bibfnamefont {P.}~\bibnamefont
  {Moller}}, \bibinfo {author} {\bibfnamefont {J.}~\bibnamefont {Nix}},
  \bibinfo {author} {\bibfnamefont {W.}~\bibnamefont {Myers}}, \ and\ \bibinfo
  {author} {\bibfnamefont {W.}~\bibnamefont {Swiatecki}},\ }\href {\doibase
  https://doi.org/10.1006/adnd.1995.1002} {\bibfield  {journal} {\bibinfo
  {journal} {Atomic Data and Nuclear Data Tables}\ }\textbf {\bibinfo {volume}
  {59}},\ \bibinfo {pages} {185 } (\bibinfo {year} {1995})}\BibitemShut
  {NoStop}%
\bibitem [{\citenamefont {Vries}\ \emph {et~al.}(1987)\citenamefont {Vries},
  \citenamefont {Jager},\ and\ \citenamefont {Vries}}]{DEVRIES1987495}%
  \BibitemOpen
  \bibfield  {author} {\bibinfo {author} {\bibfnamefont {H.~D.}\ \bibnamefont
  {Vries}}, \bibinfo {author} {\bibfnamefont {C.~D.}\ \bibnamefont {Jager}}, \
  and\ \bibinfo {author} {\bibfnamefont {C.~D.}\ \bibnamefont {Vries}},\ }\href
  {\doibase https://doi.org/10.1016/0092-640X(87)90013-1} {\bibfield  {journal}
  {\bibinfo  {journal} {Atomic Data and Nuclear Data Tables}\ }\textbf
  {\bibinfo {volume} {36}},\ \bibinfo {pages} {495 } (\bibinfo {year}
  {1987})}\BibitemShut {NoStop}%
\bibitem [{\citenamefont {Song}\ \emph {et~al.}(2011)\citenamefont {Song},
  \citenamefont {Bass}, \citenamefont {Heinz}, \citenamefont {Hirano},\ and\
  \citenamefont {Shen}}]{Song:2011hk}%
  \BibitemOpen
  \bibfield  {author} {\bibinfo {author} {\bibfnamefont {H.}~\bibnamefont
  {Song}}, \bibinfo {author} {\bibfnamefont {S.~A.}\ \bibnamefont {Bass}},
  \bibinfo {author} {\bibfnamefont {U.}~\bibnamefont {Heinz}}, \bibinfo
  {author} {\bibfnamefont {T.}~\bibnamefont {Hirano}}, \ and\ \bibinfo {author}
  {\bibfnamefont {C.}~\bibnamefont {Shen}},\ }\href {\doibase
  10.1103/PhysRevC.83.054910, 10.1103/PhysRevC.86.059903} {\bibfield  {journal}
  {\bibinfo  {journal} {Phys. Rev.}\ }\textbf {\bibinfo {volume} {C83}},\
  \bibinfo {pages} {054910} (\bibinfo {year} {2011})},\ \bibinfo {note}
  {[Erratum: Phys. Rev.C86,059903(2012)]},\ \Eprint
  {http://arxiv.org/abs/1101.4638} {arXiv:1101.4638 [nucl-th]} \BibitemShut
  {NoStop}%
\bibitem [{\citenamefont {Retinskaya}\ \emph {et~al.}(2014)\citenamefont
  {Retinskaya}, \citenamefont {Luzum},\ and\ \citenamefont
  {Ollitrault}}]{Retinskaya:2013gca}%
  \BibitemOpen
  \bibfield  {author} {\bibinfo {author} {\bibfnamefont {E.}~\bibnamefont
  {Retinskaya}}, \bibinfo {author} {\bibfnamefont {M.}~\bibnamefont {Luzum}}, \
  and\ \bibinfo {author} {\bibfnamefont {J.-Y.}\ \bibnamefont {Ollitrault}},\
  }\href {\doibase 10.1103/PhysRevC.89.014902} {\bibfield  {journal} {\bibinfo
  {journal} {Phys. Rev.}\ }\textbf {\bibinfo {volume} {C89}},\ \bibinfo {pages}
  {014902} (\bibinfo {year} {2014})},\ \Eprint {http://arxiv.org/abs/1311.5339}
  {arXiv:1311.5339 [nucl-th]} \BibitemShut {NoStop}%
\bibitem [{\citenamefont {Liu}\ \emph {et~al.}(2015)\citenamefont {Liu},
  \citenamefont {Shen},\ and\ \citenamefont {Heinz}}]{Liu:2015nwa}%
  \BibitemOpen
  \bibfield  {author} {\bibinfo {author} {\bibfnamefont {J.}~\bibnamefont
  {Liu}}, \bibinfo {author} {\bibfnamefont {C.}~\bibnamefont {Shen}}, \ and\
  \bibinfo {author} {\bibfnamefont {U.}~\bibnamefont {Heinz}},\ }\href
  {\doibase 10.1103/PhysRevC.92.049904, 10.1103/PhysRevC.91.064906} {\bibfield
  {journal} {\bibinfo  {journal} {Phys. Rev.}\ }\textbf {\bibinfo {volume}
  {C91}},\ \bibinfo {pages} {064906} (\bibinfo {year} {2015})},\ \bibinfo
  {note} {[Erratum: Phys. Rev.C92,no.4,049904(2015)]},\ \Eprint
  {http://arxiv.org/abs/1504.02160} {arXiv:1504.02160 [nucl-th]} \BibitemShut
  {NoStop}%
\bibitem [{\citenamefont {Kurkela}(2016)}]{Kurkela:2016vts}%
  \BibitemOpen
  \bibfield  {author} {\bibinfo {author} {\bibfnamefont {A.}~\bibnamefont
  {Kurkela}},\ }\bibfield  {booktitle} {\emph {\bibinfo {booktitle}
  {{Proceedings, 25th International Conference on Ultra-Relativistic
  Nucleus-Nucleus Collisions (Quark Matter 2015): Kobe, Japan, September
  27-October 3, 2015}}},\ }\href {\doibase 10.1016/j.nuclphysa.2016.01.069}
  {\bibfield  {journal} {\bibinfo  {journal} {Nucl. Phys.}\ }\textbf {\bibinfo
  {volume} {A956}},\ \bibinfo {pages} {136} (\bibinfo {year} {2016})},\ \Eprint
  {http://arxiv.org/abs/1601.03283} {arXiv:1601.03283 [hep-ph]} \BibitemShut
  {NoStop}%
\bibitem [{\citenamefont {Welsh}\ \emph {et~al.}(2016)\citenamefont {Welsh},
  \citenamefont {Singer},\ and\ \citenamefont {Heinz}}]{Welsh:2016siu}%
  \BibitemOpen
  \bibfield  {author} {\bibinfo {author} {\bibfnamefont {K.}~\bibnamefont
  {Welsh}}, \bibinfo {author} {\bibfnamefont {J.}~\bibnamefont {Singer}}, \
  and\ \bibinfo {author} {\bibfnamefont {U.~W.}\ \bibnamefont {Heinz}},\ }\href
  {\doibase 10.1103/PhysRevC.94.024919} {\bibfield  {journal} {\bibinfo
  {journal} {Phys. Rev.}\ }\textbf {\bibinfo {volume} {C94}},\ \bibinfo {pages}
  {024919} (\bibinfo {year} {2016})},\ \Eprint
  {http://arxiv.org/abs/1605.09418} {arXiv:1605.09418 [nucl-th]} \BibitemShut
  {NoStop}%
\bibitem [{\citenamefont {Moreland}\ \emph {et~al.}(2017)\citenamefont
  {Moreland}, \citenamefont {Bernhard}, \citenamefont {Ke},\ and\ \citenamefont
  {Bass}}]{Moreland:2017kdx}%
  \BibitemOpen
  \bibfield  {author} {\bibinfo {author} {\bibfnamefont {J.~S.}\ \bibnamefont
  {Moreland}}, \bibinfo {author} {\bibfnamefont {J.~E.}\ \bibnamefont
  {Bernhard}}, \bibinfo {author} {\bibfnamefont {W.}~\bibnamefont {Ke}}, \ and\
  \bibinfo {author} {\bibfnamefont {S.~A.}\ \bibnamefont {Bass}},\ }\bibfield
  {booktitle} {\emph {\bibinfo {booktitle} {{Proceedings, 26th International
  Conference on Ultra-relativistic Nucleus-Nucleus Collisions (Quark Matter
  2017): Chicago, Illinois, USA, February 5-11, 2017}}},\ }\href {\doibase
  10.1016/j.nuclphysa.2017.05.054} {\bibfield  {journal} {\bibinfo  {journal}
  {Nucl. Phys.}\ }\textbf {\bibinfo {volume} {A967}},\ \bibinfo {pages} {361}
  (\bibinfo {year} {2017})},\ \Eprint {http://arxiv.org/abs/1704.04486}
  {arXiv:1704.04486 [nucl-th]} \BibitemShut {NoStop}%
\bibitem [{\citenamefont {Schenke}\ and\ \citenamefont
  {Venugopalan}(2014{\natexlab{b}})}]{Schenke:2014gaa}%
  \BibitemOpen
  \bibfield  {author} {\bibinfo {author} {\bibfnamefont {B.}~\bibnamefont
  {Schenke}}\ and\ \bibinfo {author} {\bibfnamefont {R.}~\bibnamefont
  {Venugopalan}},\ }\bibfield  {booktitle} {\emph {\bibinfo {booktitle}
  {{Proceedings, 24th International Conference on Ultra-Relativistic
  Nucleus-Nucleus Collisions (Quark Matter 2014): Darmstadt, Germany, May
  19-24, 2014}}},\ }\href {\doibase 10.1016/j.nuclphysa.2014.08.092} {\bibfield
   {journal} {\bibinfo  {journal} {Nucl. Phys.}\ }\textbf {\bibinfo {volume}
  {A931}},\ \bibinfo {pages} {1039} (\bibinfo {year} {2014}{\natexlab{b}})},\
  \Eprint {http://arxiv.org/abs/1407.7557} {arXiv:1407.7557 [nucl-th]}
  \BibitemShut {NoStop}%
\bibitem [{\citenamefont {Schlichting}\ and\ \citenamefont
  {Schenke}(2014)}]{Schlichting:2014ipa}%
  \BibitemOpen
  \bibfield  {author} {\bibinfo {author} {\bibfnamefont {S.}~\bibnamefont
  {Schlichting}}\ and\ \bibinfo {author} {\bibfnamefont {B.}~\bibnamefont
  {Schenke}},\ }\href {\doibase 10.1016/j.physletb.2014.10.068} {\bibfield
  {journal} {\bibinfo  {journal} {Phys. Lett.}\ }\textbf {\bibinfo {volume}
  {B739}},\ \bibinfo {pages} {313} (\bibinfo {year} {2014})},\ \Eprint
  {http://arxiv.org/abs/1407.8458} {arXiv:1407.8458 [hep-ph]} \BibitemShut
  {NoStop}%
\bibitem [{\citenamefont {Adler}\ \emph {et~al.}(2014)\citenamefont {Adler}
  \emph {et~al.}}]{Adler:2013aqf}%
  \BibitemOpen
  \bibfield  {author} {\bibinfo {author} {\bibfnamefont {S.~S.}\ \bibnamefont
  {Adler}} \emph {et~al.} (\bibinfo {collaboration} {PHENIX}),\ }\href
  {\doibase 10.1103/PhysRevC.89.044905} {\bibfield  {journal} {\bibinfo
  {journal} {Phys. Rev.}\ }\textbf {\bibinfo {volume} {C89}},\ \bibinfo {pages}
  {044905} (\bibinfo {year} {2014})},\ \Eprint {http://arxiv.org/abs/1312.6676}
  {arXiv:1312.6676 [nucl-ex]} \BibitemShut {NoStop}%
\bibitem [{\citenamefont {Mitchell}\ \emph {et~al.}(2016)\citenamefont
  {Mitchell}, \citenamefont {Perepelitsa}, \citenamefont {Tannenbaum},\ and\
  \citenamefont {Stankus}}]{Mitchell:2016jio}%
  \BibitemOpen
  \bibfield  {author} {\bibinfo {author} {\bibfnamefont {J.~T.}\ \bibnamefont
  {Mitchell}}, \bibinfo {author} {\bibfnamefont {D.~V.}\ \bibnamefont
  {Perepelitsa}}, \bibinfo {author} {\bibfnamefont {M.~J.}\ \bibnamefont
  {Tannenbaum}}, \ and\ \bibinfo {author} {\bibfnamefont {P.~W.}\ \bibnamefont
  {Stankus}},\ }\href {\doibase 10.1103/PhysRevC.93.054910} {\bibfield
  {journal} {\bibinfo  {journal} {Phys. Rev.}\ }\textbf {\bibinfo {volume}
  {C93}},\ \bibinfo {pages} {054910} (\bibinfo {year} {2016})},\ \Eprint
  {http://arxiv.org/abs/1603.08836} {arXiv:1603.08836 [nucl-ex]} \BibitemShut
  {NoStop}%
\bibitem [{\citenamefont {Broniowski}\ \emph {et~al.}(2017)\citenamefont
  {Broniowski}, \citenamefont {Bozek},\ and\ \citenamefont
  {Rybczynski}}]{Broniowski:2016pvx}%
  \BibitemOpen
  \bibfield  {author} {\bibinfo {author} {\bibfnamefont {W.}~\bibnamefont
  {Broniowski}}, \bibinfo {author} {\bibfnamefont {P.}~\bibnamefont {Bozek}}, \
  and\ \bibinfo {author} {\bibfnamefont {M.}~\bibnamefont {Rybczynski}},\
  }\bibfield  {booktitle} {\emph {\bibinfo {booktitle} {{Proceedings, 10th
  International Workshop on Critical Point and Onset of Deconfinement (CPOD
  2016): Wrocław, Poland}}},\ }\href {\doibase 10.5506/APhysPolBSupp.10.513}
  {\bibfield  {journal} {\bibinfo  {journal} {Acta Phys. Polon. Supp.}\
  }\textbf {\bibinfo {volume} {10}},\ \bibinfo {pages} {513} (\bibinfo {year}
  {2017})},\ \Eprint {http://arxiv.org/abs/1611.00250} {arXiv:1611.00250
  [nucl-th]} \BibitemShut {NoStop}%
\bibitem [{\citenamefont {Bozek}\ \emph {et~al.}(2017)\citenamefont {Bozek},
  \citenamefont {Broniowski},\ and\ \citenamefont
  {Chatterjee}}]{Bozek:2017jog}%
  \BibitemOpen
  \bibfield  {author} {\bibinfo {author} {\bibfnamefont {P.}~\bibnamefont
  {Bozek}}, \bibinfo {author} {\bibfnamefont {W.}~\bibnamefont {Broniowski}}, \
  and\ \bibinfo {author} {\bibfnamefont {S.}~\bibnamefont {Chatterjee}},\
  }\bibfield  {booktitle} {\emph {\bibinfo {booktitle} {{Proceedings, 9th
  Workshop "Excited QCD" 2017: Sintra, Portugal, May 7-13, 2017}}},\ }\href
  {\doibase 10.5506/APhysPolBSupp.10.1091} {\bibfield  {journal} {\bibinfo
  {journal} {Acta Phys. Polon. Supp.}\ }\textbf {\bibinfo {volume} {10}},\
  \bibinfo {pages} {1091} (\bibinfo {year} {2017})},\ \Eprint
  {http://arxiv.org/abs/1707.04420} {arXiv:1707.04420 [nucl-th]} \BibitemShut
  {NoStop}%
\bibitem [{\citenamefont {Antchev}\ \emph {et~al.}(2011)\citenamefont {Antchev}
  \emph {et~al.}}]{Antchev:2011zz}%
  \BibitemOpen
  \bibfield  {author} {\bibinfo {author} {\bibfnamefont {G.}~\bibnamefont
  {Antchev}} \emph {et~al.} (\bibinfo {collaboration} {TOTEM}),\ }\href
  {\doibase 10.1209/0295-5075/95/41001} {\bibfield  {journal} {\bibinfo
  {journal} {EPL}\ }\textbf {\bibinfo {volume} {95}},\ \bibinfo {pages} {41001}
  (\bibinfo {year} {2011})},\ \Eprint {http://arxiv.org/abs/1110.1385}
  {arXiv:1110.1385 [hep-ex]} \BibitemShut {NoStop}%
\bibitem [{\citenamefont {Arriola}\ and\ \citenamefont
  {Broniowski}(2016)}]{Arriola2016}%
  \BibitemOpen
  \bibfield  {author} {\bibinfo {author} {\bibfnamefont {E.~R.}\ \bibnamefont
  {Arriola}}\ and\ \bibinfo {author} {\bibfnamefont {W.}~\bibnamefont
  {Broniowski}},\ }\href {\doibase 10.1007/s00601-016-1095-z} {\bibfield
  {journal} {\bibinfo  {journal} {Few-Body Systems}\ }\textbf {\bibinfo
  {volume} {57}},\ \bibinfo {pages} {485} (\bibinfo {year} {2016})}\BibitemShut
  {NoStop}%
\bibitem [{\citenamefont {Albacete}\ \emph {et~al.}(2017)\citenamefont
  {Albacete}, \citenamefont {Petersen},\ and\ \citenamefont
  {Soto-Ontoso}}]{Albacete:2016gxu}%
  \BibitemOpen
  \bibfield  {author} {\bibinfo {author} {\bibfnamefont {J.~L.}\ \bibnamefont
  {Albacete}}, \bibinfo {author} {\bibfnamefont {H.}~\bibnamefont {Petersen}},
  \ and\ \bibinfo {author} {\bibfnamefont {A.}~\bibnamefont {Soto-Ontoso}},\
  }\href {\doibase 10.1103/PhysRevC.95.064909} {\bibfield  {journal} {\bibinfo
  {journal} {Phys. Rev.}\ }\textbf {\bibinfo {volume} {C95}},\ \bibinfo {pages}
  {064909} (\bibinfo {year} {2017})},\ \Eprint
  {http://arxiv.org/abs/1612.06274} {arXiv:1612.06274 [hep-ph]} \BibitemShut
  {NoStop}%
\bibitem [{\citenamefont {Albacete}\ and\ \citenamefont
  {Soto-Ontoso}(2017)}]{Albacete:2016pmp}%
  \BibitemOpen
  \bibfield  {author} {\bibinfo {author} {\bibfnamefont {J.~L.}\ \bibnamefont
  {Albacete}}\ and\ \bibinfo {author} {\bibfnamefont {A.}~\bibnamefont
  {Soto-Ontoso}},\ }\href {\doibase 10.1016/j.physletb.2017.04.055} {\bibfield
  {journal} {\bibinfo  {journal} {Phys. Lett.}\ }\textbf {\bibinfo {volume}
  {B770}},\ \bibinfo {pages} {149} (\bibinfo {year} {2017})},\ \Eprint
  {http://arxiv.org/abs/1605.09176} {arXiv:1605.09176 [hep-ph]} \BibitemShut
  {NoStop}%
\bibitem [{\citenamefont {Mäntysaari}\ and\ \citenamefont
  {Schenke}(2016{\natexlab{a}})}]{Mantysaari:2016ykx}%
  \BibitemOpen
  \bibfield  {author} {\bibinfo {author} {\bibfnamefont {H.}~\bibnamefont
  {Mäntysaari}}\ and\ \bibinfo {author} {\bibfnamefont {B.}~\bibnamefont
  {Schenke}},\ }\href {\doibase 10.1103/PhysRevLett.117.052301} {\bibfield
  {journal} {\bibinfo  {journal} {Phys. Rev. Lett.}\ }\textbf {\bibinfo
  {volume} {117}},\ \bibinfo {pages} {052301} (\bibinfo {year}
  {2016}{\natexlab{a}})},\ \Eprint {http://arxiv.org/abs/1603.04349}
  {arXiv:1603.04349 [hep-ph]} \BibitemShut {NoStop}%
\bibitem [{\citenamefont {Mäntysaari}\ and\ \citenamefont
  {Schenke}(2016{\natexlab{b}})}]{Mantysaari:2016jaz}%
  \BibitemOpen
  \bibfield  {author} {\bibinfo {author} {\bibfnamefont {H.}~\bibnamefont
  {Mäntysaari}}\ and\ \bibinfo {author} {\bibfnamefont {B.}~\bibnamefont
  {Schenke}},\ }\href {\doibase 10.1103/PhysRevD.94.034042} {\bibfield
  {journal} {\bibinfo  {journal} {Phys. Rev.}\ }\textbf {\bibinfo {volume}
  {D94}},\ \bibinfo {pages} {034042} (\bibinfo {year} {2016}{\natexlab{b}})},\
  \Eprint {http://arxiv.org/abs/1607.01711} {arXiv:1607.01711 [hep-ph]}
  \BibitemShut {NoStop}%
\bibitem [{\citenamefont {Aaron}\ \emph {et~al.}(2010)\citenamefont {Aaron}
  \emph {et~al.}}]{Aaron:2009aa}%
  \BibitemOpen
  \bibfield  {author} {\bibinfo {author} {\bibfnamefont {F.~D.}\ \bibnamefont
  {Aaron}} \emph {et~al.} (\bibinfo {collaboration} {ZEUS, H1}),\ }\href
  {\doibase 10.1007/JHEP01(2010)109} {\bibfield  {journal} {\bibinfo  {journal}
  {JHEP}\ }\textbf {\bibinfo {volume} {01}},\ \bibinfo {pages} {109} (\bibinfo
  {year} {2010})},\ \Eprint {http://arxiv.org/abs/0911.0884} {arXiv:0911.0884
  [hep-ex]} \BibitemShut {NoStop}%
\bibitem [{\citenamefont {Abramowicz}\ \emph {et~al.}(2015)\citenamefont
  {Abramowicz} \emph {et~al.}}]{Abramowicz:2015mha}%
  \BibitemOpen
  \bibfield  {author} {\bibinfo {author} {\bibfnamefont {H.}~\bibnamefont
  {Abramowicz}} \emph {et~al.} (\bibinfo {collaboration} {ZEUS, H1}),\ }\href
  {\doibase 10.1140/epjc/s10052-015-3710-4} {\bibfield  {journal} {\bibinfo
  {journal} {Eur. Phys. J.}\ }\textbf {\bibinfo {volume} {C75}},\ \bibinfo
  {pages} {580} (\bibinfo {year} {2015})},\ \Eprint
  {http://arxiv.org/abs/1506.06042} {arXiv:1506.06042 [hep-ex]} \BibitemShut
  {NoStop}%
\bibitem [{\citenamefont {Schenke}\ \emph {et~al.}(2012)\citenamefont
  {Schenke}, \citenamefont {Tribedy},\ and\ \citenamefont
  {Venugopalan}}]{Schenke:2012wb}%
  \BibitemOpen
  \bibfield  {author} {\bibinfo {author} {\bibfnamefont {B.}~\bibnamefont
  {Schenke}}, \bibinfo {author} {\bibfnamefont {P.}~\bibnamefont {Tribedy}}, \
  and\ \bibinfo {author} {\bibfnamefont {R.}~\bibnamefont {Venugopalan}},\
  }\href {\doibase 10.1103/PhysRevLett.108.252301} {\bibfield  {journal}
  {\bibinfo  {journal} {Phys. Rev. Lett.}\ }\textbf {\bibinfo {volume} {108}},\
  \bibinfo {pages} {252301} (\bibinfo {year} {2012})},\ \Eprint
  {http://arxiv.org/abs/1202.6646} {arXiv:1202.6646 [nucl-th]} \BibitemShut
  {NoStop}%
\bibitem [{\citenamefont {Schenke}\ \emph {et~al.}(2018)\citenamefont
  {Schenke}, \citenamefont {Shen},\ and\ \citenamefont
  {Tribedy}}]{Schenke:2018fci}%
  \BibitemOpen
  \bibfield  {author} {\bibinfo {author} {\bibfnamefont {B.}~\bibnamefont
  {Schenke}}, \bibinfo {author} {\bibfnamefont {C.}~\bibnamefont {Shen}}, \
  and\ \bibinfo {author} {\bibfnamefont {P.}~\bibnamefont {Tribedy}},\ }in\
  \href@noop {} {\emph {\bibinfo {booktitle} {{27th International Conference on
  Ultrarelativistic Nucleus-Nucleus Collisions (Quark Matter 2018) Venice,
  Italy, May 14-19, 2018}}}}\ (\bibinfo {year} {2018})\ \Eprint
  {http://arxiv.org/abs/1807.05205} {arXiv:1807.05205 [nucl-th]} \BibitemShut
  {NoStop}%
\bibitem [{\citenamefont {Shen}\ \emph {et~al.}(2016)\citenamefont {Shen},
  \citenamefont {Qiu}, \citenamefont {Song}, \citenamefont {Bernhard},
  \citenamefont {Bass},\ and\ \citenamefont {Heinz}}]{Shen:2014vra}%
  \BibitemOpen
  \bibfield  {author} {\bibinfo {author} {\bibfnamefont {C.}~\bibnamefont
  {Shen}}, \bibinfo {author} {\bibfnamefont {Z.}~\bibnamefont {Qiu}}, \bibinfo
  {author} {\bibfnamefont {H.}~\bibnamefont {Song}}, \bibinfo {author}
  {\bibfnamefont {J.}~\bibnamefont {Bernhard}}, \bibinfo {author}
  {\bibfnamefont {S.}~\bibnamefont {Bass}}, \ and\ \bibinfo {author}
  {\bibfnamefont {U.}~\bibnamefont {Heinz}},\ }\href {\doibase
  10.1016/j.cpc.2015.08.039} {\bibfield  {journal} {\bibinfo  {journal}
  {Comput. Phys. Commun.}\ }\textbf {\bibinfo {volume} {19}},\ \bibinfo {pages}
  {61} (\bibinfo {year} {2016})},\ \Eprint {http://arxiv.org/abs/1409.8164}
  {arXiv:1409.8164 [nucl-th]} \BibitemShut {NoStop}%
\bibitem [{\citenamefont {Bernhard}\ \emph {et~al.}(2016)\citenamefont
  {Bernhard}, \citenamefont {Moreland}, \citenamefont {Bass}, \citenamefont
  {Liu},\ and\ \citenamefont {Heinz}}]{Bernhard:2016tnd}%
  \BibitemOpen
  \bibfield  {author} {\bibinfo {author} {\bibfnamefont {J.~E.}\ \bibnamefont
  {Bernhard}}, \bibinfo {author} {\bibfnamefont {J.~S.}\ \bibnamefont
  {Moreland}}, \bibinfo {author} {\bibfnamefont {S.~A.}\ \bibnamefont {Bass}},
  \bibinfo {author} {\bibfnamefont {J.}~\bibnamefont {Liu}}, \ and\ \bibinfo
  {author} {\bibfnamefont {U.}~\bibnamefont {Heinz}},\ }\href {\doibase
  10.1103/PhysRevC.94.024907} {\bibfield  {journal} {\bibinfo  {journal} {Phys.
  Rev.}\ }\textbf {\bibinfo {volume} {C94}},\ \bibinfo {pages} {024907}
  (\bibinfo {year} {2016})},\ \Eprint {http://arxiv.org/abs/1605.03954}
  {arXiv:1605.03954 [nucl-th]} \BibitemShut {NoStop}%
\bibitem [{\citenamefont {Moreland}\ \emph {et~al.}(2015)\citenamefont
  {Moreland}, \citenamefont {Bernhard},\ and\ \citenamefont
  {Bass}}]{Moreland:2014oya}%
  \BibitemOpen
  \bibfield  {author} {\bibinfo {author} {\bibfnamefont {J.~S.}\ \bibnamefont
  {Moreland}}, \bibinfo {author} {\bibfnamefont {J.~E.}\ \bibnamefont
  {Bernhard}}, \ and\ \bibinfo {author} {\bibfnamefont {S.~A.}\ \bibnamefont
  {Bass}},\ }\href {\doibase 10.1103/PhysRevC.92.011901} {\bibfield  {journal}
  {\bibinfo  {journal} {Phys. Rev.}\ }\textbf {\bibinfo {volume} {C92}},\
  \bibinfo {pages} {011901} (\bibinfo {year} {2015})},\ \Eprint
  {http://arxiv.org/abs/1412.4708} {arXiv:1412.4708 [nucl-th]} \BibitemShut
  {NoStop}%
\bibitem [{\citenamefont {Broniowski}\ \emph {et~al.}(2009)\citenamefont
  {Broniowski}, \citenamefont {Florkowski}, \citenamefont {Chojnacki},\ and\
  \citenamefont {Kisiel}}]{Broniowski:2008qk}%
  \BibitemOpen
  \bibfield  {author} {\bibinfo {author} {\bibfnamefont {W.}~\bibnamefont
  {Broniowski}}, \bibinfo {author} {\bibfnamefont {W.}~\bibnamefont
  {Florkowski}}, \bibinfo {author} {\bibfnamefont {M.}~\bibnamefont
  {Chojnacki}}, \ and\ \bibinfo {author} {\bibfnamefont {A.}~\bibnamefont
  {Kisiel}},\ }\href {\doibase 10.1103/PhysRevC.80.034902} {\bibfield
  {journal} {\bibinfo  {journal} {Phys. Rev.}\ }\textbf {\bibinfo {volume}
  {C80}},\ \bibinfo {pages} {034902} (\bibinfo {year} {2009})},\ \Eprint
  {http://arxiv.org/abs/0812.3393} {arXiv:0812.3393 [nucl-th]} \BibitemShut
  {NoStop}%
\bibitem [{\citenamefont {Heinz}\ and\ \citenamefont
  {Liu}(2016)}]{Heinz:2015arc}%
  \BibitemOpen
  \bibfield  {author} {\bibinfo {author} {\bibfnamefont {U.~W.}\ \bibnamefont
  {Heinz}}\ and\ \bibinfo {author} {\bibfnamefont {J.}~\bibnamefont {Liu}},\
  }\bibfield  {booktitle} {\emph {\bibinfo {booktitle} {{Proceedings, 25th
  International Conference on Ultra-Relativistic Nucleus-Nucleus Collisions
  (Quark Matter 2015): Kobe, Japan, September 27-October 3, 2015}}},\ }\href
  {\doibase 10.1016/j.nuclphysa.2016.01.065} {\bibfield  {journal} {\bibinfo
  {journal} {Nucl. Phys.}\ }\textbf {\bibinfo {volume} {A956}},\ \bibinfo
  {pages} {549} (\bibinfo {year} {2016})},\ \Eprint
  {http://arxiv.org/abs/1512.08276} {arXiv:1512.08276 [nucl-th]} \BibitemShut
  {NoStop}%
\bibitem [{\citenamefont {Bernhard}(4 19)}]{Bernhard:2018hnz}%
  \BibitemOpen
  \bibfield  {author} {\bibinfo {author} {\bibfnamefont {J.~E.}\ \bibnamefont
  {Bernhard}},\ }\emph {\bibinfo {title} {{Bayesian parameter estimation for
  relativistic heavy-ion collisions}}},\ \href
  {https://inspirehep.net/record/1669345/files/1804.06469.pdf} {Ph.D. thesis},\
  \bibinfo  {school} {Duke U.} (\bibinfo {year} {2018-04-19}),\ \Eprint
  {http://arxiv.org/abs/1804.06469} {arXiv:1804.06469 [nucl-th]} \BibitemShut
  {NoStop}%
\bibitem [{\citenamefont {Ke}\ \emph {et~al.}(2017)\citenamefont {Ke},
  \citenamefont {Moreland}, \citenamefont {Bernhard},\ and\ \citenamefont
  {Bass}}]{Ke:2016jrd}%
  \BibitemOpen
  \bibfield  {author} {\bibinfo {author} {\bibfnamefont {W.}~\bibnamefont
  {Ke}}, \bibinfo {author} {\bibfnamefont {J.~S.}\ \bibnamefont {Moreland}},
  \bibinfo {author} {\bibfnamefont {J.~E.}\ \bibnamefont {Bernhard}}, \ and\
  \bibinfo {author} {\bibfnamefont {S.~A.}\ \bibnamefont {Bass}},\ }\href
  {\doibase 10.1103/PhysRevC.96.044912} {\bibfield  {journal} {\bibinfo
  {journal} {Phys. Rev.}\ }\textbf {\bibinfo {volume} {C96}},\ \bibinfo {pages}
  {044912} (\bibinfo {year} {2017})},\ \Eprint
  {http://arxiv.org/abs/1610.08490} {arXiv:1610.08490 [nucl-th]} \BibitemShut
  {NoStop}%
\bibitem [{\citenamefont {Bozek}\ \emph {et~al.}(2011)\citenamefont {Bozek},
  \citenamefont {Broniowski},\ and\ \citenamefont {Moreira}}]{Bozek:2010vz}%
  \BibitemOpen
  \bibfield  {author} {\bibinfo {author} {\bibfnamefont {P.}~\bibnamefont
  {Bozek}}, \bibinfo {author} {\bibfnamefont {W.}~\bibnamefont {Broniowski}}, \
  and\ \bibinfo {author} {\bibfnamefont {J.}~\bibnamefont {Moreira}},\ }\href
  {\doibase 10.1103/PhysRevC.83.034911} {\bibfield  {journal} {\bibinfo
  {journal} {Phys. Rev.}\ }\textbf {\bibinfo {volume} {C83}},\ \bibinfo {pages}
  {034911} (\bibinfo {year} {2011})},\ \Eprint {http://arxiv.org/abs/1011.3354}
  {arXiv:1011.3354 [nucl-th]} \BibitemShut {NoStop}%
\bibitem [{\citenamefont {{V.V. Anisovich and Yu.M. Shabelsky and V.M.
  Shekhter}}(1978)}]{ANISOVICH1978477}%
  \BibitemOpen
  \bibfield  {author} {\bibinfo {author} {\bibnamefont {{V.V. Anisovich and
  Yu.M. Shabelsky and V.M. Shekhter}}},\ }\href {\doibase
  https://doi.org/10.1016/0550-3213(78)90237-7} {\bibfield  {journal} {\bibinfo
   {journal} {Nuclear Physics B}\ }\textbf {\bibinfo {volume} {133}},\ \bibinfo
  {pages} {477 } (\bibinfo {year} {1978})}\BibitemShut {NoStop}%
\bibitem [{\citenamefont {Bialas}\ \emph {et~al.}(1976)\citenamefont {Bialas},
  \citenamefont {Bleszynski},\ and\ \citenamefont {Czyz}}]{Bialas:1976ed}%
  \BibitemOpen
  \bibfield  {author} {\bibinfo {author} {\bibfnamefont {A.}~\bibnamefont
  {Bialas}}, \bibinfo {author} {\bibfnamefont {M.}~\bibnamefont {Bleszynski}},
  \ and\ \bibinfo {author} {\bibfnamefont {W.}~\bibnamefont {Czyz}},\ }\href
  {\doibase 10.1016/0550-3213(76)90329-1} {\bibfield  {journal} {\bibinfo
  {journal} {Nucl. Phys.}\ }\textbf {\bibinfo {volume} {B111}},\ \bibinfo
  {pages} {461} (\bibinfo {year} {1976})}\BibitemShut {NoStop}%
\bibitem [{\citenamefont {Kharzeev}\ and\ \citenamefont
  {Nardi}(2001)}]{Kharzeev:2000ph}%
  \BibitemOpen
  \bibfield  {author} {\bibinfo {author} {\bibfnamefont {D.}~\bibnamefont
  {Kharzeev}}\ and\ \bibinfo {author} {\bibfnamefont {M.}~\bibnamefont
  {Nardi}},\ }\href {\doibase 10.1016/S0370-2693(01)00457-9} {\bibfield
  {journal} {\bibinfo  {journal} {Phys. Lett.}\ }\textbf {\bibinfo {volume}
  {B507}},\ \bibinfo {pages} {121} (\bibinfo {year} {2001})},\ \Eprint
  {http://arxiv.org/abs/nucl-th/0012025} {arXiv:nucl-th/0012025 [nucl-th]}
  \BibitemShut {NoStop}%
\bibitem [{\citenamefont {Israel}\ and\ \citenamefont
  {Stewart}(1979)}]{Israel:1979wp}%
  \BibitemOpen
  \bibfield  {author} {\bibinfo {author} {\bibfnamefont {W.}~\bibnamefont
  {Israel}}\ and\ \bibinfo {author} {\bibfnamefont {J.~M.}\ \bibnamefont
  {Stewart}},\ }\href {\doibase 10.1016/0003-4916(79)90130-1} {\bibfield
  {journal} {\bibinfo  {journal} {Annals Phys.}\ }\textbf {\bibinfo {volume}
  {118}},\ \bibinfo {pages} {341} (\bibinfo {year} {1979})}\BibitemShut
  {NoStop}%
\bibitem [{\citenamefont {Israel}\ and\ \citenamefont
  {Stewart}(1976)}]{Israel:1976aa}%
  \BibitemOpen
  \bibfield  {author} {\bibinfo {author} {\bibfnamefont {W.}~\bibnamefont
  {Israel}}\ and\ \bibinfo {author} {\bibfnamefont {J.}~\bibnamefont
  {Stewart}},\ }\href {\doibase 10.1016/0375-9601(76)90075-X} {\bibfield
  {journal} {\bibinfo  {journal} {Phys. Lett.}\ }\textbf {\bibinfo {volume}
  {A58}},\ \bibinfo {pages} {213} (\bibinfo {year} {1976})}\BibitemShut
  {NoStop}%
\bibitem [{\citenamefont {Denicol}\ \emph {et~al.}(2012)\citenamefont
  {Denicol}, \citenamefont {Niemi}, \citenamefont {Molnar},\ and\ \citenamefont
  {Rischke}}]{Denicol:2012cn}%
  \BibitemOpen
  \bibfield  {author} {\bibinfo {author} {\bibfnamefont {G.~S.}\ \bibnamefont
  {Denicol}}, \bibinfo {author} {\bibfnamefont {H.}~\bibnamefont {Niemi}},
  \bibinfo {author} {\bibfnamefont {E.}~\bibnamefont {Molnar}}, \ and\ \bibinfo
  {author} {\bibfnamefont {D.~H.}\ \bibnamefont {Rischke}},\ }\href {\doibase
  10.1103/PhysRevD.85.114047, 10.1103/PhysRevD.91.039902} {\bibfield  {journal}
  {\bibinfo  {journal} {Phys. Rev.}\ }\textbf {\bibinfo {volume} {D85}},\
  \bibinfo {pages} {114047} (\bibinfo {year} {2012})},\ \bibinfo {note}
  {[Erratum: Phys. Rev.D91,no.3,039902(2015)]},\ \Eprint
  {http://arxiv.org/abs/1202.4551} {arXiv:1202.4551 [nucl-th]} \BibitemShut
  {NoStop}%
\bibitem [{\citenamefont {Denicol}\ \emph {et~al.}(2010)\citenamefont
  {Denicol}, \citenamefont {Koide},\ and\ \citenamefont
  {Rischke}}]{Denicol:2010xn}%
  \BibitemOpen
  \bibfield  {author} {\bibinfo {author} {\bibfnamefont {G.~S.}\ \bibnamefont
  {Denicol}}, \bibinfo {author} {\bibfnamefont {T.}~\bibnamefont {Koide}}, \
  and\ \bibinfo {author} {\bibfnamefont {D.~H.}\ \bibnamefont {Rischke}},\
  }\href {\doibase 10.1103/PhysRevLett.105.162501} {\bibfield  {journal}
  {\bibinfo  {journal} {Phys. Rev. Lett.}\ }\textbf {\bibinfo {volume} {105}},\
  \bibinfo {pages} {162501} (\bibinfo {year} {2010})},\ \Eprint
  {http://arxiv.org/abs/1004.5013} {arXiv:1004.5013 [nucl-th]} \BibitemShut
  {NoStop}%
\bibitem [{\citenamefont {Denicol}\ \emph {et~al.}(2014)\citenamefont
  {Denicol}, \citenamefont {Jeon},\ and\ \citenamefont
  {Gale}}]{Denicol:2014vaa}%
  \BibitemOpen
  \bibfield  {author} {\bibinfo {author} {\bibfnamefont {G.~S.}\ \bibnamefont
  {Denicol}}, \bibinfo {author} {\bibfnamefont {S.}~\bibnamefont {Jeon}}, \
  and\ \bibinfo {author} {\bibfnamefont {C.}~\bibnamefont {Gale}},\ }\href
  {\doibase 10.1103/PhysRevC.90.024912} {\bibfield  {journal} {\bibinfo
  {journal} {Phys. Rev.}\ }\textbf {\bibinfo {volume} {C90}},\ \bibinfo {pages}
  {024912} (\bibinfo {year} {2014})},\ \Eprint {http://arxiv.org/abs/1403.0962}
  {arXiv:1403.0962 [nucl-th]} \BibitemShut {NoStop}%
\bibitem [{\citenamefont {Bazavov}\ \emph {et~al.}(2014)\citenamefont {Bazavov}
  \emph {et~al.}}]{Bazavov:2014pvz}%
  \BibitemOpen
  \bibfield  {author} {\bibinfo {author} {\bibfnamefont {A.}~\bibnamefont
  {Bazavov}} \emph {et~al.} (\bibinfo {collaboration} {HotQCD}),\ }\href
  {\doibase 10.1103/PhysRevD.90.094503} {\bibfield  {journal} {\bibinfo
  {journal} {Phys. Rev.}\ }\textbf {\bibinfo {volume} {D90}},\ \bibinfo {pages}
  {094503} (\bibinfo {year} {2014})},\ \Eprint {http://arxiv.org/abs/1407.6387}
  {arXiv:1407.6387 [hep-lat]} \BibitemShut {NoStop}%
\bibitem [{\citenamefont {Borsanyi}\ \emph {et~al.}(2016)\citenamefont
  {Borsanyi} \emph {et~al.}}]{Borsanyi:2016ksw}%
  \BibitemOpen
  \bibfield  {author} {\bibinfo {author} {\bibfnamefont {S.}~\bibnamefont
  {Borsanyi}} \emph {et~al.},\ }\href {\doibase 10.1038/nature20115} {\bibfield
   {journal} {\bibinfo  {journal} {Nature}\ }\textbf {\bibinfo {volume}
  {539}},\ \bibinfo {pages} {69} (\bibinfo {year} {2016})},\ \Eprint
  {http://arxiv.org/abs/1606.07494} {arXiv:1606.07494 [hep-lat]} \BibitemShut
  {NoStop}%
\bibitem [{\citenamefont {Noronha-Hostler}\ and\ \citenamefont
  {Ratti}(2018)}]{Noronha-Hostler:2018zxc}%
  \BibitemOpen
  \bibfield  {author} {\bibinfo {author} {\bibfnamefont {J.}~\bibnamefont
  {Noronha-Hostler}}\ and\ \bibinfo {author} {\bibfnamefont {C.}~\bibnamefont
  {Ratti}},\ }\href@noop {} {\  (\bibinfo {year} {2018})},\ \Eprint
  {http://arxiv.org/abs/1804.10661} {arXiv:1804.10661 [nucl-th]} \BibitemShut
  {NoStop}%
\bibitem [{\citenamefont {Song}\ and\ \citenamefont
  {Heinz}(2008)}]{Song:2007ux}%
  \BibitemOpen
  \bibfield  {author} {\bibinfo {author} {\bibfnamefont {H.}~\bibnamefont
  {Song}}\ and\ \bibinfo {author} {\bibfnamefont {U.~W.}\ \bibnamefont
  {Heinz}},\ }\href {\doibase 10.1103/PhysRevC.77.064901} {\bibfield  {journal}
  {\bibinfo  {journal} {Phys. Rev.}\ }\textbf {\bibinfo {volume} {C77}},\
  \bibinfo {pages} {064901} (\bibinfo {year} {2008})},\ \Eprint
  {http://arxiv.org/abs/0712.3715} {arXiv:0712.3715 [nucl-th]} \BibitemShut
  {NoStop}%
\bibitem [{\citenamefont {Cooper}\ and\ \citenamefont
  {Frye}(1974)}]{PhysRevD.10.186}%
  \BibitemOpen
  \bibfield  {author} {\bibinfo {author} {\bibfnamefont {F.}~\bibnamefont
  {Cooper}}\ and\ \bibinfo {author} {\bibfnamefont {G.}~\bibnamefont {Frye}},\
  }\href {\doibase 10.1103/PhysRevD.10.186} {\bibfield  {journal} {\bibinfo
  {journal} {Phys. Rev. D}\ }\textbf {\bibinfo {volume} {10}},\ \bibinfo
  {pages} {186} (\bibinfo {year} {1974})}\BibitemShut {NoStop}%
\bibitem [{\citenamefont {Sollfrank}\ \emph {et~al.}(1991)\citenamefont
  {Sollfrank}, \citenamefont {Koch},\ and\ \citenamefont
  {Heinz}}]{Sollfrank:1991xm}%
  \BibitemOpen
  \bibfield  {author} {\bibinfo {author} {\bibfnamefont {J.}~\bibnamefont
  {Sollfrank}}, \bibinfo {author} {\bibfnamefont {P.}~\bibnamefont {Koch}}, \
  and\ \bibinfo {author} {\bibfnamefont {U.~W.}\ \bibnamefont {Heinz}},\ }\href
  {\doibase 10.1007/BF01562334} {\bibfield  {journal} {\bibinfo  {journal} {Z.
  Phys.}\ }\textbf {\bibinfo {volume} {C52}},\ \bibinfo {pages} {593} (\bibinfo
  {year} {1991})}\BibitemShut {NoStop}%
\bibitem [{\citenamefont {Huovinen}\ \emph {et~al.}(2017)\citenamefont
  {Huovinen}, \citenamefont {Lo}, \citenamefont {Marczenko}, \citenamefont
  {Morita}, \citenamefont {Redlich},\ and\ \citenamefont
  {Sasaki}}]{Huovinen:2016xxq}%
  \BibitemOpen
  \bibfield  {author} {\bibinfo {author} {\bibfnamefont {P.}~\bibnamefont
  {Huovinen}}, \bibinfo {author} {\bibfnamefont {P.~M.}\ \bibnamefont {Lo}},
  \bibinfo {author} {\bibfnamefont {M.}~\bibnamefont {Marczenko}}, \bibinfo
  {author} {\bibfnamefont {K.}~\bibnamefont {Morita}}, \bibinfo {author}
  {\bibfnamefont {K.}~\bibnamefont {Redlich}}, \ and\ \bibinfo {author}
  {\bibfnamefont {C.}~\bibnamefont {Sasaki}},\ }\href {\doibase
  10.1016/j.physletb.2017.03.060} {\bibfield  {journal} {\bibinfo  {journal}
  {Phys. Lett.}\ }\textbf {\bibinfo {volume} {B769}},\ \bibinfo {pages} {509}
  (\bibinfo {year} {2017})},\ \Eprint {http://arxiv.org/abs/1608.06817}
  {arXiv:1608.06817 [hep-ph]} \BibitemShut {NoStop}%
\bibitem [{\citenamefont {Vovchenko}\ \emph {et~al.}(2018)\citenamefont
  {Vovchenko}, \citenamefont {Gorenstein},\ and\ \citenamefont
  {Stoecker}}]{Vovchenko:2018fmh}%
  \BibitemOpen
  \bibfield  {author} {\bibinfo {author} {\bibfnamefont {V.}~\bibnamefont
  {Vovchenko}}, \bibinfo {author} {\bibfnamefont {M.~I.}\ \bibnamefont
  {Gorenstein}}, \ and\ \bibinfo {author} {\bibfnamefont {H.}~\bibnamefont
  {Stoecker}},\ }\href@noop {} {\  (\bibinfo {year} {2018})},\ \Eprint
  {http://arxiv.org/abs/1807.02079} {arXiv:1807.02079 [nucl-th]} \BibitemShut
  {NoStop}%
\bibitem [{\citenamefont {Patrignani}\ and\ \citenamefont
  {Group}(2016)}]{PDG:2017}%
  \BibitemOpen
  \bibfield  {author} {\bibinfo {author} {\bibfnamefont {C.}~\bibnamefont
  {Patrignani}}\ and\ \bibinfo {author} {\bibfnamefont {P.~D.}\ \bibnamefont
  {Group}},\ }\href {http://stacks.iop.org/1674-1137/40/i=10/a=100001}
  {\bibfield  {journal} {\bibinfo  {journal} {Chinese Physics C}\ }\textbf
  {\bibinfo {volume} {40}},\ \bibinfo {pages} {100001} (\bibinfo {year}
  {2016})}\BibitemShut {NoStop}%
\bibitem [{\citenamefont {Pratt}\ and\ \citenamefont
  {Torrieri}(2010)}]{Pratt:2010jt}%
  \BibitemOpen
  \bibfield  {author} {\bibinfo {author} {\bibfnamefont {S.}~\bibnamefont
  {Pratt}}\ and\ \bibinfo {author} {\bibfnamefont {G.}~\bibnamefont
  {Torrieri}},\ }\href {\doibase 10.1103/PhysRevC.82.044901} {\bibfield
  {journal} {\bibinfo  {journal} {Phys. Rev.}\ }\textbf {\bibinfo {volume}
  {C82}},\ \bibinfo {pages} {044901} (\bibinfo {year} {2010})},\ \Eprint
  {http://arxiv.org/abs/1003.0413} {arXiv:1003.0413 [nucl-th]} \BibitemShut
  {NoStop}%
\bibitem [{\citenamefont {Bass}\ \emph {et~al.}(1998)\citenamefont {Bass} \emph
  {et~al.}}]{Bass:1998ca}%
  \BibitemOpen
  \bibfield  {author} {\bibinfo {author} {\bibfnamefont {S.~A.}\ \bibnamefont
  {Bass}} \emph {et~al.},\ }\href {\doibase 10.1016/S0146-6410(98)00058-1}
  {\bibfield  {journal} {\bibinfo  {journal} {Prog. Part. Nucl. Phys.}\
  }\textbf {\bibinfo {volume} {41}},\ \bibinfo {pages} {255} (\bibinfo {year}
  {1998})},\ \bibinfo {note} {[Prog. Part. Nucl. Phys.41,225(1998)]},\ \Eprint
  {http://arxiv.org/abs/nucl-th/9803035} {arXiv:nucl-th/9803035 [nucl-th]}
  \BibitemShut {NoStop}%
\bibitem [{\citenamefont {Bleicher}\ \emph {et~al.}(1999)\citenamefont
  {Bleicher} \emph {et~al.}}]{Bleicher:1999xi}%
  \BibitemOpen
  \bibfield  {author} {\bibinfo {author} {\bibfnamefont {M.}~\bibnamefont
  {Bleicher}} \emph {et~al.},\ }\href {\doibase 10.1088/0954-3899/25/9/308}
  {\bibfield  {journal} {\bibinfo  {journal} {J. Phys.}\ }\textbf {\bibinfo
  {volume} {G25}},\ \bibinfo {pages} {1859} (\bibinfo {year} {1999})},\ \Eprint
  {http://arxiv.org/abs/hep-ph/9909407} {arXiv:hep-ph/9909407 [hep-ph]}
  \BibitemShut {NoStop}%
\bibitem [{\citenamefont {Adam}\ \emph
  {et~al.}(2016{\natexlab{a}})\citenamefont {Adam} \emph
  {et~al.}}]{Adam:2015ptt}%
  \BibitemOpen
  \bibfield  {author} {\bibinfo {author} {\bibfnamefont {J.}~\bibnamefont
  {Adam}} \emph {et~al.} (\bibinfo {collaboration} {ALICE}),\ }\href {\doibase
  10.1103/PhysRevLett.116.222302} {\bibfield  {journal} {\bibinfo  {journal}
  {Phys. Rev. Lett.}\ }\textbf {\bibinfo {volume} {116}},\ \bibinfo {pages}
  {222302} (\bibinfo {year} {2016}{\natexlab{a}})},\ \Eprint
  {http://arxiv.org/abs/1512.06104} {arXiv:1512.06104 [nucl-ex]} \BibitemShut
  {NoStop}%
\bibitem [{\citenamefont {Adam}\ \emph {et~al.}(2015)\citenamefont {Adam} \emph
  {et~al.}}]{Adam:2014qja}%
  \BibitemOpen
  \bibfield  {author} {\bibinfo {author} {\bibfnamefont {J.}~\bibnamefont
  {Adam}} \emph {et~al.} (\bibinfo {collaboration} {ALICE}),\ }\href {\doibase
  10.1103/PhysRevC.91.064905} {\bibfield  {journal} {\bibinfo  {journal} {Phys.
  Rev.}\ }\textbf {\bibinfo {volume} {C91}},\ \bibinfo {pages} {064905}
  (\bibinfo {year} {2015})},\ \Eprint {http://arxiv.org/abs/1412.6828}
  {arXiv:1412.6828 [nucl-ex]} \BibitemShut {NoStop}%
\bibitem [{\citenamefont {Adam}\ \emph
  {et~al.}(2016{\natexlab{b}})\citenamefont {Adam} \emph
  {et~al.}}]{Adam:2016izf}%
  \BibitemOpen
  \bibfield  {author} {\bibinfo {author} {\bibfnamefont {J.}~\bibnamefont
  {Adam}} \emph {et~al.} (\bibinfo {collaboration} {ALICE}),\ }\href {\doibase
  10.1103/PhysRevLett.116.132302} {\bibfield  {journal} {\bibinfo  {journal}
  {Phys. Rev. Lett.}\ }\textbf {\bibinfo {volume} {116}},\ \bibinfo {pages}
  {132302} (\bibinfo {year} {2016}{\natexlab{b}})},\ \Eprint
  {http://arxiv.org/abs/1602.01119} {arXiv:1602.01119 [nucl-ex]} \BibitemShut
  {NoStop}%
\bibitem [{\citenamefont {Chatrchyan}\ \emph
  {et~al.}(2013{\natexlab{b}})\citenamefont {Chatrchyan} \emph
  {et~al.}}]{Chatrchyan:2013nka}%
  \BibitemOpen
  \bibfield  {author} {\bibinfo {author} {\bibfnamefont {S.}~\bibnamefont
  {Chatrchyan}} \emph {et~al.} (\bibinfo {collaboration} {CMS}),\ }\href
  {\doibase 10.1016/j.physletb.2013.06.028} {\bibfield  {journal} {\bibinfo
  {journal} {Phys. Lett.}\ }\textbf {\bibinfo {volume} {B724}},\ \bibinfo
  {pages} {213} (\bibinfo {year} {2013}{\natexlab{b}})},\ \Eprint
  {http://arxiv.org/abs/1305.0609} {arXiv:1305.0609 [nucl-ex]} \BibitemShut
  {NoStop}%
\bibitem [{\citenamefont {Abelev}\ \emph
  {et~al.}(2013{\natexlab{b}})\citenamefont {Abelev} \emph
  {et~al.}}]{Abelev:2013bla}%
  \BibitemOpen
  \bibfield  {author} {\bibinfo {author} {\bibfnamefont {B.~B.}\ \bibnamefont
  {Abelev}} \emph {et~al.} (\bibinfo {collaboration} {ALICE}),\ }\href
  {\doibase 10.1016/j.physletb.2013.10.054} {\bibfield  {journal} {\bibinfo
  {journal} {Phys. Lett.}\ }\textbf {\bibinfo {volume} {B727}},\ \bibinfo
  {pages} {371} (\bibinfo {year} {2013}{\natexlab{b}})},\ \Eprint
  {http://arxiv.org/abs/1307.1094} {arXiv:1307.1094 [nucl-ex]} \BibitemShut
  {NoStop}%
\bibitem [{\citenamefont {O’Hagan}(2006)}]{OHagan:2006ba}%
  \BibitemOpen
  \bibfield  {author} {\bibinfo {author} {\bibfnamefont {A.}~\bibnamefont
  {O’Hagan}},\ }\href {\doibase 10.1016/j.ress.2005.11.025} {\bibfield
  {journal} {\bibinfo  {journal} {Rel.Engin.Sys.Safety}\ }\textbf {\bibinfo
  {volume} {91}},\ \bibinfo {pages} {1290} (\bibinfo {year}
  {2006})}\BibitemShut {NoStop}%
\bibitem [{\citenamefont {Higdon}\ \emph {et~al.}(2008)\citenamefont {Higdon},
  \citenamefont {Gattiker}, \citenamefont {Williams},\ and\ \citenamefont
  {Rightley}}]{Higdon:2008cmc}%
  \BibitemOpen
  \bibfield  {author} {\bibinfo {author} {\bibfnamefont {D.}~\bibnamefont
  {Higdon}}, \bibinfo {author} {\bibfnamefont {J.}~\bibnamefont {Gattiker}},
  \bibinfo {author} {\bibfnamefont {B.}~\bibnamefont {Williams}}, \ and\
  \bibinfo {author} {\bibfnamefont {M.}~\bibnamefont {Rightley}},\ }\href
  {\doibase 10.1198/016214507000000888} {\bibfield  {journal} {\bibinfo
  {journal} {J.Amer.Stat.Assoc.}\ }\textbf {\bibinfo {volume} {103}},\ \bibinfo
  {pages} {570} (\bibinfo {year} {2008})}\BibitemShut {NoStop}%
\bibitem [{\citenamefont {Higdon}\ \emph {et~al.}(2015)\citenamefont {Higdon},
  \citenamefont {McDonnell}, \citenamefont {Schunck}, \citenamefont {Sarich},\
  and\ \citenamefont {Wild}}]{Higdon:2014tva}%
  \BibitemOpen
  \bibfield  {author} {\bibinfo {author} {\bibfnamefont {D.}~\bibnamefont
  {Higdon}}, \bibinfo {author} {\bibfnamefont {J.~D.}\ \bibnamefont
  {McDonnell}}, \bibinfo {author} {\bibfnamefont {N.}~\bibnamefont {Schunck}},
  \bibinfo {author} {\bibfnamefont {J.}~\bibnamefont {Sarich}}, \ and\ \bibinfo
  {author} {\bibfnamefont {S.~M.}\ \bibnamefont {Wild}},\ }\href {\doibase
  10.1088/0954-3899/42/3/034009} {\bibfield  {journal} {\bibinfo  {journal}
  {J.Phys.}\ }\textbf {\bibinfo {volume} {G42}},\ \bibinfo {pages} {034009}
  (\bibinfo {year} {2015})},\ \Eprint {http://arxiv.org/abs/1407.3017}
  {arXiv:1407.3017 [physics.data-an]} \BibitemShut {NoStop}%
\bibitem [{\citenamefont {Morris}\ and\ \citenamefont
  {Mitchell}(1995)}]{Morris:1995lh}%
  \BibitemOpen
  \bibfield  {author} {\bibinfo {author} {\bibfnamefont {M.~D.}\ \bibnamefont
  {Morris}}\ and\ \bibinfo {author} {\bibfnamefont {T.~J.}\ \bibnamefont
  {Mitchell}},\ }\href {\doibase 10.1016/0378-3758(94)00035-T} {\bibfield
  {journal} {\bibinfo  {journal} {J.Stat.Plan.Inf.}\ }\textbf {\bibinfo
  {volume} {43}},\ \bibinfo {pages} {381} (\bibinfo {year} {1995})}\BibitemShut
  {NoStop}%
\bibitem [{\citenamefont {Rasmussen}\ and\ \citenamefont
  {Williams}(2006)}]{Rasmussen:2006gp}%
  \BibitemOpen
  \bibfield  {author} {\bibinfo {author} {\bibfnamefont {C.~E.}\ \bibnamefont
  {Rasmussen}}\ and\ \bibinfo {author} {\bibfnamefont {C.~K.~I.}\ \bibnamefont
  {Williams}},\ }\href {http://www.gaussianprocess.org/gpml} {\emph {\bibinfo
  {title} {Gaussian Processes for Machine Learning}}}\ (\bibinfo  {publisher}
  {MIT Press},\ \bibinfo {address} {Cambridge, MA},\ \bibinfo {year}
  {2006})\BibitemShut {NoStop}%
\bibitem [{\citenamefont {Tipping}\ and\ \citenamefont
  {Bishop}(1999)}]{Tipping:1999}%
  \BibitemOpen
  \bibfield  {author} {\bibinfo {author} {\bibfnamefont {M.~E.}\ \bibnamefont
  {Tipping}}\ and\ \bibinfo {author} {\bibfnamefont {C.~M.}\ \bibnamefont
  {Bishop}},\ }\href {\doibase 10.1162/089976699300016728} {\bibfield
  {journal} {\bibinfo  {journal} {Neural Comput.}\ }\textbf {\bibinfo {volume}
  {11}},\ \bibinfo {pages} {443} (\bibinfo {year} {1999})}\BibitemShut
  {NoStop}%
\bibitem [{\citenamefont {Goodman}\ and\ \citenamefont
  {Weare}(2010)}]{Goodman:2010en}%
  \BibitemOpen
  \bibfield  {author} {\bibinfo {author} {\bibfnamefont {J.}~\bibnamefont
  {Goodman}}\ and\ \bibinfo {author} {\bibfnamefont {J.}~\bibnamefont
  {Weare}},\ }\href {\doibase 10.2140/camcos.2010.5.65} {\bibfield  {journal}
  {\bibinfo  {journal} {Comm.App.Math.Comp.Sc.}\ }\textbf {\bibinfo {volume}
  {5}},\ \bibinfo {pages} {65} (\bibinfo {year} {2010})}\BibitemShut {NoStop}%
\bibitem [{\citenamefont {{Foreman-Mackey}}\ \emph {et~al.}(2013)\citenamefont
  {{Foreman-Mackey}}, \citenamefont {{Hogg}}, \citenamefont {{Lang}},\ and\
  \citenamefont {{Goodman}}}]{FM:2013mc}%
  \BibitemOpen
  \bibfield  {author} {\bibinfo {author} {\bibfnamefont {D.}~\bibnamefont
  {{Foreman-Mackey}}}, \bibinfo {author} {\bibfnamefont {D.~W.}\ \bibnamefont
  {{Hogg}}}, \bibinfo {author} {\bibfnamefont {D.}~\bibnamefont {{Lang}}}, \
  and\ \bibinfo {author} {\bibfnamefont {J.}~\bibnamefont {{Goodman}}},\ }\href
  {\doibase 10.1086/670067} {\bibfield  {journal} {\bibinfo  {journal} {PASP}\
  }\textbf {\bibinfo {volume} {125}},\ \bibinfo {pages} {306} (\bibinfo {year}
  {2013})},\ \Eprint {http://arxiv.org/abs/1202.3665} {arXiv:1202.3665
  [astro-ph.IM]} \BibitemShut {NoStop}%
\bibitem [{\citenamefont {Bernauer}\ \emph {et~al.}(2010)\citenamefont
  {Bernauer} \emph {et~al.}}]{Bernauer:2010wm}%
  \BibitemOpen
  \bibfield  {author} {\bibinfo {author} {\bibfnamefont {J.~C.}\ \bibnamefont
  {Bernauer}} \emph {et~al.} (\bibinfo {collaboration} {A1}),\ }\href {\doibase
  10.1103/PhysRevLett.105.242001} {\bibfield  {journal} {\bibinfo  {journal}
  {Phys. Rev. Lett.}\ }\textbf {\bibinfo {volume} {105}},\ \bibinfo {pages}
  {242001} (\bibinfo {year} {2010})},\ \Eprint {http://arxiv.org/abs/1007.5076}
  {arXiv:1007.5076 [nucl-ex]} \BibitemShut {NoStop}%
\bibitem [{\citenamefont {Bzdak}\ \emph {et~al.}(2013)\citenamefont {Bzdak},
  \citenamefont {Schenke}, \citenamefont {Tribedy},\ and\ \citenamefont
  {Venugopalan}}]{Bzdak:2013zma}%
  \BibitemOpen
  \bibfield  {author} {\bibinfo {author} {\bibfnamefont {A.}~\bibnamefont
  {Bzdak}}, \bibinfo {author} {\bibfnamefont {B.}~\bibnamefont {Schenke}},
  \bibinfo {author} {\bibfnamefont {P.}~\bibnamefont {Tribedy}}, \ and\
  \bibinfo {author} {\bibfnamefont {R.}~\bibnamefont {Venugopalan}},\ }\href
  {\doibase 10.1103/PhysRevC.87.064906} {\bibfield  {journal} {\bibinfo
  {journal} {Phys. Rev.}\ }\textbf {\bibinfo {volume} {C87}},\ \bibinfo {pages}
  {064906} (\bibinfo {year} {2013})},\ \Eprint {http://arxiv.org/abs/1304.3403}
  {arXiv:1304.3403 [nucl-th]} \BibitemShut {NoStop}%
\bibitem [{\citenamefont {Schenke}(2017)}]{Schenke:2017bog}%
  \BibitemOpen
  \bibfield  {author} {\bibinfo {author} {\bibfnamefont {B.}~\bibnamefont
  {Schenke}},\ }\bibfield  {booktitle} {\emph {\bibinfo {booktitle}
  {{Proceedings, 26th International Conference on Ultra-relativistic
  Nucleus-Nucleus Collisions (Quark Matter 2017): Chicago, Illinois, USA,
  February 5-11, 2017}}},\ }\href {\doibase 10.1016/j.nuclphysa.2017.05.017}
  {\bibfield  {journal} {\bibinfo  {journal} {Nucl. Phys.}\ }\textbf {\bibinfo
  {volume} {A967}},\ \bibinfo {pages} {105} (\bibinfo {year} {2017})},\ \Eprint
  {http://arxiv.org/abs/1704.03914} {arXiv:1704.03914 [nucl-th]} \BibitemShut
  {NoStop}%
\bibitem [{\citenamefont {Bożek}\ and\ \citenamefont
  {Broniowski}(2017)}]{Bozek:2017elk}%
  \BibitemOpen
  \bibfield  {author} {\bibinfo {author} {\bibfnamefont {P.}~\bibnamefont
  {Bożek}}\ and\ \bibinfo {author} {\bibfnamefont {W.}~\bibnamefont
  {Broniowski}},\ }\href {\doibase 10.1103/PhysRevC.96.014904} {\bibfield
  {journal} {\bibinfo  {journal} {Phys. Rev.}\ }\textbf {\bibinfo {volume}
  {C96}},\ \bibinfo {pages} {014904} (\bibinfo {year} {2017})},\ \Eprint
  {http://arxiv.org/abs/1701.09105} {arXiv:1701.09105 [nucl-th]} \BibitemShut
  {NoStop}%
\bibitem [{\citenamefont {Bilandzic}\ \emph {et~al.}(2014)\citenamefont
  {Bilandzic}, \citenamefont {Christensen}, \citenamefont {Gulbrandsen},
  \citenamefont {Hansen},\ and\ \citenamefont {Zhou}}]{Bilandzic:2013kga}%
  \BibitemOpen
  \bibfield  {author} {\bibinfo {author} {\bibfnamefont {A.}~\bibnamefont
  {Bilandzic}}, \bibinfo {author} {\bibfnamefont {C.~H.}\ \bibnamefont
  {Christensen}}, \bibinfo {author} {\bibfnamefont {K.}~\bibnamefont
  {Gulbrandsen}}, \bibinfo {author} {\bibfnamefont {A.}~\bibnamefont {Hansen}},
  \ and\ \bibinfo {author} {\bibfnamefont {Y.}~\bibnamefont {Zhou}},\ }\href
  {\doibase 10.1103/PhysRevC.89.064904} {\bibfield  {journal} {\bibinfo
  {journal} {Phys. Rev.}\ }\textbf {\bibinfo {volume} {C89}},\ \bibinfo {pages}
  {064904} (\bibinfo {year} {2014})},\ \Eprint {http://arxiv.org/abs/1312.3572}
  {arXiv:1312.3572 [nucl-ex]} \BibitemShut {NoStop}%
\bibitem [{\citenamefont {Adam}\ \emph
  {et~al.}(2016{\natexlab{c}})\citenamefont {Adam} \emph
  {et~al.}}]{ALICE:2016kpq}%
  \BibitemOpen
  \bibfield  {author} {\bibinfo {author} {\bibfnamefont {J.}~\bibnamefont
  {Adam}} \emph {et~al.} (\bibinfo {collaboration} {ALICE}),\ }\href {\doibase
  10.1103/PhysRevLett.117.182301} {\bibfield  {journal} {\bibinfo  {journal}
  {Phys. Rev. Lett.}\ }\textbf {\bibinfo {volume} {117}},\ \bibinfo {pages}
  {182301} (\bibinfo {year} {2016}{\natexlab{c}})},\ \Eprint
  {http://arxiv.org/abs/1604.07663} {arXiv:1604.07663 [nucl-ex]} \BibitemShut
  {NoStop}%
\bibitem [{\citenamefont {Niemi}\ \emph {et~al.}(2016)\citenamefont {Niemi},
  \citenamefont {Eskola},\ and\ \citenamefont {Paatelainen}}]{Niemi:2015qia}%
  \BibitemOpen
  \bibfield  {author} {\bibinfo {author} {\bibfnamefont {H.}~\bibnamefont
  {Niemi}}, \bibinfo {author} {\bibfnamefont {K.~J.}\ \bibnamefont {Eskola}}, \
  and\ \bibinfo {author} {\bibfnamefont {R.}~\bibnamefont {Paatelainen}},\
  }\href {\doibase 10.1103/PhysRevC.93.024907} {\bibfield  {journal} {\bibinfo
  {journal} {Phys. Rev.}\ }\textbf {\bibinfo {volume} {C93}},\ \bibinfo {pages}
  {024907} (\bibinfo {year} {2016})},\ \Eprint
  {http://arxiv.org/abs/1505.02677} {arXiv:1505.02677 [hep-ph]} \BibitemShut
  {NoStop}%
\bibitem [{\citenamefont {Gale}\ \emph {et~al.}(2013)\citenamefont {Gale},
  \citenamefont {Jeon}, \citenamefont {Schenke}, \citenamefont {Tribedy},\ and\
  \citenamefont {Venugopalan}}]{Gale:2012rq}%
  \BibitemOpen
  \bibfield  {author} {\bibinfo {author} {\bibfnamefont {C.}~\bibnamefont
  {Gale}}, \bibinfo {author} {\bibfnamefont {S.}~\bibnamefont {Jeon}}, \bibinfo
  {author} {\bibfnamefont {B.}~\bibnamefont {Schenke}}, \bibinfo {author}
  {\bibfnamefont {P.}~\bibnamefont {Tribedy}}, \ and\ \bibinfo {author}
  {\bibfnamefont {R.}~\bibnamefont {Venugopalan}},\ }\href {\doibase
  10.1103/PhysRevLett.110.012302} {\bibfield  {journal} {\bibinfo  {journal}
  {Phys. Rev. Lett.}\ }\textbf {\bibinfo {volume} {110}},\ \bibinfo {pages}
  {012302} (\bibinfo {year} {2013})},\ \Eprint {http://arxiv.org/abs/1209.6330}
  {arXiv:1209.6330 [nucl-th]} \BibitemShut {NoStop}%
\bibitem [{\citenamefont {Rezaeian}\ \emph {et~al.}(2013)\citenamefont
  {Rezaeian}, \citenamefont {Siddikov}, \citenamefont {Van~de Klundert},\ and\
  \citenamefont {Venugopalan}}]{Rezaeian:2012ji}%
  \BibitemOpen
  \bibfield  {author} {\bibinfo {author} {\bibfnamefont {A.~H.}\ \bibnamefont
  {Rezaeian}}, \bibinfo {author} {\bibfnamefont {M.}~\bibnamefont {Siddikov}},
  \bibinfo {author} {\bibfnamefont {M.}~\bibnamefont {Van~de Klundert}}, \ and\
  \bibinfo {author} {\bibfnamefont {R.}~\bibnamefont {Venugopalan}},\ }\href
  {\doibase 10.1103/PhysRevD.87.034002} {\bibfield  {journal} {\bibinfo
  {journal} {Phys. Rev.}\ }\textbf {\bibinfo {volume} {D87}},\ \bibinfo {pages}
  {034002} (\bibinfo {year} {2013})},\ \Eprint {http://arxiv.org/abs/1212.2974}
  {arXiv:1212.2974 [hep-ph]} \BibitemShut {NoStop}%
\bibitem [{\citenamefont {Moreland}\ \emph {et~al.}(2016)\citenamefont
  {Moreland}, \citenamefont {Bernhard},\ and\ \citenamefont
  {Bass}}]{trento:code}%
  \BibitemOpen
  \bibfield  {author} {\bibinfo {author} {\bibfnamefont {J.~S.}\ \bibnamefont
  {Moreland}}, \bibinfo {author} {\bibfnamefont {J.~E.}\ \bibnamefont
  {Bernhard}}, \ and\ \bibinfo {author} {\bibfnamefont {S.~A.}\ \bibnamefont
  {Bass}},\ }\href@noop {} {\enquote {\bibinfo {title}
  {{{T\raisebox{-0.5ex}{R}ENTo} initial condition model with nucleon
  substructure}},}\ }\bibinfo {howpublished}
  {{\url{https://github.com/morelandjs/trento-substructure}}} (\bibinfo {year}
  {2016})\BibitemShut {NoStop}%
\bibitem [{\citenamefont {Bernhard}(2015)}]{freestream:code}%
  \BibitemOpen
  \bibfield  {author} {\bibinfo {author} {\bibfnamefont {J.~E.}\ \bibnamefont
  {Bernhard}},\ }\href@noop {} {\enquote {\bibinfo {title} {{Freestreaming and
  Landau matching for boost-invariant hydrodynamic initial conditions}},}\
  }\bibinfo {howpublished} {{\url{https://github.com/Duke-QCD/freestream}}}
  (\bibinfo {year} {2015})\BibitemShut {NoStop}%
\bibitem [{\citenamefont {Song}\ \emph {et~al.}(2016)\citenamefont {Song},
  \citenamefont {Qiu}, \citenamefont {Shen}, \citenamefont {Liu},\ and\
  \citenamefont {Heinz}}]{osuhydro:code}%
  \BibitemOpen
  \bibfield  {author} {\bibinfo {author} {\bibfnamefont {H.}~\bibnamefont
  {Song}}, \bibinfo {author} {\bibfnamefont {Z.}~\bibnamefont {Qiu}}, \bibinfo
  {author} {\bibfnamefont {C.}~\bibnamefont {Shen}}, \bibinfo {author}
  {\bibfnamefont {J.}~\bibnamefont {Liu}}, \ and\ \bibinfo {author}
  {\bibfnamefont {U.}~\bibnamefont {Heinz}},\ }\href@noop {} {\enquote
  {\bibinfo {title} {{The Ohio State University viscous hydrodynamics code for
  relativistic heavy-ion collisions}},}\ }\bibinfo {howpublished}
  {{\url{https://u.osu.edu/vishnu/}}} (\bibinfo {year} {2016}),\ \bibinfo
  {note} {{We use a slightly modified version of the hydro code by J. Bernhard,
  \url{https://github.com/jbernhard/osu-hydro}.}}\BibitemShut {Stop}%
\bibitem [{\citenamefont {Bernhard}(2016)}]{frzout:code}%
  \BibitemOpen
  \bibfield  {author} {\bibinfo {author} {\bibfnamefont {J.~E.}\ \bibnamefont
  {Bernhard}},\ }\href@noop {} {\enquote {\bibinfo {title} {{Particlization
  model (Cooper-Frye sampler) for relativistic heavy-ion collisions}},}\
  }\bibinfo {howpublished} {{\url{https://github.com/Duke-QCD/frzout}}}
  (\bibinfo {year} {2016})\BibitemShut {NoStop}%
\bibitem [{\citenamefont {{{UrQMD} collaboration}}(2015)}]{urqmd:code}%
  \BibitemOpen
  \bibfield  {author} {\bibinfo {author} {\bibnamefont {{{UrQMD}
  collaboration}}},\ }\href@noop {} {\enquote {\bibinfo {title}
  {{Ultra-relativistic quantum molecular dynamics}},}\ }\bibinfo {howpublished}
  {{\url{http://urqmd.org}}} (\bibinfo {year} {2015}),\ \bibinfo {note} {{UrQMD
  is a large, complex model written by many contributors. Please see the
  license file on the official website for more information. We use a slightly
  modified version of the software by J. Bernhard,
  \url{https://github.com/jbernhard/urqmd-afterburner}.}}\BibitemShut {Stop}%
\bibitem [{\citenamefont {Moreland}\ and\ \citenamefont
  {Bernhard}(2017{\natexlab{a}})}]{eventgen:code}%
  \BibitemOpen
  \bibfield  {author} {\bibinfo {author} {\bibfnamefont {J.~S.}\ \bibnamefont
  {Moreland}}\ and\ \bibinfo {author} {\bibfnamefont {J.~E.}\ \bibnamefont
  {Bernhard}},\ }\href@noop {} {\enquote {\bibinfo {title} {{Nuclear collision
  event generator}},}\ }\bibinfo {howpublished}
  {{\url{https://github.com/morelandjs/hic-eventgen}}} (\bibinfo {year}
  {2017}{\natexlab{a}}),\ \bibinfo {note} {{This is a forked and modified
  version of original code written by J. Bernhard.}}\BibitemShut {Stop}%
\bibitem [{\citenamefont {Moreland}\ and\ \citenamefont
  {Bernhard}(2017{\natexlab{b}})}]{bayesian:code}%
  \BibitemOpen
  \bibfield  {author} {\bibinfo {author} {\bibfnamefont {J.~S.}\ \bibnamefont
  {Moreland}}\ and\ \bibinfo {author} {\bibfnamefont {J.~E.}\ \bibnamefont
  {Bernhard}},\ }\href@noop {} {\enquote {\bibinfo {title} {{Bayesian parameter
  estimation of nuclear collisions with nucleon substructure}},}\ }\bibinfo
  {howpublished} {\url{https://github.com/morelandjs/hic-param-est-qm18}}
  (\bibinfo {year} {2017}{\natexlab{b}}),\ \bibinfo {note} {{This is a forked
  and modified version of original code written by J. Bernhard.}}\BibitemShut
  {Stop}%
\end{thebibliography}%

\appendix

\section{Event-by-event grid resizing}
\label{app:adaptive_grid}

\begin{figure}
  \begin{tikzpicture}[scale=.9]
    \draw [black, thin, step=0.3cm] (0.9, 0.9) grid +(4.5, 4.5);
    \draw [gray, thin, step=0.9cm] (0, 0) grid +(6.3, 6.3);
    \draw [theblue] (3.15, 3.15) circle (2.25cm);
  \end{tikzpicture}
  \caption{
    \label{fig:adaptive_grid}
    Diagram of the adaptive grid resizing algorithm (not drawn to scale).
    Each initial condition event is first run on a very large coarse-grained mesh (large gray grid) of one-third the spatial resolution otherwise required to measure hydrodynamic observables.
    We then measure the maximum transverse radius $R_\text{max}$ (blue circle) of the hypersurface defined by the temperature isotherm $T = T(e_\text{min})$, where $e_\text{min}$ is the largest energy density which can be truncated without modifying the hydrodynamic observables calculated from the event.
    Finally, the initial condition event is rerun on a smaller and finer mesh (smaller black grid) with three-times the cell density of the pre-run event and a smaller transverse extent $-R_\text{max} < x < R_\text{max}$.
  }
\end{figure}
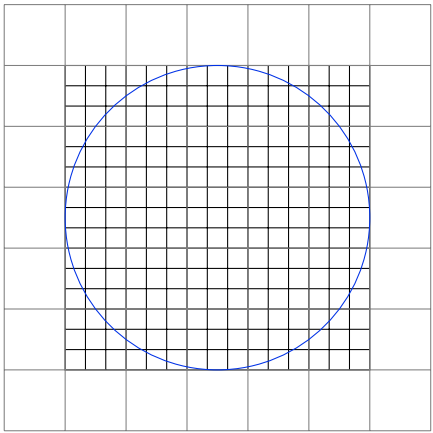

\begin{figure}[t]
  \includegraphics{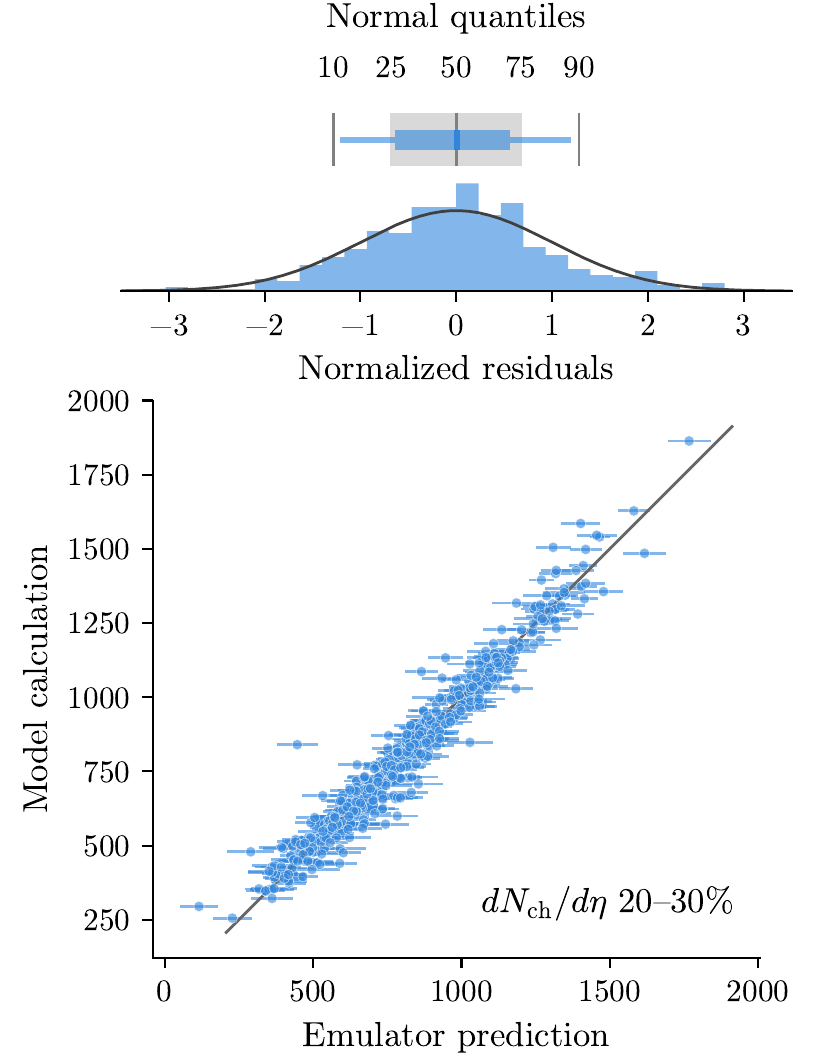}
  \caption{
    \label{fig:validation_example}
    Example emulator validation for one observable, the Pb-Pb charged-particle yield $d\nch/d\eta$ in the 20-30\% centrality class.
    We use the k-fold cross validation method (explained in the text) to partition the model inputs $X$ and outputs $Y$ into training and validation data.
    The scatterplot on the left shows the emulator predictions and one sigma error bars (x-axis) against explicit model calculations (y-axis).
    Perfect emulator/model agreement is indicated by the black like $y_\text{pred}=y_\text{obs}$.
    The histogram on the right shows that the errors are properly accounted for, i.e.\ the normalized residuals follow a normal distribution with unit variance and zero mean.
  }
\end{figure}

The boost-invariant VISH2+1 hydrodynamics code used in this work \cite{Song:2007ux, Shen:2014vra} runs on a Cartesian transverse grid specified by a maximum grid size $x_\text{max}$ and grid step width $dx$ which fix the transverse grid extent ${-x_\text{max} < x < x_\text{max}}$ and number of grid cells along each dimension ${n_x = 2\, x_\text{max} / dx}$.
In general, the maximum grid size $x_\text{max}$ should be set large enough to contain the full spacetime evolution of the event.
This means that the truncation of $T^{\mu\nu}$ at the boundaries of the grid should never modify the final-state observables.
We enforce this requirement by finding an energy density cutoff $e_\text{min}$ for which the matter $e < e_\text{min}$ can be effectively discarded without significantly modifying the simulation observables.
We then fix the maximum grid size $x_\text{max}$ such that it fully encloses the isotherm $T=T(e_\text{min})$ for the full lifetime of the fireball.

We find that we can quickly estimate the maximum radius $R_\text{max} = |\x_\text{max}|^2$ of the spacetime hypersurface ${T = T(e_\text{min})}$ by running the event on a coarse-grained spatial grid with one-third the spatial resolution we would otherwise require to resolve typical hydrodynamic observables such as mean $p_T$ and flows.
The simulation time of a single VISH2+1 event scales like ${\sim}n_x^3$ since $dx \propto d\tau$, and thus our ``pre-run'' event requires only ${\sim}1/27$th the time of a production event.
We therefore run a coarse-grained pre-event on an excessively large grid for \emph{every} minimum-bias event to estimate $R_\text{max}$, then rerun the same event on a thrice finer grid with a trimmed spatial extent $x_\text{max} \equiv R_\text{max}$.
See Fig.~\ref{fig:adaptive_grid} for a simple diagram of the procedure.

In practice, we find that event-by-event grid resizing leads to a massive speed increase for minimum bias events compared to using a single fixed grid for the entire minimum bias sample.
This is because the maximum transverse size of each event varies dramatically, from a few fm in peripheral Pb-Pb collisions to 50 fm or more in central Pb-Pb collisions.
The procedure should generalize to other hydrodynamic codes, including those with $3+1$ spacetime dimensions, where the time savings could be even more pronounced.

\section{Emulator validation}
\label{app:validation}

\begin{figure*}
  \includegraphics{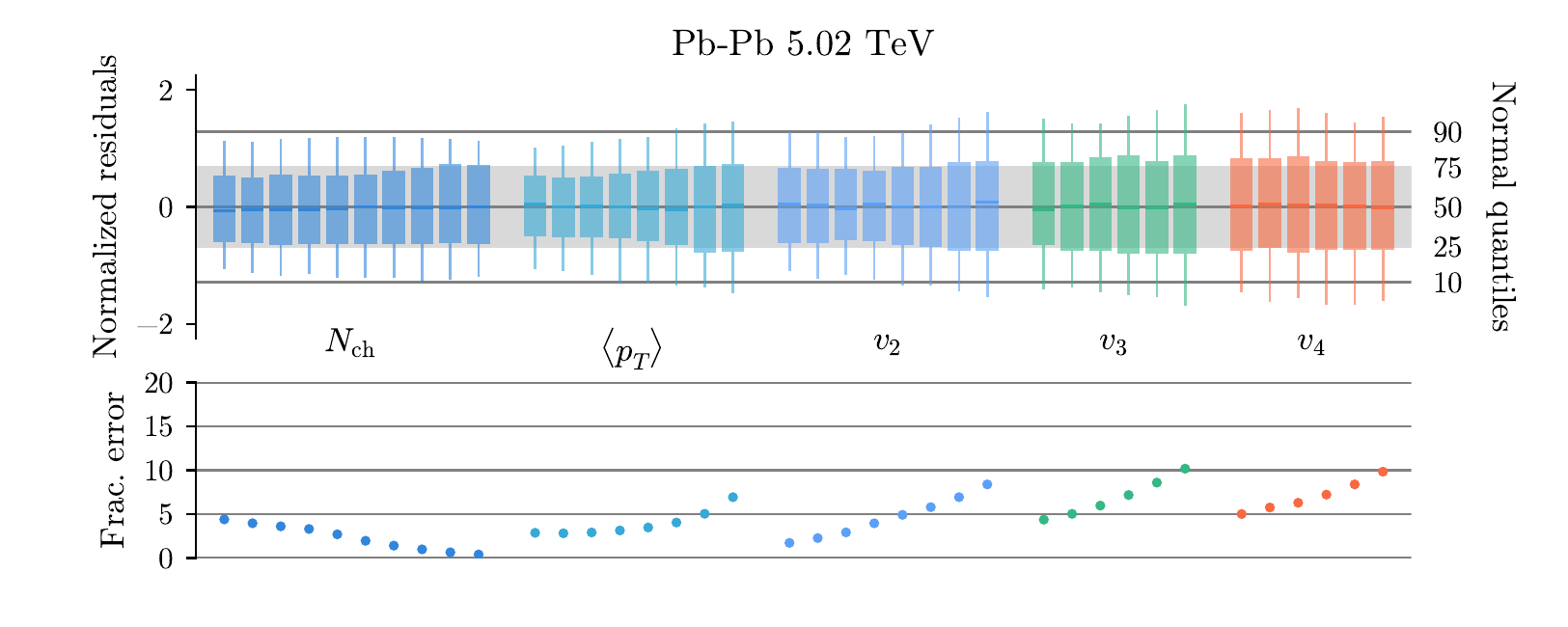}\\
  \includegraphics{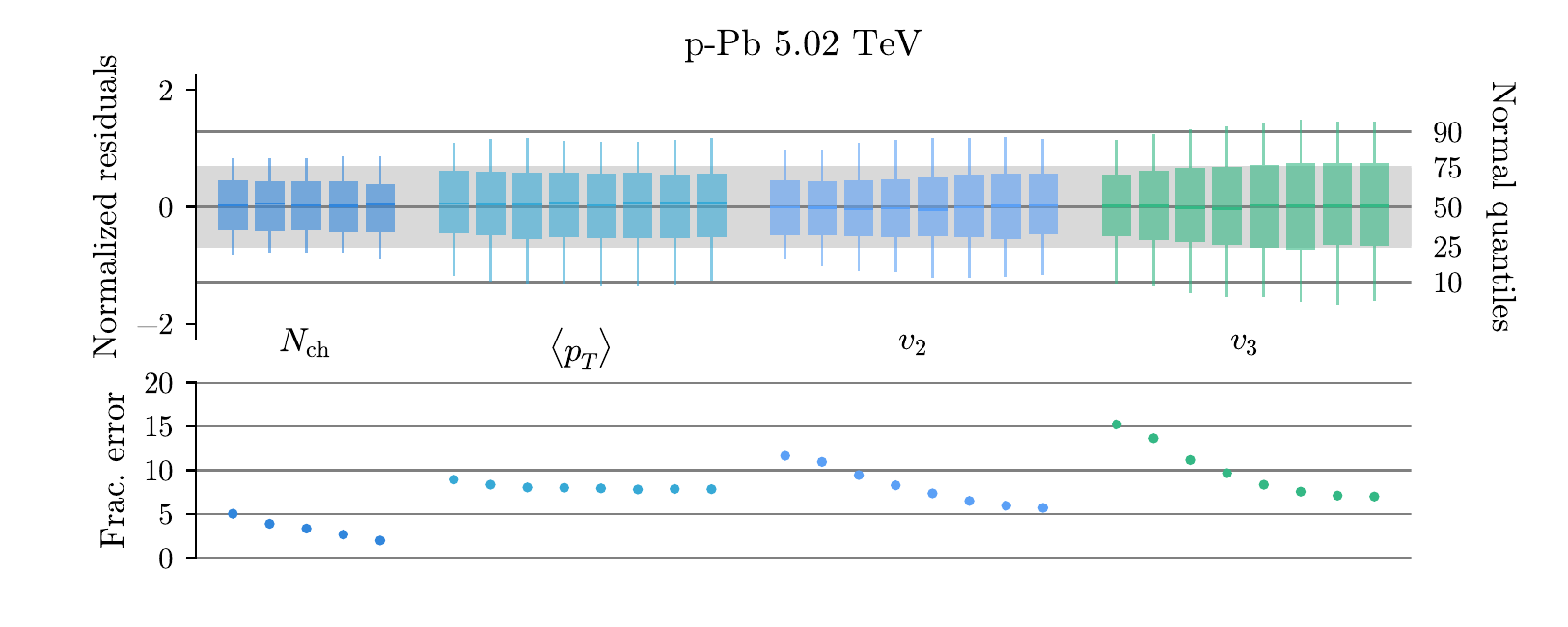}
  \caption{
    \label{fig:validation_all}
    Emulator validation for the Pb-Pb collision system (top) and $p$-Pb collision system (bottom) at $\sqrts=5.02$~TeV.
    The ``piano keys'' in the top row of each figure are horizontally stacked box plots for the normalized residuals of each model observable.
    The boxes are 50\% interquartile ranges and whiskers are the 90\% interquantiles.
    The bottom row of each figure is the RMS fractional error defined by Eq.~\eqref{eq:frac_error}.
  }
\end{figure*}

The emulator is a surrogate for the full physics simulation which generates probabilistic predictions for the model observables $\y_m$ at a given point $\x$.
Here we validate these probabilistic predictions using a method known as k-fold cross validation.
We first randomly partition our $d=500$ training points into $k=20$ equal sized subsamples or ``folds''.
One of the subsamples is used to validate the emulator and the remaining $k-1$ subsamples are used to train it.
The process is then repeated for each of the subsamples so that we end up validating on all of the training data.

Figure~\ref{fig:validation_example} shows a scatterplot of the emulator predictions with one-sigma error bars (x-axis) against explicit model calculations (y-axis).
Perfect emulator and model agreement is indicated by the black line $y_\text{pred} = y_\text{obs}$.
If the emulator errors are properly accounted for, then the normalized residuals ${z=(y_\text{pred} - y_\text{obs})/\sigma_\text{pred}}$ sample a unit normal distribution:
\begin{equation}
  \label{eq:frac_error}
  P(z) \sim \mathcal{N}(\mu=0,\sigma=1).
\end{equation}

This comparison is shown by the histogram and box plot on the right side of Fig.~\ref{fig:validation_example}.
The emulator error is clearly significant, but it is also properly modeled, as indicated by the agreement between the normalized residuals and the unit normal distribution on the right (black curve).
Moreover, since we include this uncertainty in the likelihood covariance matrix \eqref{eq:likelihood}, we expect our results to be robust to the emulator limitations.
This is an important point that bears repeating.
The emulator uncertainty does not erode the veracity of the posterior distribution if it is correctly modeled and accounted for.

More generally, we can perform the validation test in Fig.~\ref{fig:validation_example} for \emph{every} observable $y \in \y_m$ and check that each observable's normalized residuals ${z=(y_\text{pred} - y_\text{obs})/\sigma_\text{pred}}$ follow a unit normal distribution.
This test is applied to the $p$-Pb and Pb-Pb collision systems in Fig.~\ref{fig:validation_all}.
The top row of each figure shows a box-plot for the normalized residuals of each observable compared to the quantiles of a unit normal distribution.
The thin horizontal black lines correspond to the 10th and 90th percentiles of a unit normal distribution, and the gray band its interquartile range.
These visual references should be compared to the whiskers and interquartile range respectively of each box plot, analogous to the comparison test of Fig.~\ref{fig:validation_example}.
The emulators generally behave as expected, although the validation is somewhat better for the Pb-Pb system than the $p$-Pb system.
For instance, the $p$-Pb charged particle yield $d\nch/d\eta$ uncertainties are over predicted.
It is not immediately clear why this would be the case, but the MAP observables in Fig.~\ref{fig:obs_map} are in good agreement with their emulator predictions which suggests it should not be a grave concern.

We also show in Fig.~\ref{fig:validation_all} an estimate of the emulator error magnitude.
This error is expressed in terms of the unitless variable
\begin{equation}
  \hat{z} = \frac{y_\text{pred} - y_\text{obs}}{(\Delta y)_{99\%}},
\end{equation}
where $(\Delta y)_{99\%}$ is 99\% of the full variability of $y$ across the design.
Thus $\hat{z}$ can be thought of as a fractional emulator error relative to the full design variability.
The bottom row of each figure shows the root-mean-square value of $\hat{z}$.
We see that $\text{RMS}\{\hat{z}\}$ ranges from a few percent for most observables to a maximum value of 15\% for the $p$-Pb triangular flow $v_3\{2\}$ in the lowest multiplicity bin.
This suggests that the present analysis would benefit the most from more $p$-Pb events, in particular, from more multiplicity triggered events which are used to calculate the flows.

\end{document}